\documentclass[traditabstract]{aa}
\usepackage{graphicx}
\usepackage{txfonts}
\usepackage{natbib}
\bibpunct{(}{)}{;}{a}{}{,}
\newcommand{\vcas}  {4U~0115+63}
\newcommand{\exo}   {EXO~2030+375}
\newcommand{\bq}    {V0332+53}
\newcommand{\ks}    {KS~1947+300}
\newcommand{\hen}    {1A~1118--616}
\newcommand{\xte}   {XTE~J0658--073}
\newcommand{\mxb}   {XTE~J0658--073}
\newcommand{\uhu}    {4U~1901+03}
\newcommand{\sw}    {Swift~J1656.6--5156}
\newcommand{\gro}    {GRO~J1008--57}
\newcommand{\vtau}    {1A~0535+262}
\newcommand{\bex}    {BeXB}

\def\simless{\mathbin{\lower 3pt\hbox
     {$\rlap{\raise 5pt\hbox{$\char'074$}}\mathchar"7218$}}}   
\def\simmore{\mathbin{\lower 3pt\hbox
     {$\rlap{\raise 5pt\hbox{$\char'076$}}\mathchar"7218$}}}   

\def\msun{~{\rm M}_\odot}

\begin{document}

   \title{Patterns of variability in Be/X-ray pulsars during giant outbursts \\
   }

   \subtitle{}

   \author{P. Reig
          \inst{1,2}
\and
	E. Nespoli\inst{3,4}
          }

\authorrunning{Reig et~al.}
\titlerunning{Patterns of variability in Be/X-ray pulsars}

   \offprints{P. Reig}

   \institute{IESL, Foundation for Research and Technology-Hellas, 71110,
   Heraklion, Greece.
   \email{pau@physics.uoc.gr}
   \and
   	Institute of Theoretical \& Computational 
Physics, University of Crete, PO Box 2208, GR-710 03, Heraklion, Crete,
Greece.
   \and
        Observatorio Astron\'omico de la Universidad de Valencia, 
	46980 Paterna, Valencia, Spain. 
          \email{elisa.nespoli@uv.es}  
   \and   
   	European Space Astronomy Centre (ESA/ESAC), Science Operations 
	Department, Villanueva de la Ca\~nada (Madrid), Spain
	  }

   \date{Received ; accepted}

\abstract{
The discovery of source states in the X-ray emission of black-hole binaries
and neutron-star low-mass X-ray binaries constituted a major step forward
in the understanding of the physics of accretion onto compact objects.
While there are numerous studies on the correlated timing and spectral
variability of these systems, very little work has been done on high-mass
X-ray binaries, the third major type of X-ray binaries. The main goal of
this work is to investigate whether Be accreting X-ray pulsars display
source states and characterise those states through their spectral and
timing properties. We have made a systematic study of the power spectra,
energy spectra and X-ray hardness-intensity diagrams of nine Be/X-ray
pulsars. The evolution of the timing and spectral parameters were monitored
through changes over two orders of magnitude in luminosity. We find that
Be/X-ray pulsars trace two different branches in the hardness-intensity
diagram: the horizontal branch corresponds to a low-intensity state of the
source and it is characterised by fast colour and spectral changes and high
X-ray variability. The diagonal branch is a high-intensity state that
emerges when the X-ray luminosity exceeds a critical limit. The photon
index anticorrelates with X-ray flux in the horizontal branch but
correlates with it in the diagonal branch. The correlation between QPO
frequency and X-ray flux reported in some pulsars is also observed if the
peak frequency of the broad-band noise that accounts for the aperiodic
variability is used. The two branches may reflect two different accretion modes,
depending on whether the luminosity of the source is above or below a
critical value. This critical luminosity is mainly determined by the
magnetic field strength, hence it differs for different sources. }


   \maketitle

\begin{table*}
\caption{Optical and X-ray information of the systems and the outbursts analysed 
in this work. The X-ray luminosity quoted is for the 3--30 keV energy
range.  The critical luminosity depends on the energy of the cyclotron line
\citep[see][and Sect.~\ref{soustat}]{becker12}. A "--" sign indicates that
either the parameter has not been detected or it is not known.}
\label{info}      
\centering          
\begin{tabular}{@{~~}l@{~~}c@{~~}c@{~~}c@{~~}c@{~~}c@{~~}c@{~~}c@{~~}c@{~~}c@{~~}c@{~~}c@{~~}l}
\hline\hline
Source		&Spectral&P$_{\rm spin}$ &P$_{\rm orb}$ &$e$ &$E_{\rm cyc}$	&QPO	&Distance &Outburst	&$L_{x}^{\rm peak}$ &$L_{x}^{\rm peak}$/$L_{\rm Edd}$	&$L_{x}^{\rm peak}$/$L_{\rm crit}$ &References  \\
name		&type		&(s)   	&(days)		&	&(keV)		&(mHz)	&(kpc)	&duration (d) 	&(erg s$^{-1}$)     \\
\hline
V 0332+53	&O8-9V		&4.4	&34.2	&0.42 	&25	&51,223	&7	&100	&$3.4\times10^{38}$	&2.0	&10		&1,2,3	\\
EXO 2030+375	&B0.5III-V	&41.8	&46.0	&0.41	&11	&200	&7.1	&155	&$1.6\times10^{38}$	&0.9	&11		&4,5,6\\
4U 0115+63	&B0.2V		&3.6	&24.3	&0.34 	&12	&2,62	&8.1	&55	&$1.4\times10^{38}$	&0.8	&9		&7,8,9\\
KS 1947+300	&B0V		&18.7	&40.4	&0.03 	&--	&20	&10	&165	&$7.1\times10^{37}$	&0.4	&$\sim7^{\dag}$	&10,11\\
1A 0535+262	&B0III		&105	&111	&0.47	&46	&27-72	&2.4	&45	&$6.8\times10^{37}$	&0.4	&$\simless$1	&12,13,14\\
Swift J1626.6--5156&B0V		&15.4	&133	&0.08	&--	&--	&$\sim$10&$\sim$100&$5.2\times10^{37}$	&0.3	&--		&15,16,17\\
XTE J0658--073	&O9.7V		&160.7	&101	&--	&33	&--	&3.9	&105	&$3.6\times10^{37}$	&0.2	&$\simless$1	&18,19,20\\
1A 1118--616	&O9.5III-V	&405	&24	&0.0	&60	&80	&5	&30	&$2.9\times10^{37}$	&0.2	&0.4		&21,22,23,24\\
GRO J1008--57	&B1-B2III-V	&93.5	&247.8	&0.68	&--	&--	&5	&50	&$2.0\times10^{37}$	&0.1	&--		&25,26,27\\
\hline\hline
\multicolumn{13}{l}{$\dag$:  Estimated from the break observed in the $\Gamma-L_x$ diagram, and not from $E_{\rm cyc}$ (see Sect.~\ref{soustat}).} \\
\multicolumn{2}{l}{[1] \citet{raichur10}} & \multicolumn{4}{l}{[2] \citet{negueruela99}}& \multicolumn{3}{l}{[3] \citet{qu05}}		& \multicolumn{3}{l}{[4] \citet{angelini89}} \\
\multicolumn{2}{l}{[5] \citet{wilson02}} & \multicolumn{4}{l}{[6] \citet{wilson08}} 	& \multicolumn{3}{l}{[7] \citet{li12}}		& \multicolumn{3}{l}{[8] \citet{reigetal07}}	\\
\multicolumn{2}{l}{[9] \citet{heindl99}} & \multicolumn{4}{l}{[10] \citet{negueruela03}} & \multicolumn{3}{l}{[11] \citet{galloway04}}& \multicolumn{3}{l}{[12] \citet{caballero07}}	\\
\multicolumn{2}{l}{[13] \citet{clark98}} & \multicolumn{4}{l}{[14] \citet{finger96}} 	& \multicolumn{3}{l}{[15] \citet{baykal10}} 	& \multicolumn{3}{l}{[16] \citet{reig11b}}  \\
\multicolumn{2}{l}{[17] \citet{icdem11}} & \multicolumn{4}{l}{[18] \citet{mcbride06}} 	& \multicolumn{3}{l}{[19] \citet{yan12}} 	& \multicolumn{3}{l}{[20] \citet{nespoli12}}   \\
\multicolumn{2}{l}{[21] \citet{coe94a}}& \multicolumn{4}{l}{[22] \citet{doroshenko10}} 	& \multicolumn{3}{l}{[23] \citet{staubert11}} & \multicolumn{3}{l}{[24] \citet{nespoli11}} \\
\multicolumn{2}{l}{[25] \citet{coe94b}}& \multicolumn{4}{l}{[26] \citet{shrader99}}	& \multicolumn{3}{l}{[27] \citet{naik11}} 	&\\
\end{tabular}
\end{table*}
   \begin{figure*}
   \centering
   \begin{tabular}{ccc}
   \includegraphics[width=5cm]{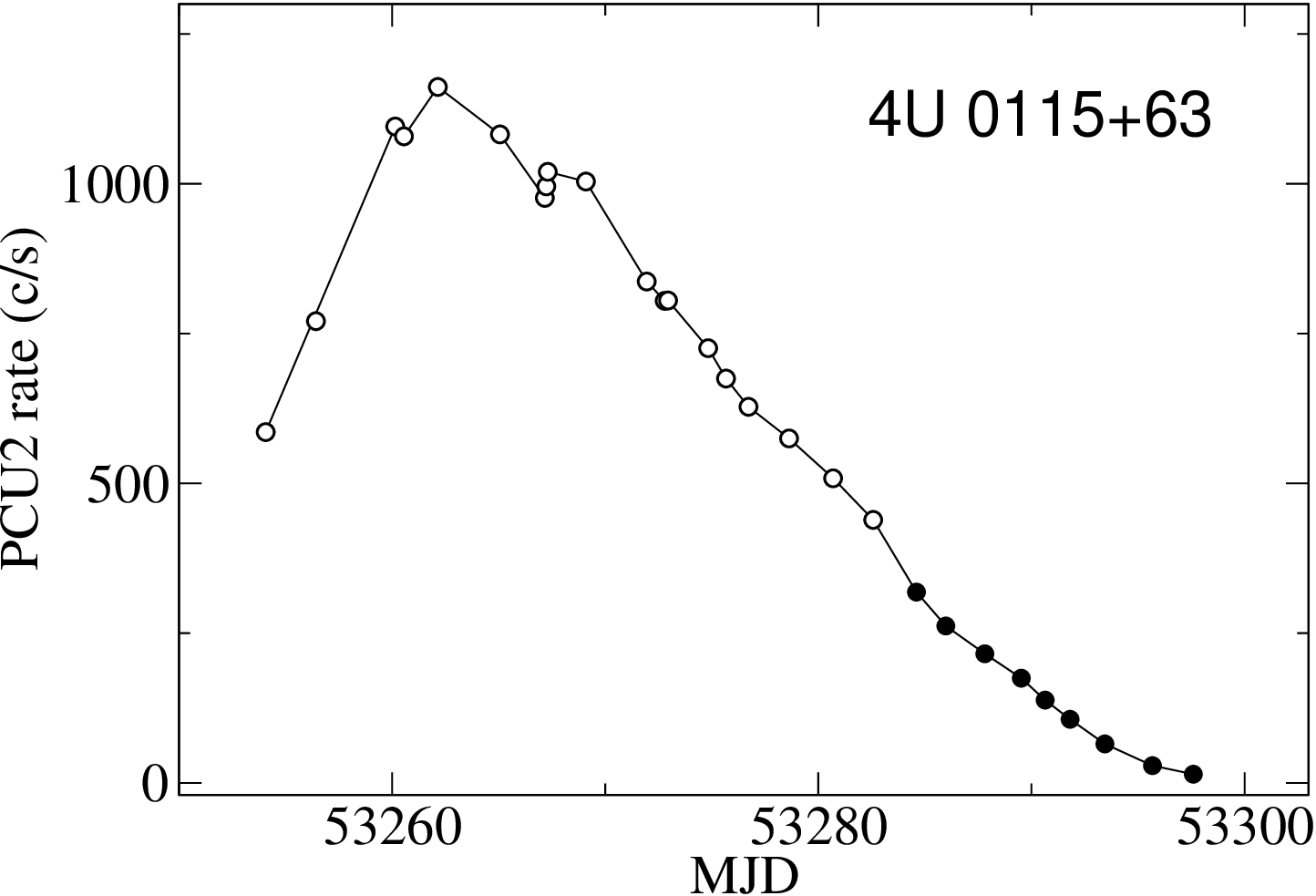}  &
   \includegraphics[width=5cm]{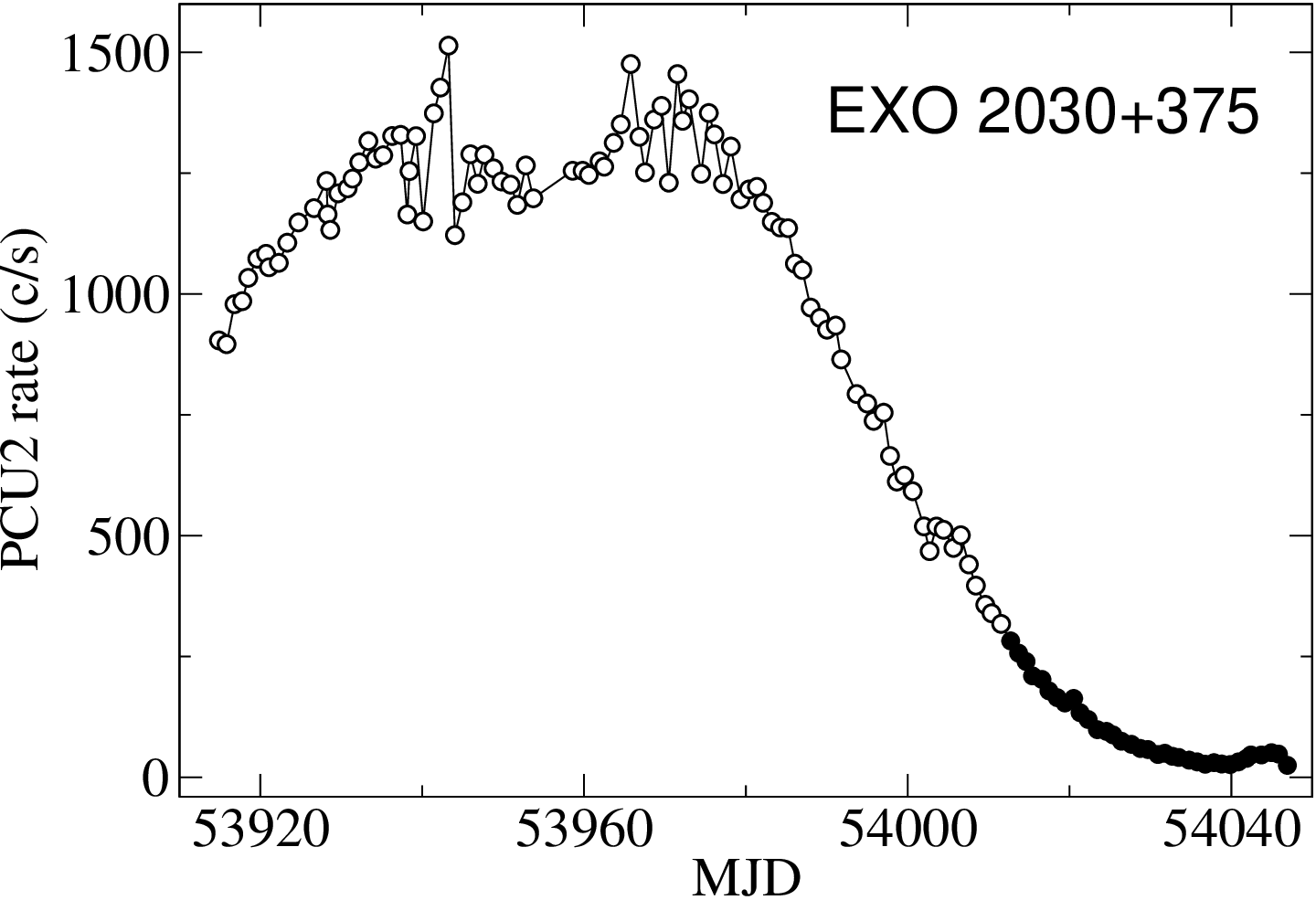} &
   \includegraphics[width=5cm]{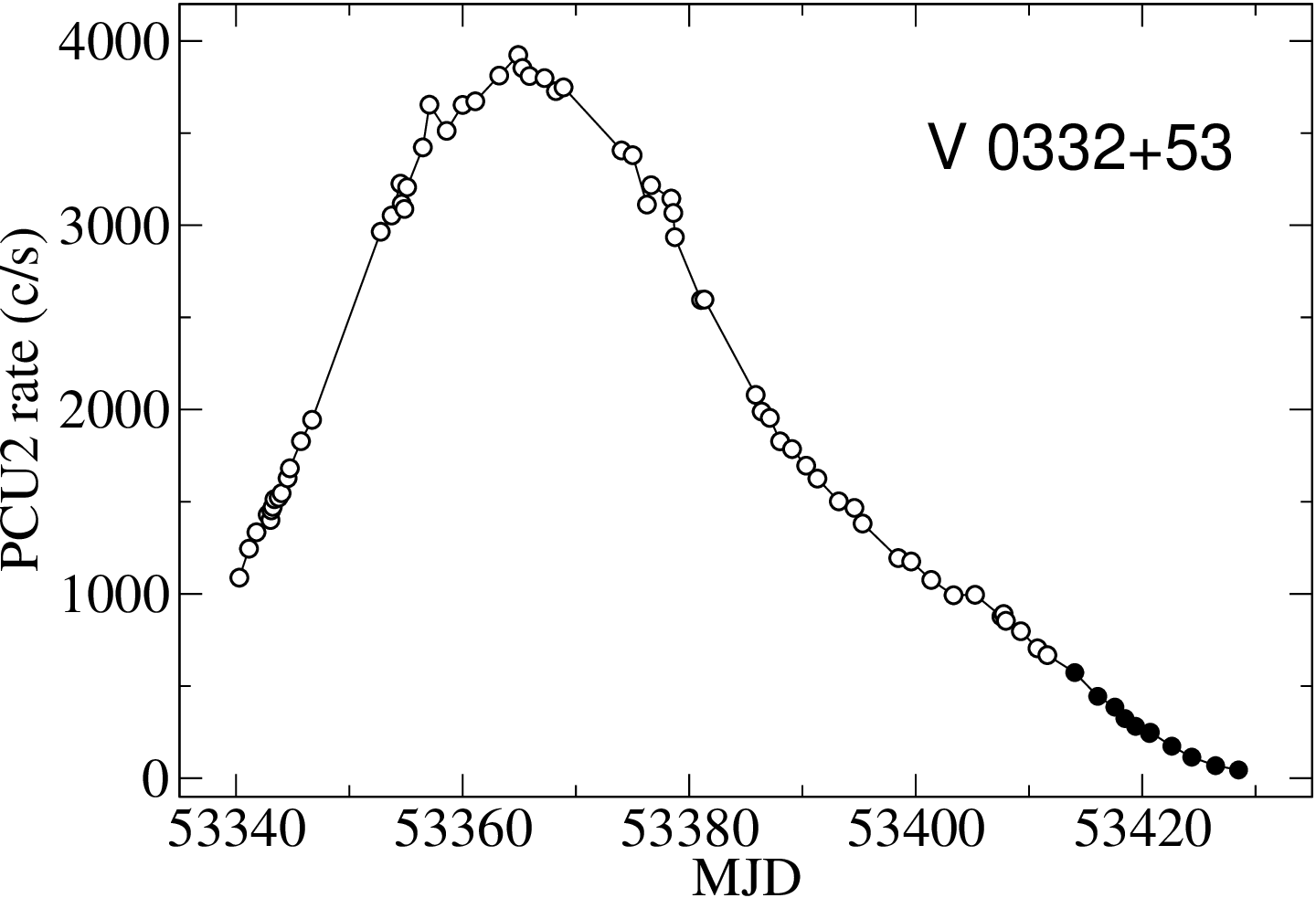}  \\
   \includegraphics[width=5cm]{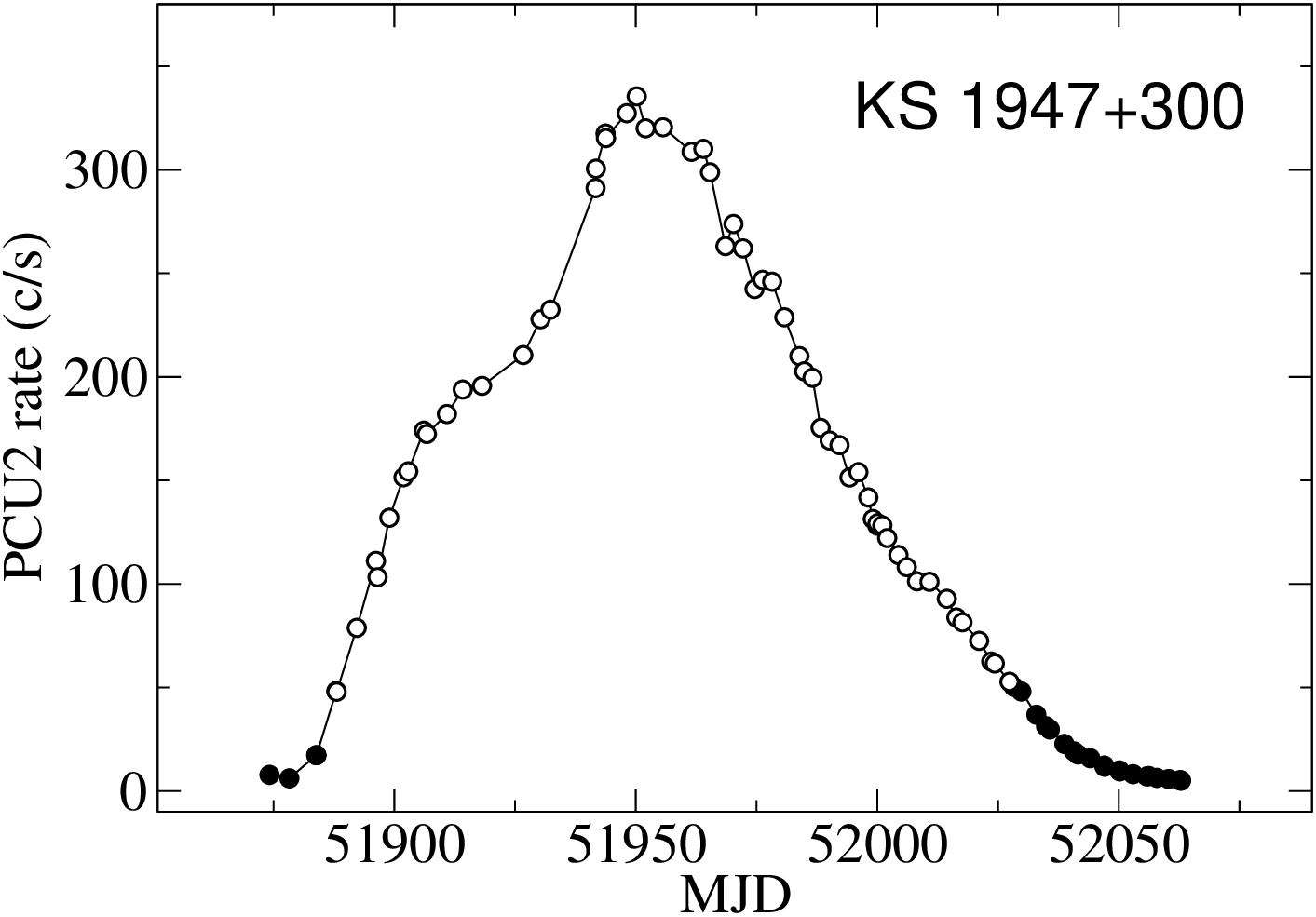}  &
   \includegraphics[width=5cm]{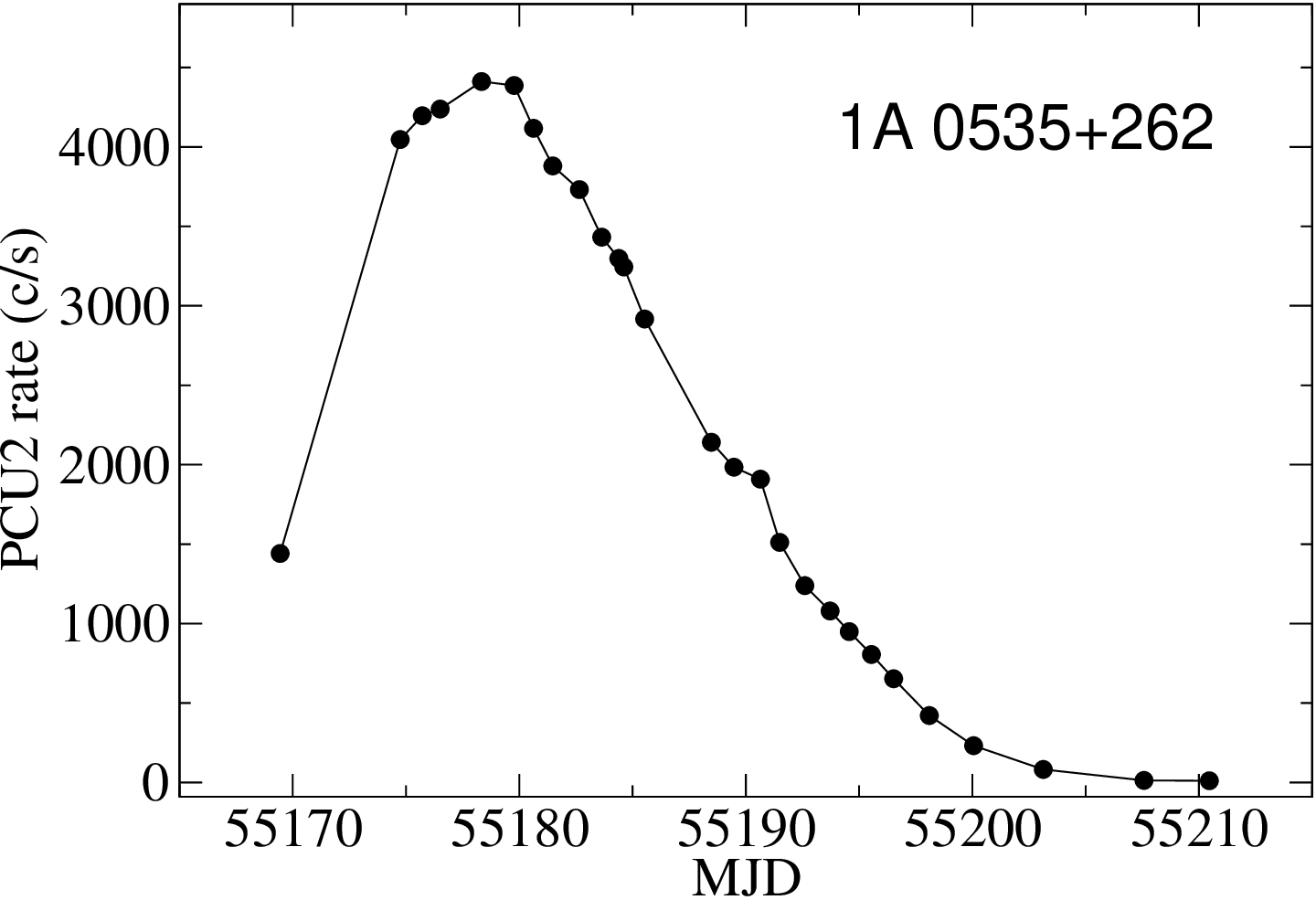}  &
   \includegraphics[width=5cm]{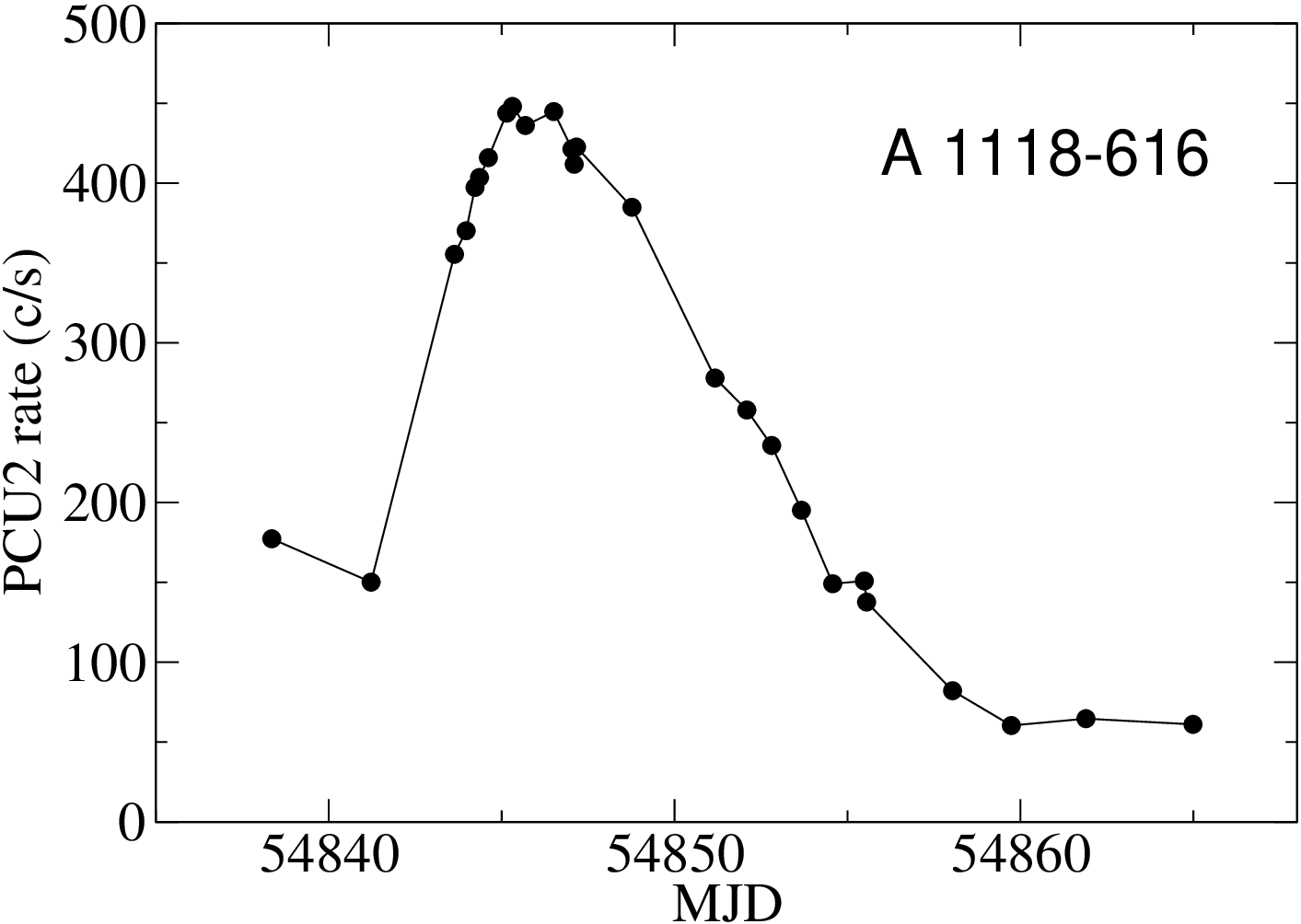}   \\
   \includegraphics[width=5cm]{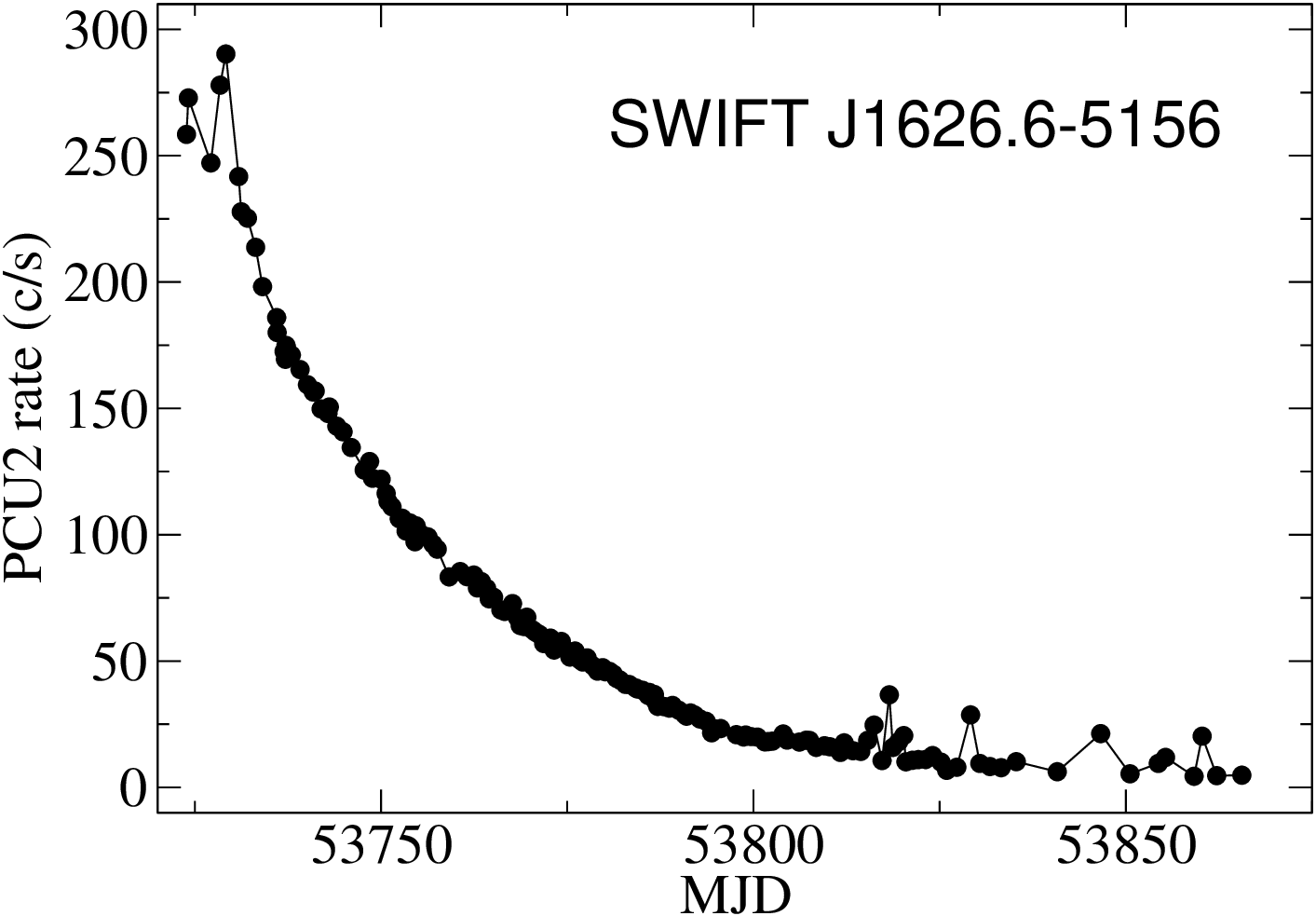}   &  
   \includegraphics[width=5cm]{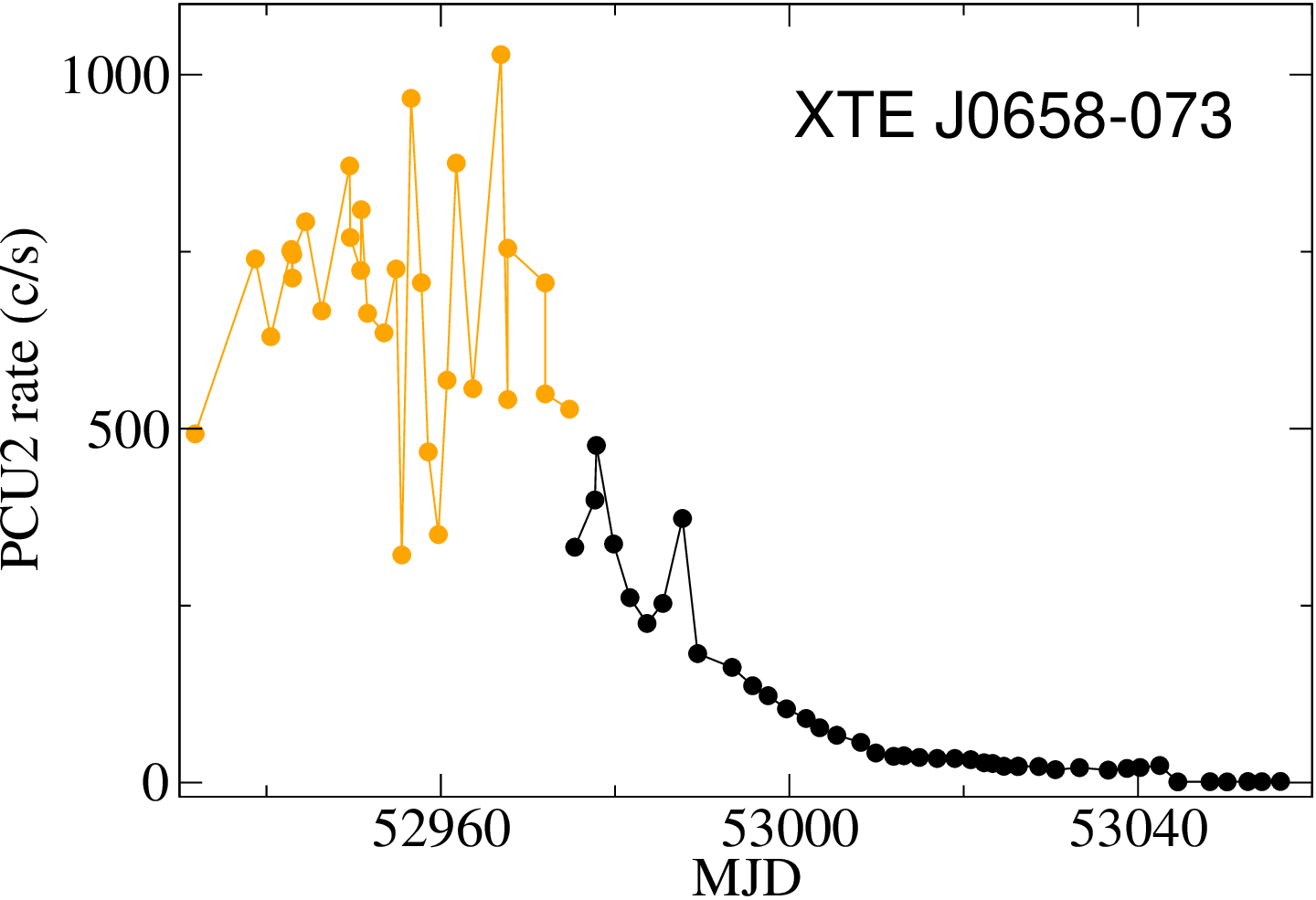}  &
   \includegraphics[width=5cm]{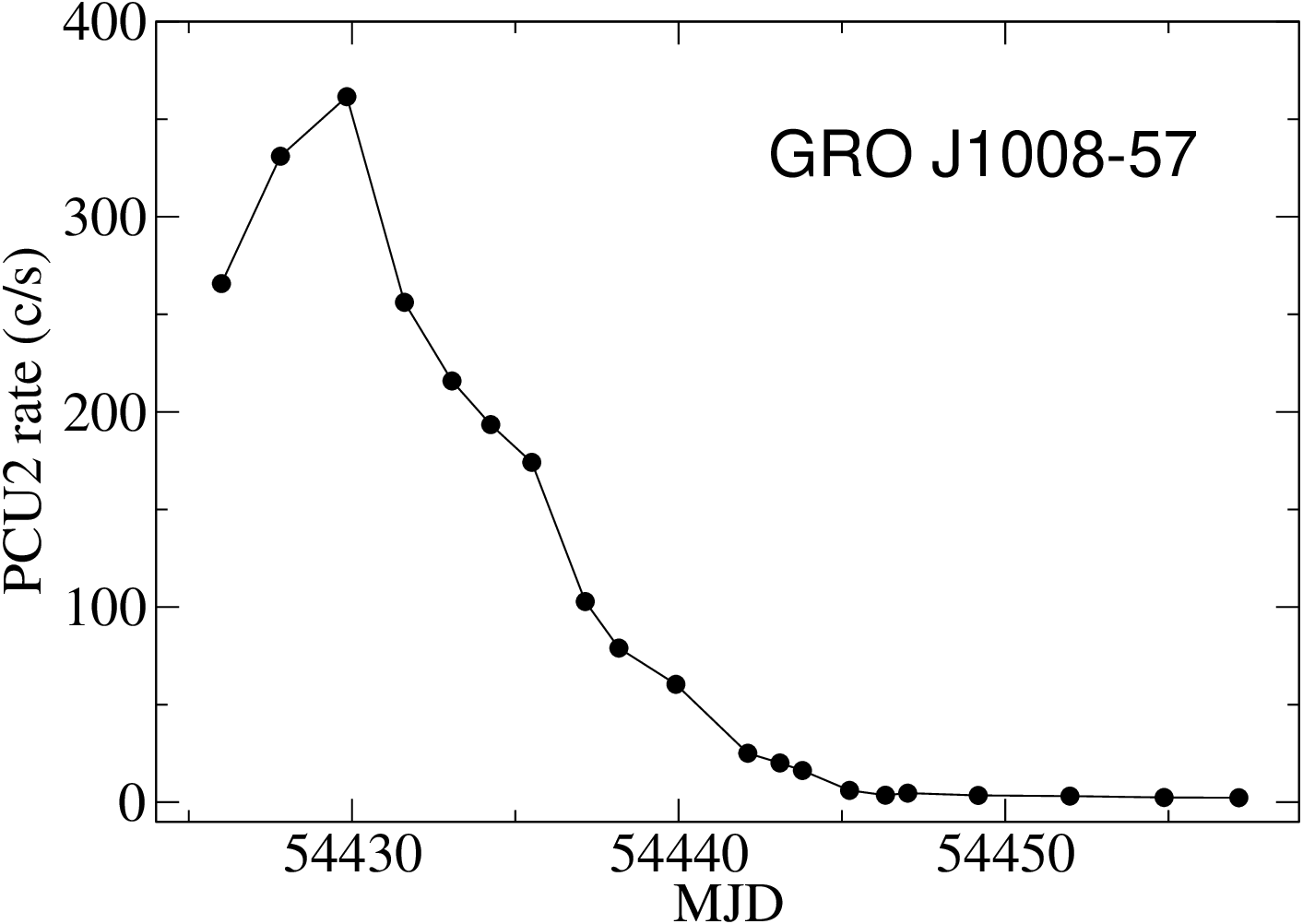}   \\   
   \end{tabular}
      \caption{The outburst profiles as seen by {\it RXTE}/PCA. Each point
      represents the average over an observation interval (typically 
      1000-3000 s). The count rate was obtained using the PCU2 in the energy 
      band 4--30 keV. Different symbols denote different source states: open 
      circles represent the diagonal branch (DB), while filled circles 
      correspond to the horizontal branch (HB). The flaring activity during 
      the peak of the outburst in \xte\ was analysed separately from the 
      smoother decay.}
  \label{rate-time}
  \end{figure*}

\section{Introduction}

The definition of {\em source states} in low-mass X-ray binaries (LMXB) and
black-hole binaries (BHB) has been a very useful way to describe the rich
phenomenology exhibited by these systems in the X-ray band  
\citep{hasinger89,muno02,gierlinski02,homan05,klis06,klein08,kording08,dunn10,belloni10}.
A state is defined by the appearance of a spectral (e.g., power-law,
blackbody) or variability component (e.g., Lorentzian) associated with a
particular and well-defined position of the source in the colour-colour
(CD) or hardness-intensity (HID) diagram. The physical interpretation of
these components ({\it i.e.}, thermal and non-thermal Comptonisation,
emission from an accretion disc, quasi-periodic oscillations, etc.) allows
the formulation of theoretical models and their confrontation with the
observations. Different theoretical models differ in the details of the
physical location and combination of these components.

As the amount of data provided by X-ray space missions increased, so did
the complexity of the observed phenomenology. Although the real picture can
be quite complex, two are the basic states found in neutron-star and
black-hole binaries: a low-intensity, spectrally-hard state and a
high-intensity, spectrally-soft state.  The current understanding is that
the two states reflect the relative contribution of the accretion disc on
the X-ray emission \citep{done07,lin07}.  When the disc emission
dominates, the soft state is seen. Matter in the disc moves in
near-Keplerian orbits. The innermost stable orbit of an accretion disc
around a non-rotating black hole is $\sim3 R_{\rm Schw}$ or $\sim$ 100 km
for a 10 $\msun$ black hole, which is of the same order of magnitude as the
magnetosphere of a weakly magnetized ($B\simless 10^9$ G)  neutron star.
Therefore, the accretion flows around a low-magnetic neutron star and a
stellar-mass black hole in an X-ray binary are expected to be very similar
\citep{klis94}. 

On the other hand, the accretion flow around a strongly magnetized ($B\sim
10^{12}$ G) neutron star may be very different because in this case the
magnetic field begins to dominate the dynamics of the accreting flow at
much larger distance ($r\gg R_{\rm Schw}$) and the details of how the
magnetic field affects the accretion flow are not fully understood 
\citep{becker07}. High magnetic field neutron stars are found in binary
systems orbiting a massive (spectral type late O or early B) companion. 

While there have been numerous studies on the characterisation of source
states in terms of timing and spectral parameters in LMXBs \citep[][and
references therein]{klis06} and BHBs \citep[][and references
therein]{belloni10}, very little work of this kind has been done on
high-mass X-ray binaries (HMXB), despite the fact that they represent a
significant fraction of the galactic X-ray binary population. Although the
discovery of rapid aperiodic variability (flickering) in the Be X-ray
pulsar \bq\ dates back to EXOSAT times \citep{stella85}, timing analysis on
accreting pulsars have concentrated on the properties and variability of
the coherent X-ray emission (pulsations) and the detection of
quasi-periodic oscillations  \citep[][and references therein]{james10}.
In contrast, the study of the aperiodic variability and broad-band noise
components is very scarce  \citep{belloni90a,revnivtsev09}.


Massive X-ray binaries are classified according to the luminosity class of
the optical component into Be/X-ray binaries (\bex), in which the optical
companion is a dwarf or subgiant star, and supergiant (luminosity class
I-II) X-ray binaries. The former tend to be transient systems while the
latter are persistent sources. Be stars are non-supergiant fast-rotating, 
B-type and luminosity class III-V stars which at some point of their lives
have shown spectral lines in emission \citep{port03}. In the infrared, they
are brighter than their non-emitting  counterparts of the same spectral
type. The line emission and infrared excess originate in extended
circumstellar envelopes of ionized gas surrounding the equator of the B
star \citep[see][for a recent review]{reig11a}. 

\bex s represent one of the most extreme case of X-ray variability in HMXBs, 
displaying changes in X-ray intensity that span up to four orders of
magnitude. In the X-ray band, \bex s are most of the time in a quiescence
state, below the detection level of X-ray detectors. When active, the X-ray
variability of \bex s is characterised by two type of outbursts:

\begin{table}
\caption{Journal of the RXTE observations.}             
\label{xobs}      
\centering          
\begin{tabular}{@{~~}l@{~~}c@{~~}c@{~~}c@{~~}c}
\hline\hline
Source		&Proposal&Time range	&Num.	&Exp. \\
name		&ID	&MJD		&obs 	&time (ks) \\
\hline
4U 0115+63	&90089	&53254.1--53280.3	&19	&58.8	  \\		
		&90014	&53282.6--53304.6	&17	&37.6	  \\
KS 1947+300	&50425	&51874.2--51968.6	&33	&82.0	  \\	
		&60402	&51970.2--52078.0	&53	&62.9	  \\
EXO 2030+375	&91089	&53914.9--53998.7	&83	&83.5	  \\	
		&92067	&53999.6--54069.9	&71	&56.6	  \\
V 0332+53	&90089	&53340.3--53365.9	&30	&82.2	  \\	
		&90427	&53367.2--53376.6	&8	&13.1	  \\
		&90014	&53378.4--53430.5	&42	&79.2	  \\
1A 1118--616	&94412	&54838.3--54838.4	&1	&1.3	  \\
		&94032	&54841.2--54865.0	&25	&8.9	  \\
XTE J0658--073	&80067	&52932.7--52975.4	&31	&128.0	  \\	
		&80430	&52965.7--53056.4	&45	&73.9	\\
Swift J1656.6--5156&91094&53723.9--53736.0	&11	&28.9	  \\
		&91081	&53736.0--53766.6	&18	&68.8	  \\
		&91082	&53747.7--53835.3	&119	&111.1	  \\
		&92412	&53840.8--54269.9	&112	&104.2	  \\
1A 0535+262	&94323	&55169.4--55210.5	&27	&126.7	  \\
GRO J1008--57	&93032	&54426.0--54447.0	&16	&21.2	\\
		&93423	&54449.2--54457.2	&4	&6.3	\\
\hline
\end{tabular}
\end{table}

\begin{itemize}

\item Type I outbursts are regular and (quasi)periodic events,
normally peaking at or close to periastron passage of the neutron star.
They are short-lived, {\it i.e.}, tend to cover a relatively small fraction of
the orbital period (typically 0.2-0.3 $P_{\rm orb}$). The peak X-ray
luminosity during this type of outbursts is typically $\simless 0.2$ the
Eddington luminosity for a neutron star ($L_{\rm Edd}= 1.7 \times 10^{38}$\,
 erg s$^{-1}$)

\item Type II outbursts represent major increases of the X-ray flux. The
luminosity at the peak of the outburst is close to the Eddington limit.
These outbursts do not show any preferred orbital phase and last for a
large fraction of an orbital period or even for several orbital periods.
The presence of quasi-periodic oscillations and the large and steady
spin-up rates measured during giant outbursts support the formation of
an accretion disc during type II outbursts.

\end{itemize}

   \begin{figure}
   \centering
   \includegraphics[width=8cm]{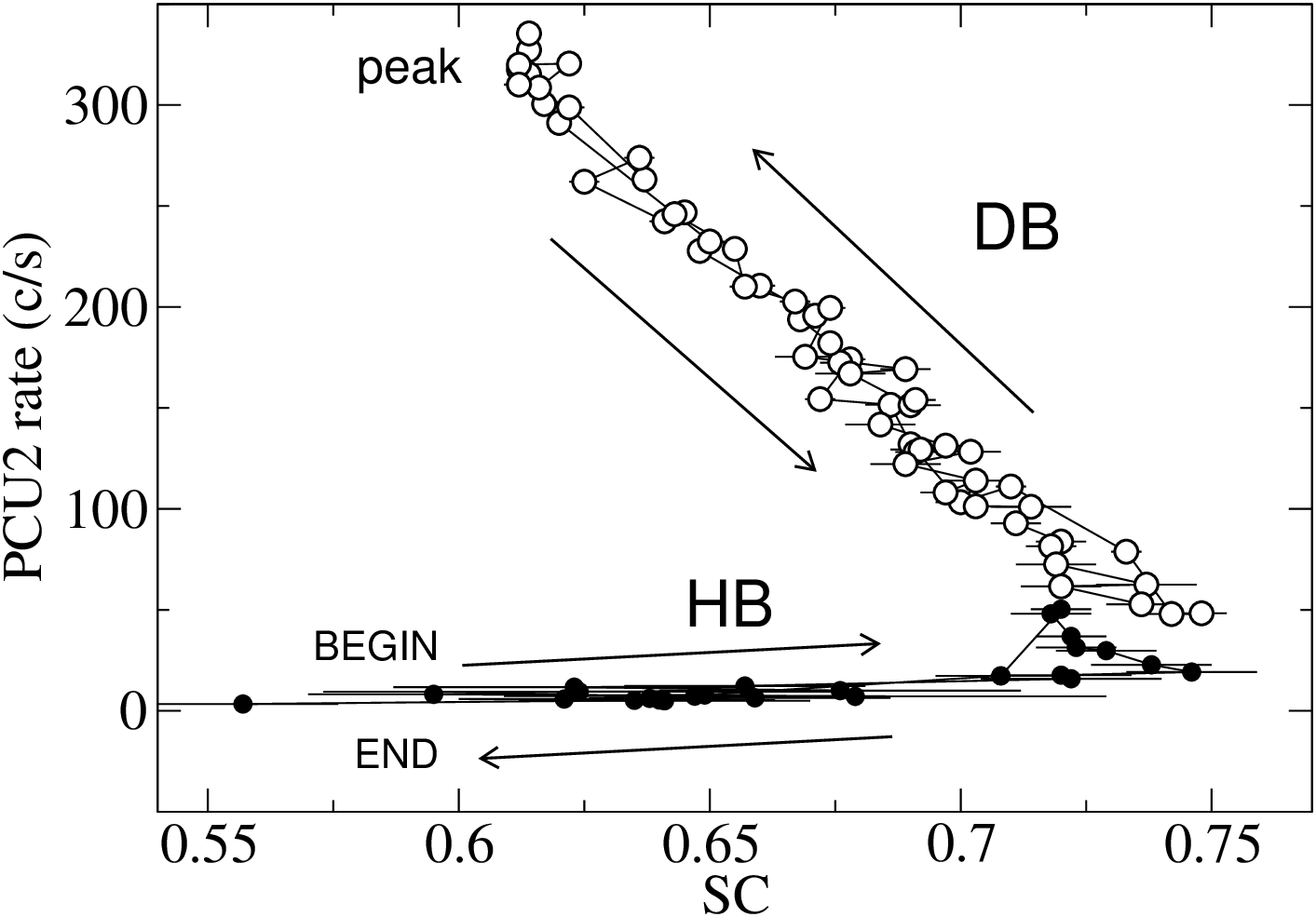}
  \caption{Hardness-intensity diagram of \ks\ showing two spectral branches: 
  a low-intensity horizontal branch (HB, filled circles) and a high-intensity 
  diagonal branch (DB, open circles). Arrows mark the flow of time.}
  \label{hid}
  \end{figure}

The purpose of this work is to investigate the timing and spectral
variability in \bex s that display large X--ray intensity changes (type II
outbursts). Because virtually all \bex s are pulsars, this work seeks to
shed light on the X-ray phenomenology of accretion-powered pulsars.

An attempt to define source states  through CD/HID and timing analysis was
carried out by \citet[][hereafter Paper I]{reig06} on the hard transient
X-ray pulsar \bq. Subsequently, this kind of analysis was extended to three
other sources: \exo, \vcas, and \ks\ \citep[][hereafter Paper II]{reig08b}
in an attempt to generalise the results of Paper I. These works clearly
identified the existence of two spectral branches that were called
horizontal (HB) and diagonal (DB) branches according to the motion of the
source in the HID. The HB is associated with low-flux and highly variable
states, whereas the DB  appears at high-flux states. 

This paper builds up on the results of Paper II and constitutes a
systematic study of the evolution of the spectral and timing variability of
all the \bex\ detected by {\it RXTE} that went into type II outburst in the
period 1996-2011. Based on the evolution and correlations of the spectral
and timing parameters and the position of the sources in the CD/HID
diagrams, we aim to define and characterise source states in X-ray pulsars.
The main difference with respect to Paper II, besides doubling the number
of systems investigated, is the inclusion of broad-band spectral analysis
and the finer sampling of the data. In Paper II, the X-ray outbursts were
divided into several (between 7--9) intervals and averaged power spectra
for each interval were extracted. Each interval covered typically a few
days. Here, we obtain power spectra {\em and} energy spectra (PCA+HEXTE)
for each pointing, which is typically few thousand seconds long. This order
of magnitude increase in sampling resolution allows us to track much faster
changes, likely to be associated with accretion, whereas the acquisition of
energy spectra allows the search for correlations between the spectral and
timing parameters.

Section~\ref{obs} summarizes the observations and the characteristics of the
instruments used in the analysis. Section~\ref{red} describes the methodology
and techniques used in the data analysis. A detailed description of the
results is given in Sect.~\ref{res}. Our interpretation of the results and
the implications of this work are discussed in Sect.~\ref{discussion}.   
Finally, conclusions are drawn in Sect.~\ref{con}.

   \begin{figure*}
   \centering
   \begin{tabular}{ccc}
   \includegraphics[width=5cm]{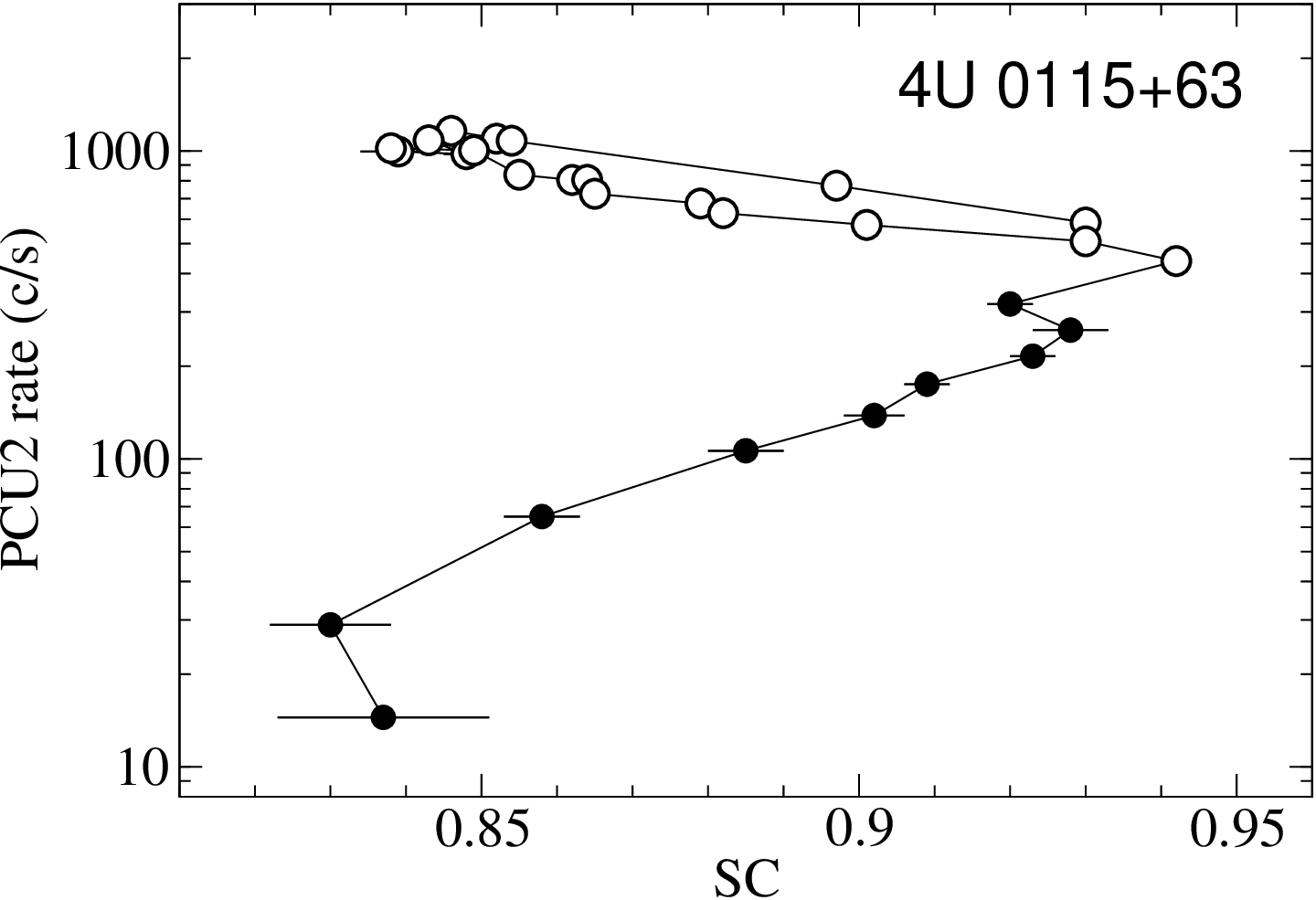}  &
   \includegraphics[width=5cm]{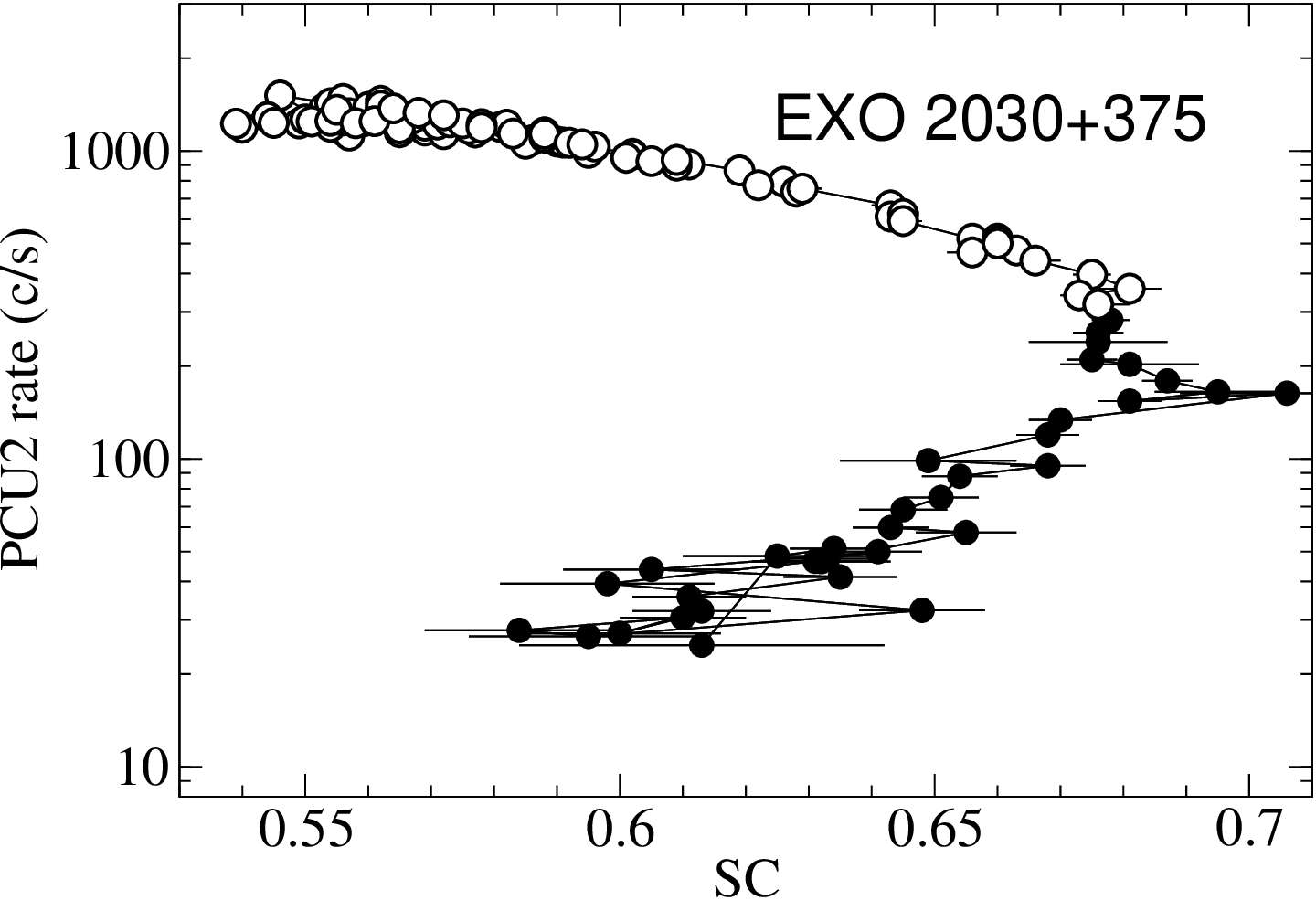} &
   \includegraphics[width=5cm]{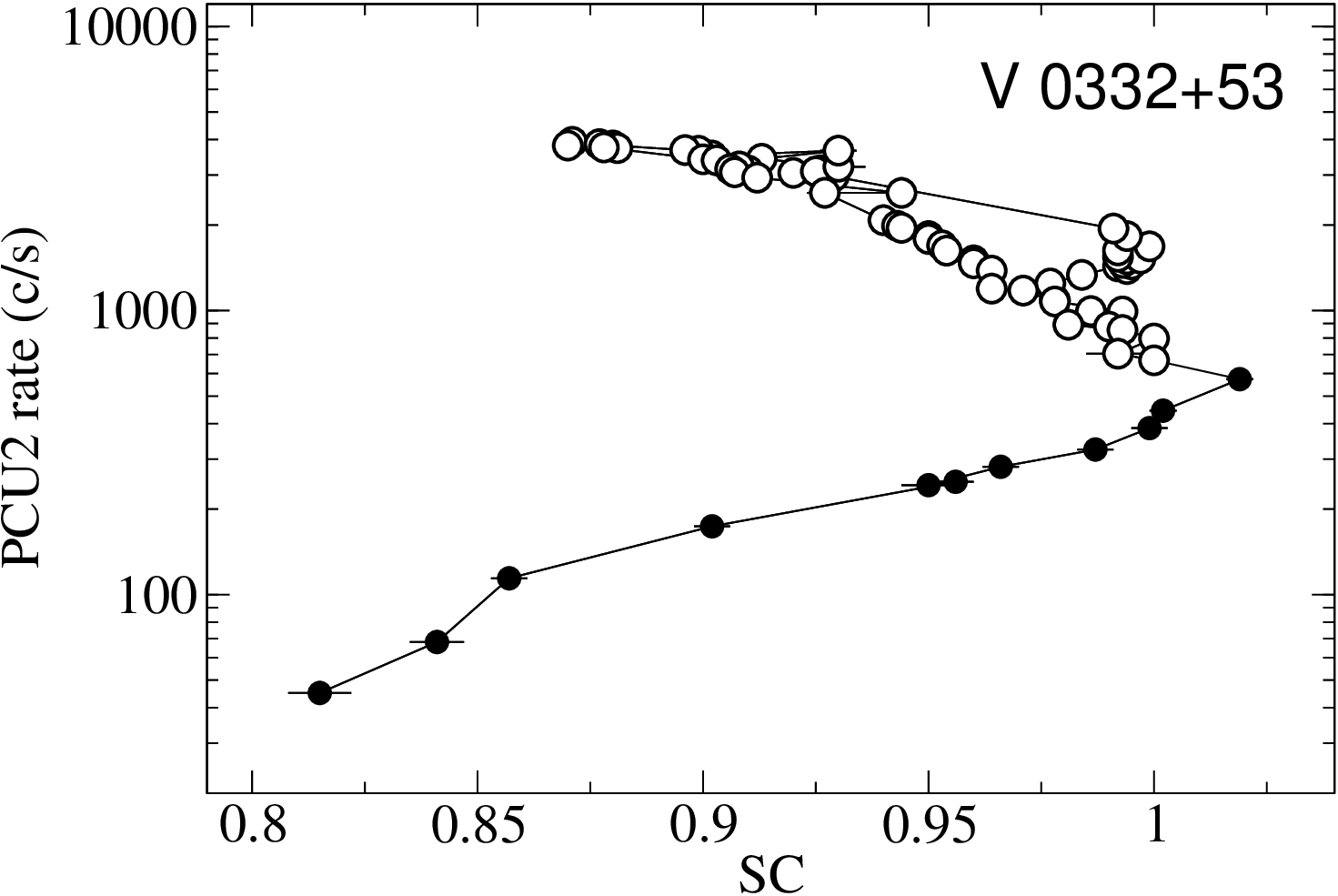}  \\
   \includegraphics[width=5cm]{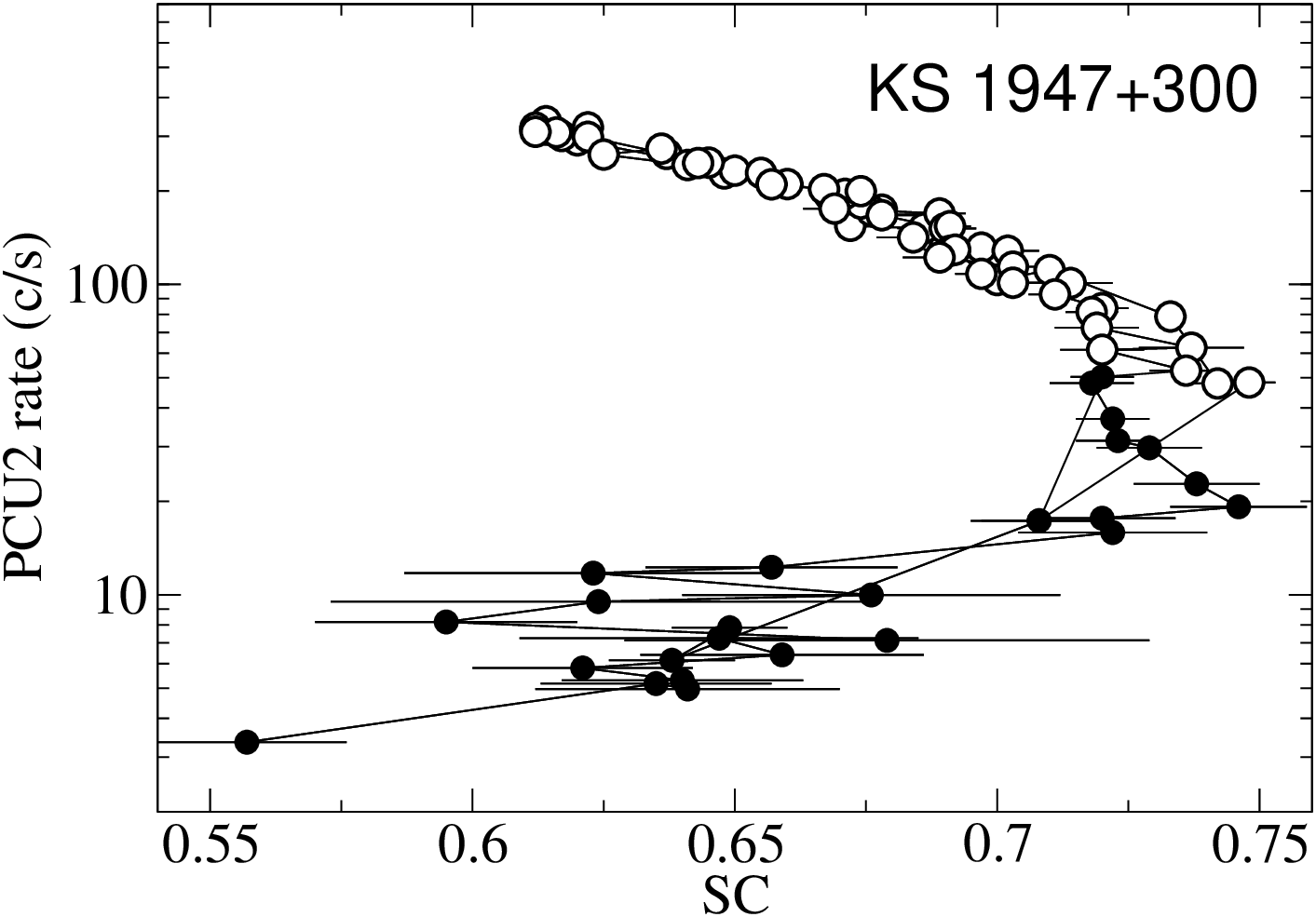}  &
   \includegraphics[width=5cm]{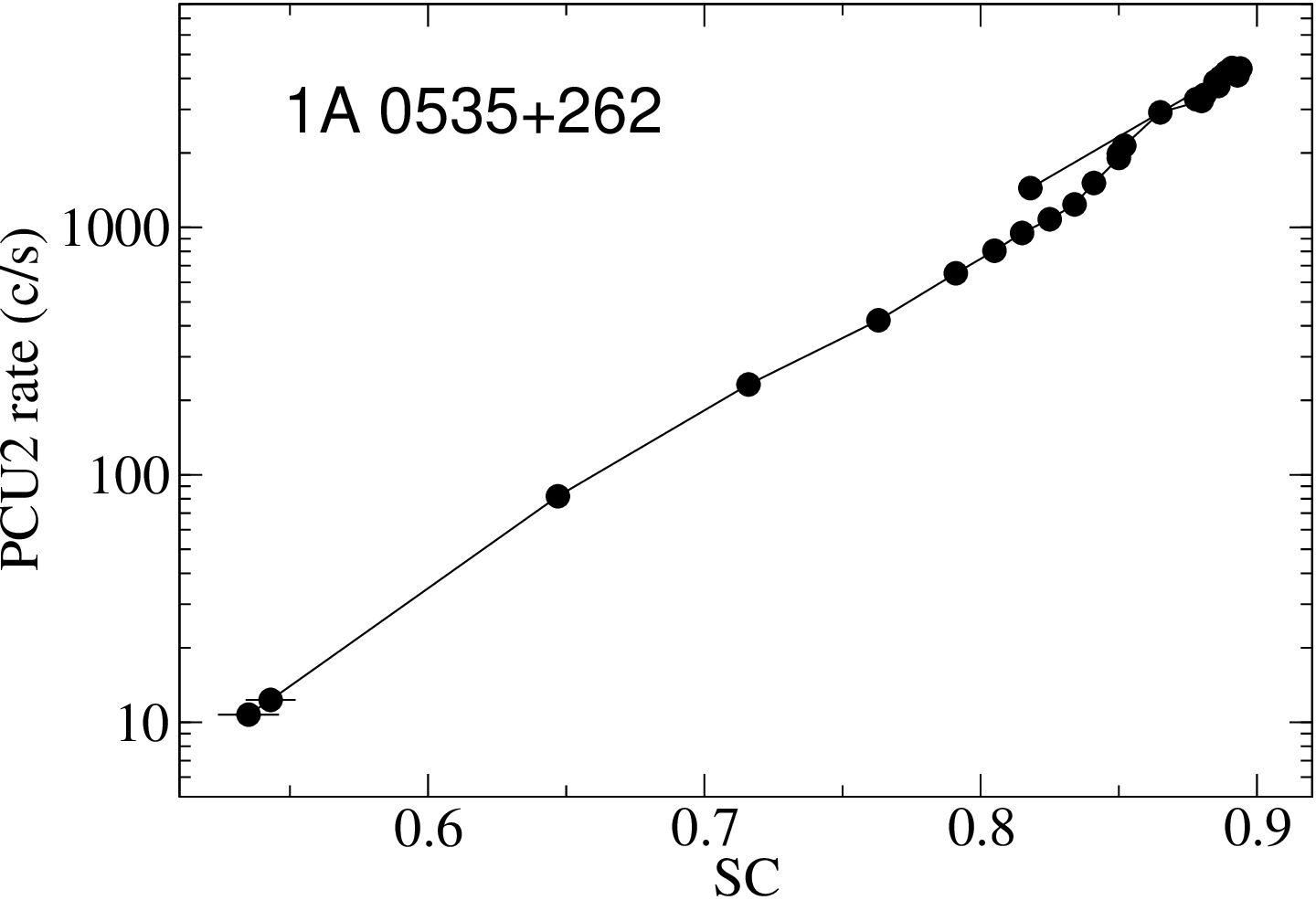}  &
   \includegraphics[width=5cm]{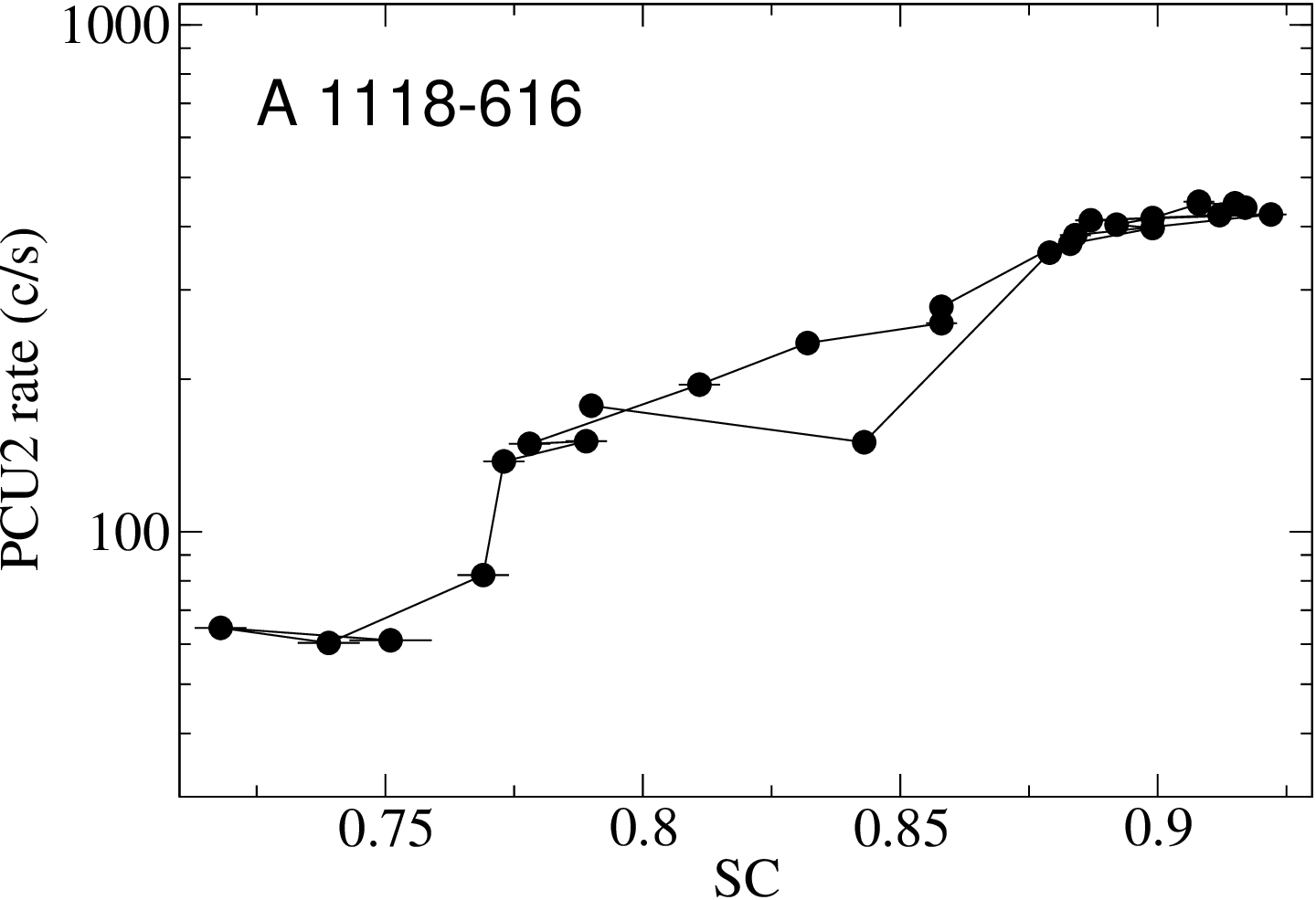}   \\
   \includegraphics[width=5cm]{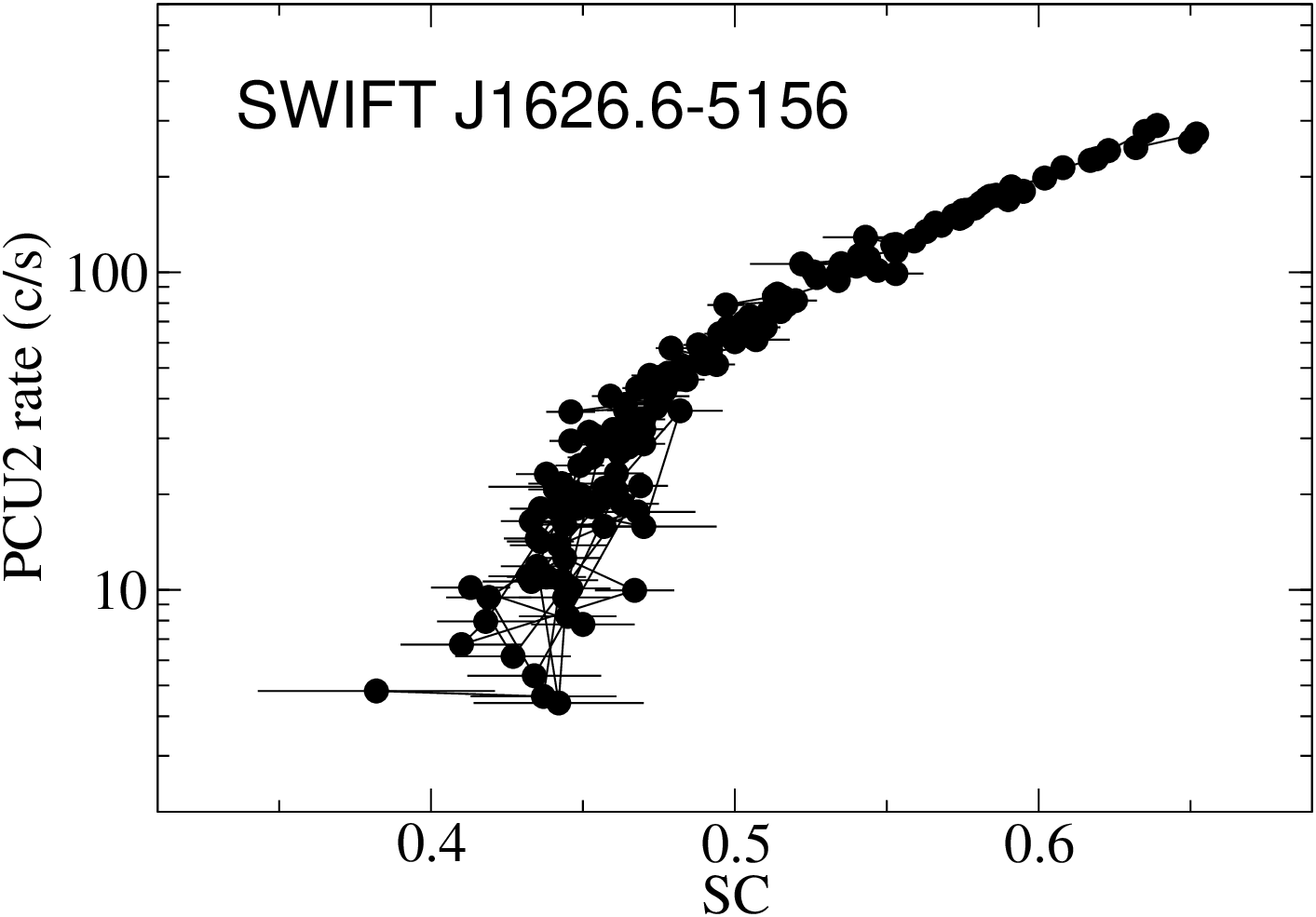}   &  
   \includegraphics[width=5cm]{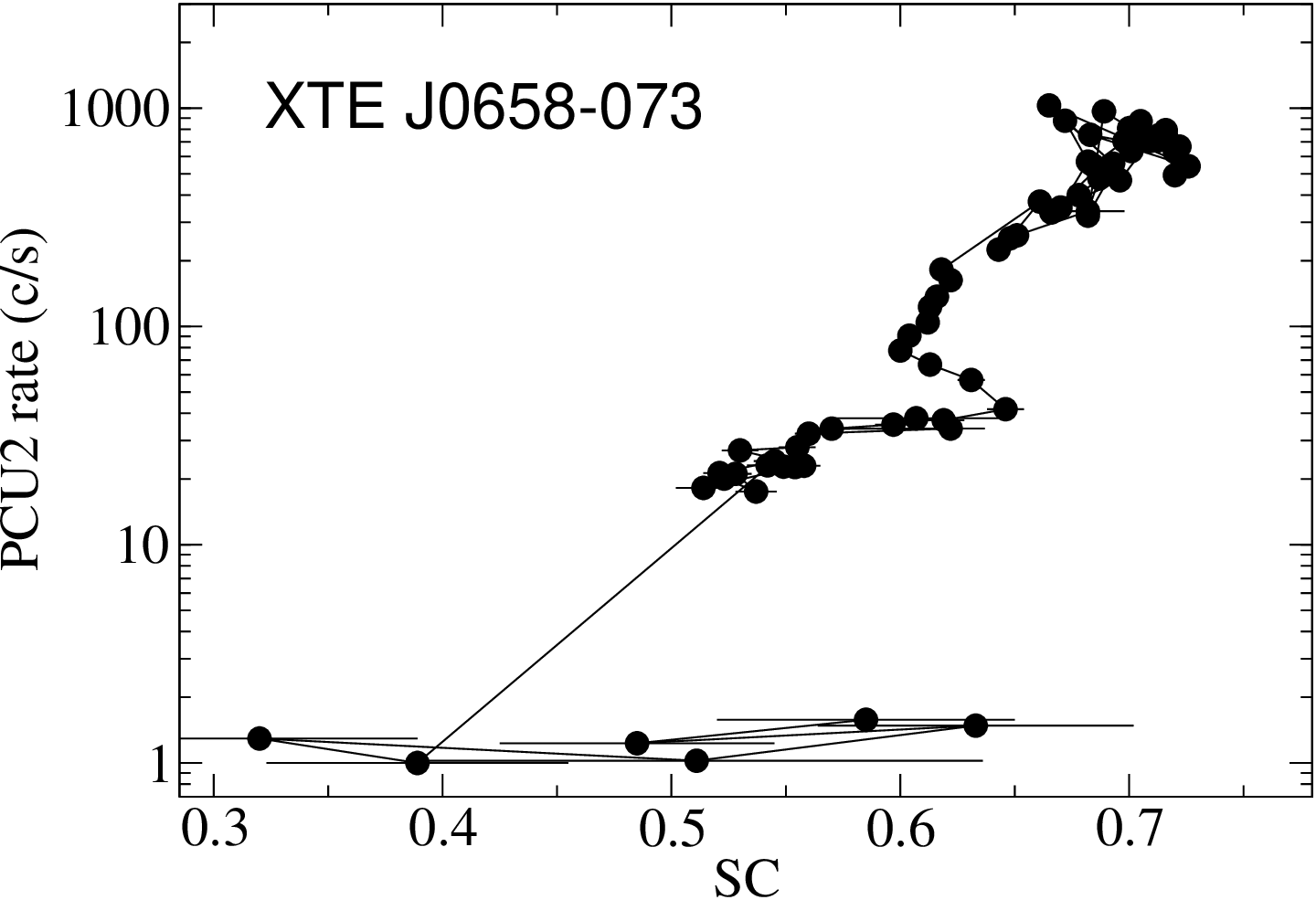}  &
   \includegraphics[width=5cm]{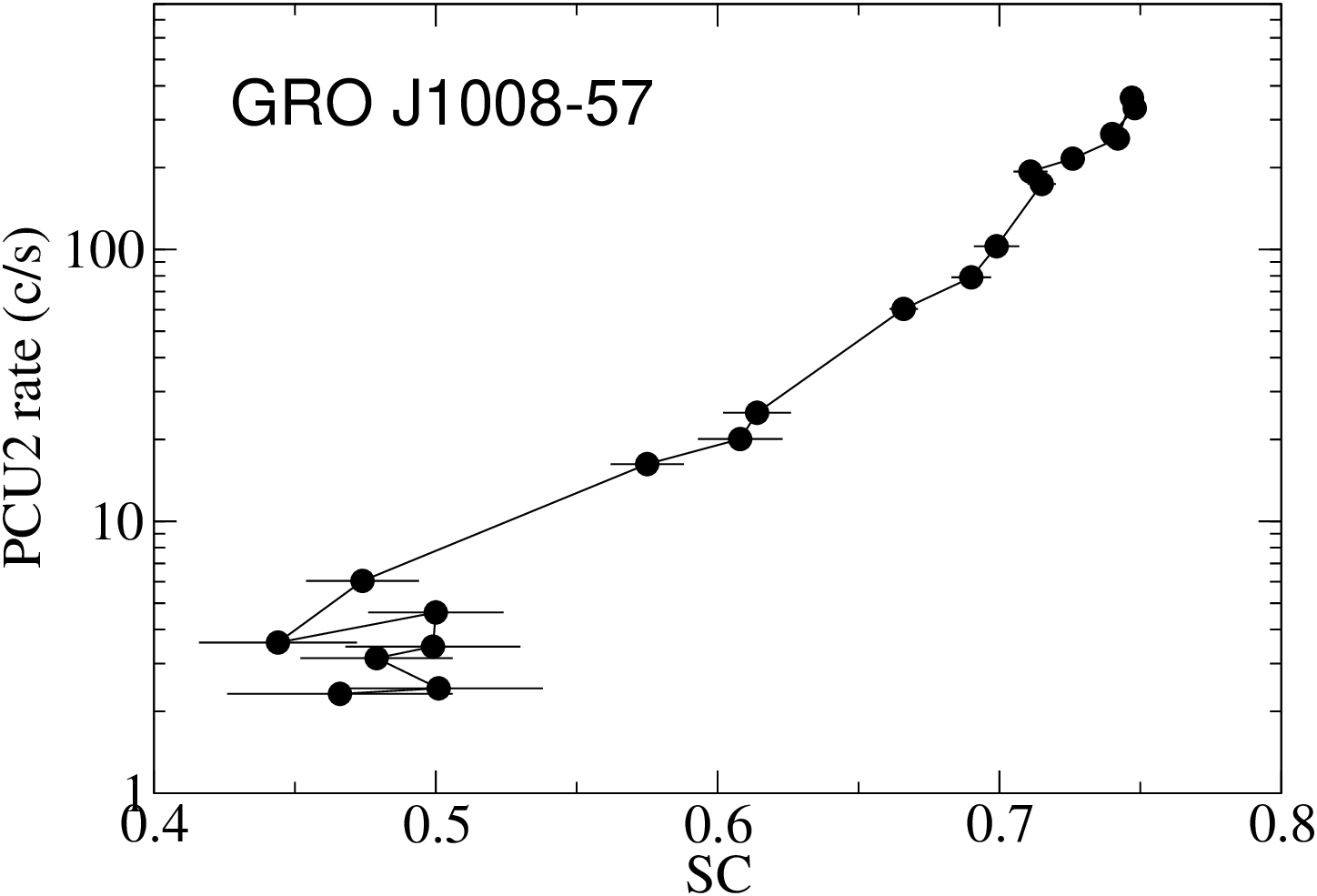}   \\   
   \end{tabular}
      \caption{Hardness (soft colour)-intensity diagram. The soft colour was 
      defined as the ratio 7-10 keV / 4-7 keV. The count rate was obtained
      for the 4-30 keV energy band. Open circles designate
      points in the DB, while filled circles correspond to the HB. A
      logarithmic scale is used for the count rate (confront with
      Fig.~\ref{hid}).}
  \label{rate-SC}
  \end{figure*}

\section{Observations}
\label{obs}

Our aim is to study the characteristics of the population of X-ray pulsars
with Be companions as a whole. In order to have consistent and homogeneous
data sets, we used data from one single X-ray observatory. Thanks to the
All-sky Monitor (ASM) and fast reaction to unexpected events, the Rossi
X-ray Timing Explorer ({\it RXTE}) has been providing  large amount of data
of hard X-ray transients. Its high time resolution, moderate spectral
resolution and  relatively broad-band response (2-150 keV) makes it the
perfect observatory to undergo this type of study. 

We analysed archived data obtained by all three instruments on board
{\it RXTE}  \citep{bradt93}. The data provided by the All Sky Monitor (ASM)
consists of daily flux averages in the energy range 1.3-12.1 keV 
\citep{levine96}. The Proportional Counter Array (PCA) covers the energy
range 2--60 keV, and  consists of five identical coaligned gas-filled
proportional units giving a total collecting area of 6500 cm$^{2}$ and
provides an energy resolution of 18\% at 6 keV  \citep{jahoda96}. The High
Energy X-ray Timing Experiment (HEXTE) is constituted by 2 clusters of 4
NaI/CsI scintillation counters, with a total collecting area of 2 $\times$
800 cm$^2$, sensitive in the 15--250 keV band with a nominal energy
resolution of 15\% at 60 keV  \citep{rothschild98}. 

Due to {\em RXTE}'s low-Earth orbit, the observations consist of a number
of contiguous data intervals or "pointings" (typically 0.5-1 hr long)
interspersed with observational gaps produced by Earth occultations of the
source and passages of the satellite through the South Atlantic Anomaly.
Data taken during satellite slews, passage through the South Atlantic
Anomaly, Earth occultation,  and high voltage breakdown were filtered out
\footnote{Data were filtered out when the difference between the source
position and the pointing of the satellite was greater than 0.02$^\circ$,
the elevation angle was smaller than 8$^\circ$ (timing) and 10$^\circ$
(spectra), and contamination from electrons trapped in the Earth's
magnetosphere or from solar flare activity was above 0.1. Good Time
Intervals which excluded the times of PCA breakdowns were used.}, 
following the recommendations of the {\em RXTE}
team\footnote{http://heasarc.gsfc.nasa.gov/docs/xte/abc/screening.html}.

We have studied all \bex s that went into outburst (the X-ray intensity
increased by $\simmore 100$) during the lifetime of {\it RXTE} and have
enough number of observations to allow a meaningful analysis. We did not
include the \bex\ candidate \uhu\ because no optical counterpart is known
for this source and no reliable estimate of its distance is available
\citep{galloway05}.

Because type II outbursts are rare events, there are not many studies reporting
the evolution of the spectral and/or timing parameters over the entire
outburst in an accreting neutron star. An exception is \vcas, which
displayed four giant outbursts during the period of time analysed in this
work, in 1999, 2000, 2004, and 2008. The 2000 outburst lacks good data
coverage. Here we present results for the 2004 event only. However, it is
worth noticing that the source traced very similar HIDs in all outbursts.
The two branches, HB and DB, are clearly distinguished in correspondence to
the X-ray flux level. For an analysis of the 2008 outburst the reader is
referred to \citet{li12} and \citet{muller12}. The 1999 and 2004 events
have also been studied by \citet{nakajima06} and \citet{tsygankov07},
although these studies focus on the dependence of the cyclotron line energy
with flux. Spectral and/or timing studies covering most of the same giant
outbursts analysed in this work have also been performed for the following
sources: \bq\ \citep{tsygankov10,nakajima10}, \ks\ \citep{galloway04},
\exo\ \citep{wilson08}, \hen\ \citep{nespoli11,devasia11}, \vtau\
\citep{acciari11}, \mxb\ \citep{mcbride06,nespoli12}, \gro\ \citep{naik11},
\sw\ \citep{icdem11,reig11b}.

Figure \ref{rate-time} shows the outburst profiles for each source, created
with the count rates of PCU2 in the 4-30 keV band. Table~\ref{info}
summarises some of the properties of the systems and of the outbursts analysed in
this work. Table~\ref{xobs} gives the proposal ID, time span, the
total number of "pointings" or observation intervals, and the on-source
exposure time of the observations.


\section{Data reduction and analysis}
\label{red}

In this section we describe the different techniques used in the data
analysis. Heasarc FTOOLS version 6.6.3 was employed to perform data
reduction, while XSPEC v12.6 \citep{arnaud96} was used for spectral
analysis. Power spectra were obtained using an FFT algorithm implemented in
a Fortran code created by us,  based on subroutines from
\citet{press96}.
Data reduction and model fitting were automated so that each observation
was treated in exactly the same way. The same type of spectral and noise
components were used for all sources. Therefore, this work constitutes the
first attempt to perform a consistent, homogeneous and systematic study of
the variability of accreting pulsars during giant outbursts.

\subsection{Colour analysis}

In analogy with optical photometry, one can quantify the broad-band X-ray
spectral shape by defining X-ray colours. An X-ray colour is a hardness ratio
between the photon counts in two broad bands. X-ray colours were directly
obtained from the background-subtracted {\em Standard 2} PCA light curves in the
following energy bands: soft color (SC): 7-10 keV / 4-7 keV and hard color (HC):
15-30 keV / 10-15 keV. These ratios are expected to be insensitive to
interstellar absorption effects because the hydrogen column densities to the
systems, obtained from model fits to the X-ray spectra, are in the range
$0.6-3\times 10^{22}$ cm$^{-2}$. These values only affect significantly the
X-ray spectral continuum below $\sim 2$ keV.

\begin{table}
\caption{Model photon distributions used in the spectral analysis.
The spectra were fitted with the function $f(E)=A(E)\times (P(E)+G(E))\times (C_f(E)+C_h(E))$ }
\label{models}      
\centering          
\begin{tabular}{ll}
\hline\hline
Model name as in XSPEC	&		 \\
\hline
PHABS				&$A(E)=e^{-N_{\rm H} \sigma_{MM}(E)}$ \\
CUTOFFPL			&$P(E)=K\,E^{-\Gamma} e^{-E/E_{\rm cut}}$	\\
POWERLAW$\times$HIGHECUT$^\dag$	&$P(E)=K\,E^{-\Gamma} e^{(E_{\rm cut}-E)/E_{\rm fold}}$	\\
GAUSS				&$G(E)= K \frac{1}{\sigma \sqrt{2\pi}} e^{-0.5\left(\frac{E-E_l}{\sigma}\right)^2}$\\
CYCLABS$^{\dag \dag}$		&$C_i(E)=e^{-D_i\left(\frac{W_i^2(E/E_i)^2}{(E-E_i)^2+W_i^2}\right)}$  \\
GABS$^{\dag \dag \dag}$		&$G_{\rm abs}(E)=e^{\left(-\tau_0\,e^{-0.5\left(\frac{E-E_0}{\sigma_0}\right)^2}\right)}$   \\
\hline
\multicolumn{2}{l}{$\sigma_{MM}(E)$: 0.03-10 keV interstellar photoelectric absorption} \\
\multicolumn{2}{l}{\hspace{1.1cm} cross-section.} \\
\multicolumn{2}{l}{$N_{\rm H}$: equivalent hydrogen column.} \\
\multicolumn{2}{l}{K: normalization in photons keV$^{-1}$ cm$^{-2}$
s$^{-1}$ at 1 keV.} \\
\multicolumn{2}{l}{$\Gamma$: power-law photon index.} 	\\
\multicolumn{2}{l}{$E_{\rm cut}$: cut-off energy in keV.} 	\\
\multicolumn{2}{l}{$E_{\rm fold}$: folding energy in keV.} 	\\
\multicolumn{2}{l}{$E_l$: iron line energy in keV.}	\\
\multicolumn{2}{l}{$\sigma$: iron line width in keV.}	\\
\multicolumn{2}{l}{$E_{i}$: energy of the cyclotron line in keV.}	\\
\multicolumn{2}{l}{$D_i$: depth of the cyclotron line.}	\\
\multicolumn{2}{l}{$W_i$: width of the cyclotron line in keV.}	\\
\multicolumn{2}{l}{$E_{0}$: central energy of the absorption line in keV.}	\\
\multicolumn{2}{l}{$\tau_0$: optical depth of the absorption line.}	\\
\multicolumn{2}{l}{$\sigma_0$: width of the absorption line in keV.}	\\
\multicolumn{2}{l}{$^{\dag}$: Only used in \vcas.} \\
\multicolumn{2}{l}{$^{\dag \dag}$: $i$ refers to the fundamental ($f$) or 
harmonic ($h$).} \\ 
\multicolumn{2}{l}{$^{\dag \dag \dag}$: Used in \vcas, \xte\ and \vtau\ } \\
\multicolumn{2}{l}{\hspace{0.5cm} to fit the "10-keV feature".} \\
\end{tabular}
\end{table}

\subsection{Spectral analysis}
\label{specana}

The spectral analysis was carried out on each individual observation using
PCA (PCU2 only) \emph{Standard 2} mode data and \emph{Standard (archive)}
mode from the HEXTE (Cluster A generally, and Cluster B for data after
December 2005, when Cluster A stopped rocking between source and
background). Both instruments configurations provide a time resolution of
16s and cover their energy ranges with 129 channels. All spectra were
background-subtracted and dead-time corrected. For each observation, the
two spectra were fitted simultaneously, covering an overall 3--100 keV
energy range.  A systematic error of 0.6\% was added in quadrature to
the PCA spectra, slightly larger than the recommended 0.5\% by the
instrument
team\footnote{http://www.universe.nasa.gov/xrays/programs/rxte/pca/doc/rmf/pcarmf-11.7}
to obtain more conservative estimates of the spectral parameters. No
systematic error was added to the HEXTE spectra because uncertainties  are
dominated by statistical fluctuations in this instrument. In the case
of \vtau, the system which showed the most recent outburst, only PCA data
were employed, covering the 3--60 keV energy range, because cluster B
stopped rocking in December 2009. The model developed to obtain background
counts from cluster B to be used with cluster A data is known to produce
spurious line-like features around 63 keV, which for this source lies near
a harmonic of a cyclotron resonant scattering feature. Also, because the
signal-to-noise of the HEXTE spectra of \gro\ was very low, only PCA data
were used for this source.

The lack of adequate theoretical continuum models for accreting neutron
stars implies the use of empirical models to describe the observations. To
compare the results from all sources consistently, we used the same (or
very similar, if leading to better fit) spectral components. 
Table~\ref{models} gives the photon distribution for the different models.
To fit the spectral continuum we used a model composed by a combination of
photoelectric absorption  \citep[][PHABS in XSPEC]{balucinska92} and a
power law with high-energy exponential cutoff (CUTOFFPL).  In \vcas, the
CUTOFFPL model left significant residuals in the energy range 10--20 keV.
For this source the POWERLAW $\times$ HIGHECUT  \citep[see][and
references therein]{white83} provided better fits (see Table~\ref{models}
for the differences between CUTOFFPL and HIGHECUT models). For other
choices of the spectral continuum the reader is referred to \citet[][and
references therein]{kreykenbohm99,coburn02,maitra12}. In general, the value
of the hydrogen column density $N_{\rm H}$ cannot be well constrained by
the PCA, whose energy threshold is $\sim$3 keV. When known from previous
works with higher sensitive instruments below 3 keV, $N_{\rm H}$ was fixed
to the known value. To account for uncertainties in the absolute flux
normalisation between PCA and HEXTE we introduced  a multiplicative factor
which was fixed to 1 for the PCA and let vary freely for HEXTE.

On top of the spectral continuum a number of discrete components are
clearly present in the spectra of accreting pulsars. A Gaussian line
profile (GAUSS) at 6.4 keV was used to account for Fe K  fluorescence,
whereas cyclotron resonant scattering features (CRSF), which we will refer
to as "cyclotron lines", were accounted for with a Lorentzian profile
(CYCLABS)  \citep{tanaka86,makishima90}.  The  width of the
iron line were initially let free, but once we checked that there was no
significant trend with flux, it was fixed to a value of 0.5 keV.


The sources \vcas, \xte\ and \vtau\ show significant residuals around 7--10
keV. This feature is known as the "10-keV feature", and its origin is
uncertain \citep{coburn02}. It may appear in emission, such as in \vcas\
\citep{ferrigno09,muller12} and \exo\ \citep{klochkov07} and then it is
modelled with a broad Gaussian emission line,  and it is referred to as
the "bump" model, or in absorption, such as in \xte\
\citep{mcbride06,nespoli12}. In this case, a Gaussian absorption profile
(GABS) is normally used\footnote{Note that the GABS model was initially
proposed to fit CRSFs \citep{soong90, coburn02}}. In \exo, 
this extra broad emission component is not needed if a CRSF is added
at $\sim$10 keV \citep{klochkov07,wilson08}.  In \vcas, the
POWERLAW$\times$HIGHECUT model gives comparable fits to the combination of
the CUTOFFPL model with a broad gaussian emission profile. Although
the use of the POWERLAW$\times$HIGHECUT model has been widely used to
describe the exponential decay in the spectral continuum of many accreting
pulsars, including \vcas\ \citep{tsygankov07,li12}, several studies
\citep[see e.g.][]{kreykenbohm99} have shown that the cutoff energy in
HIGHECUT may produce an unphysical break, which may be interpreted as a
line feature. Because the $E_{\rm cut}$ and the energy of the fundamental
cyclotron line occupy the same region of the spectrum in \vcas, the
application of the POWERLAW$\times$HIGHECUT model to the spectrum of this
source can lead to  erroneous conclusions regarding the variability of the
CRSF \citep{muller12}. In view of this, the spectral analysis of \vcas\ was
done using the two models.



\subsection{Timing analysis}

For each observation, a light curve in the energy range 2--20 keV (PCA
channels 0--49) or 2--15 keV (PCA channels 0--35), depending on the data
mode available, was extracted with a time resolution of $2^{-6}$ s. The
light curve was then divided into 128-s segments and a Fast Fourier
Transform was computed for each segment. The final power spectrum is the
average of all the power spectra obtained for each segment. The final power
spectra were logarithmically rebinned in frequency and corrected for dead
time effects according to the prescription given in \citet{nowak99}. Power
spectra were normalized such that the integral gives the squared rms
fractional variability \citep{belloni90b,miyamoto91}. To have a unified
phenomenological description of the timing features within a source and
across different sources, we fitted the noise components with a function
consisting of one or multiple Lorentzians, each denoted as $L_i$, where $i$
determines the number of the component. The characteristic frequency
$\nu_{\rm max}$ of $L_i$ was denoted as $\nu_i$. This is the frequency where
the component contributes most of its variance per logarithmic frequency
interval and is defined as \citep{belloni02} 

\begin{equation}
\nu_{\rm max}=\sqrt{(\nu_0^2+(FWHM/2)^2)}= \nu_0\sqrt{1+1/4Q^2}
\end{equation}

\noindent where $\nu_0$ is the centroid frequency and FWHM is the
Lorentzian full-width at half maximum. Note that $\nu_{\rm max} \neq
\nu_0$. The broader the noise component, the larger the difference between
$\nu_{\rm max}$ and $\nu_0$. The quality factor $Q$ is defined as $Q =
\nu_0/FWHM$, and is used as a measure of the coherence of the variability
feature. Peaked noise is considered to be a quasi-periodic oscillation
(QPO) if $Q>2$.

The peaks from the neutron star pulsations were fitted to Lorentzian
functions with the frequency fixed at the expected value and width equal to
0.001 Hz (approximately, the inverse of the timing resolution of the power
spectra). The results from the use of Lorentzians are best visualized using
the  $\nu \times P_{\nu}$ representation, where each power is multiplied by
the corresponding frequency.

   \begin{figure}
   \centering
   \begin{tabular}{c}
   \includegraphics[width=8.5cm]{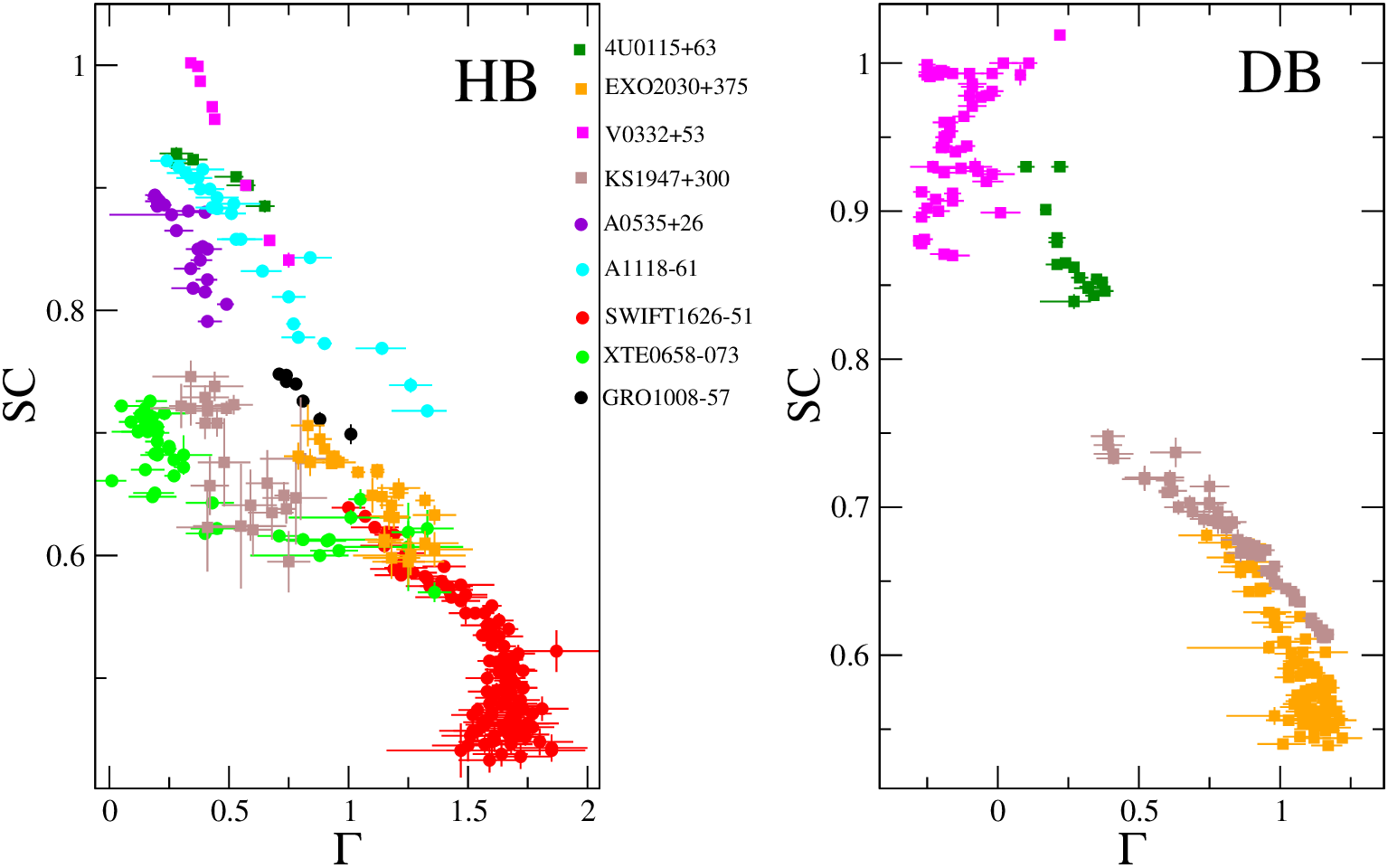} \\
   \includegraphics[width=8.5cm]{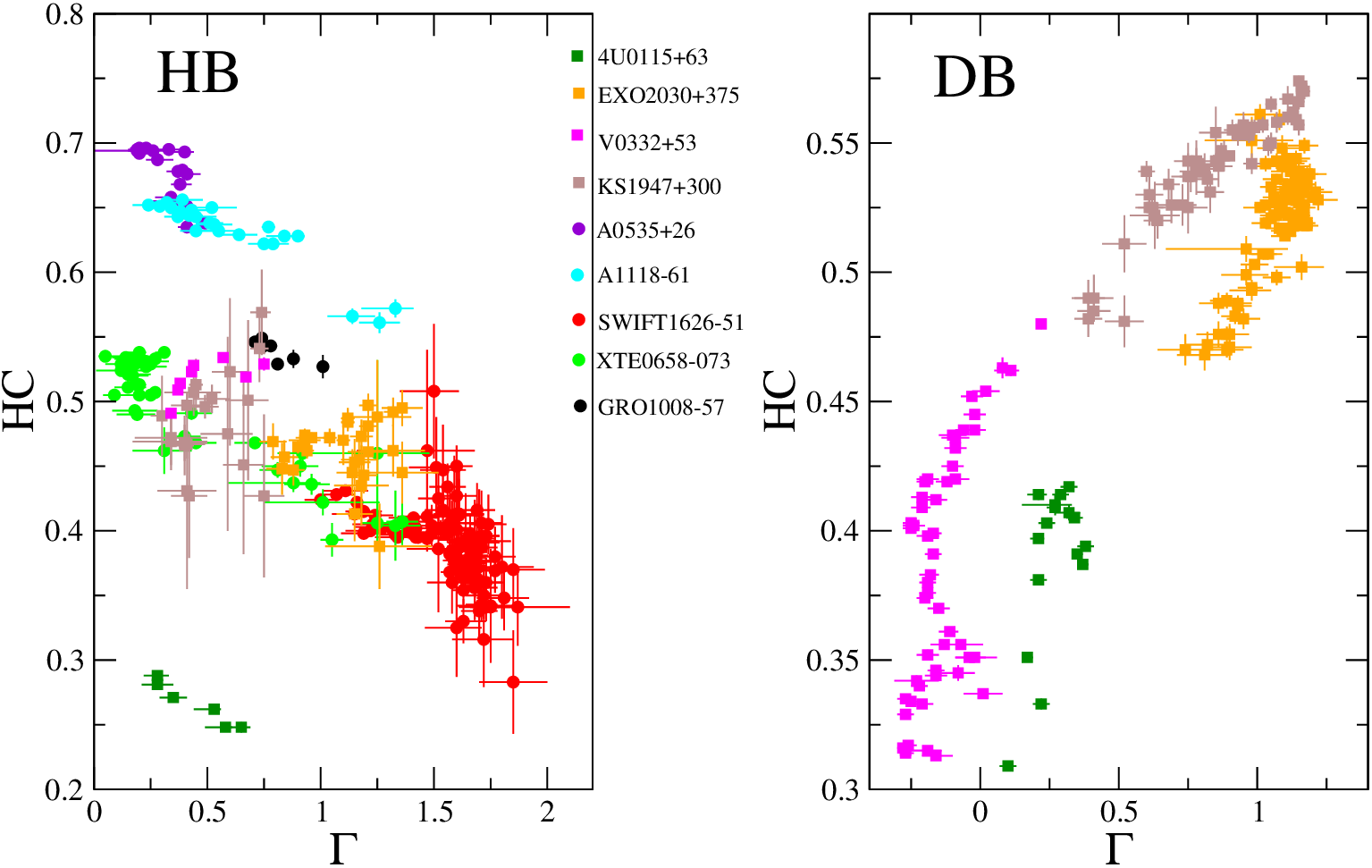}
   \end{tabular}
  \caption{Relationship between the X-ray colours (SC= 7--10 keV/4--7 keV, 
  HC=15-30 keV/10-15 keV) and 
  photon index. The SC and the HC of the horizontal branch decrease as the spectrum becomes 
  softer, as expected. In contrast, a positive correlation is seen in the
  HC of the diagonal branch. }
  \label{gamma-HC}
  \end{figure}
   \begin{figure}
   \centering
   \includegraphics[width=9cm]{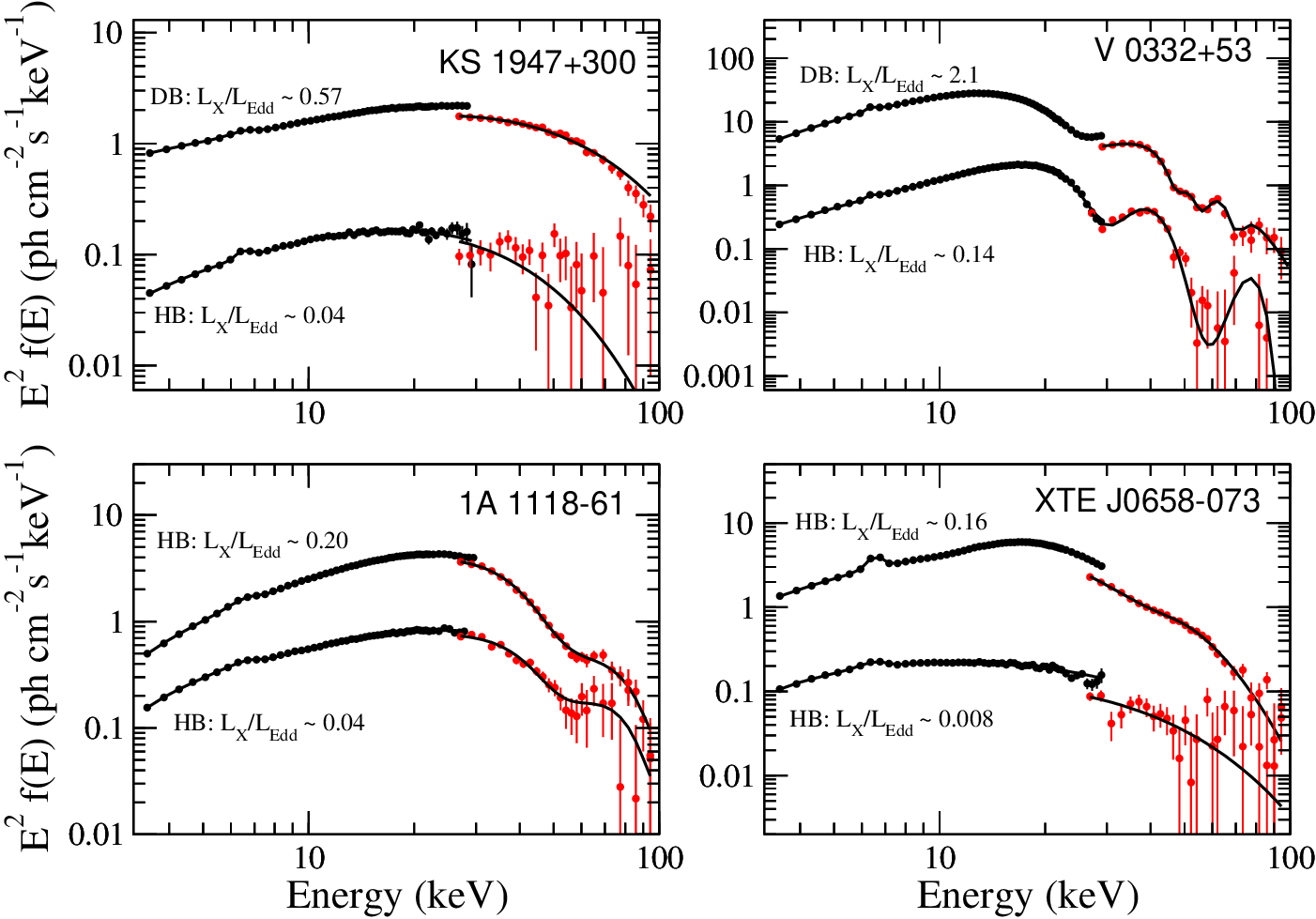}   
  \caption{Average energy spectra of the horizontal and diagonal
  branches at various flux levels. Black and red symbols represent PCA and 
  HEXTE data, respectively. The quoted X-ray luminosity corresponds to the 
  2-100 keV range.}
  \label{spectra}
  \end{figure}

   \begin{figure*}
   \centering
   \begin{tabular}{ccc}
   \includegraphics[width=5cm]{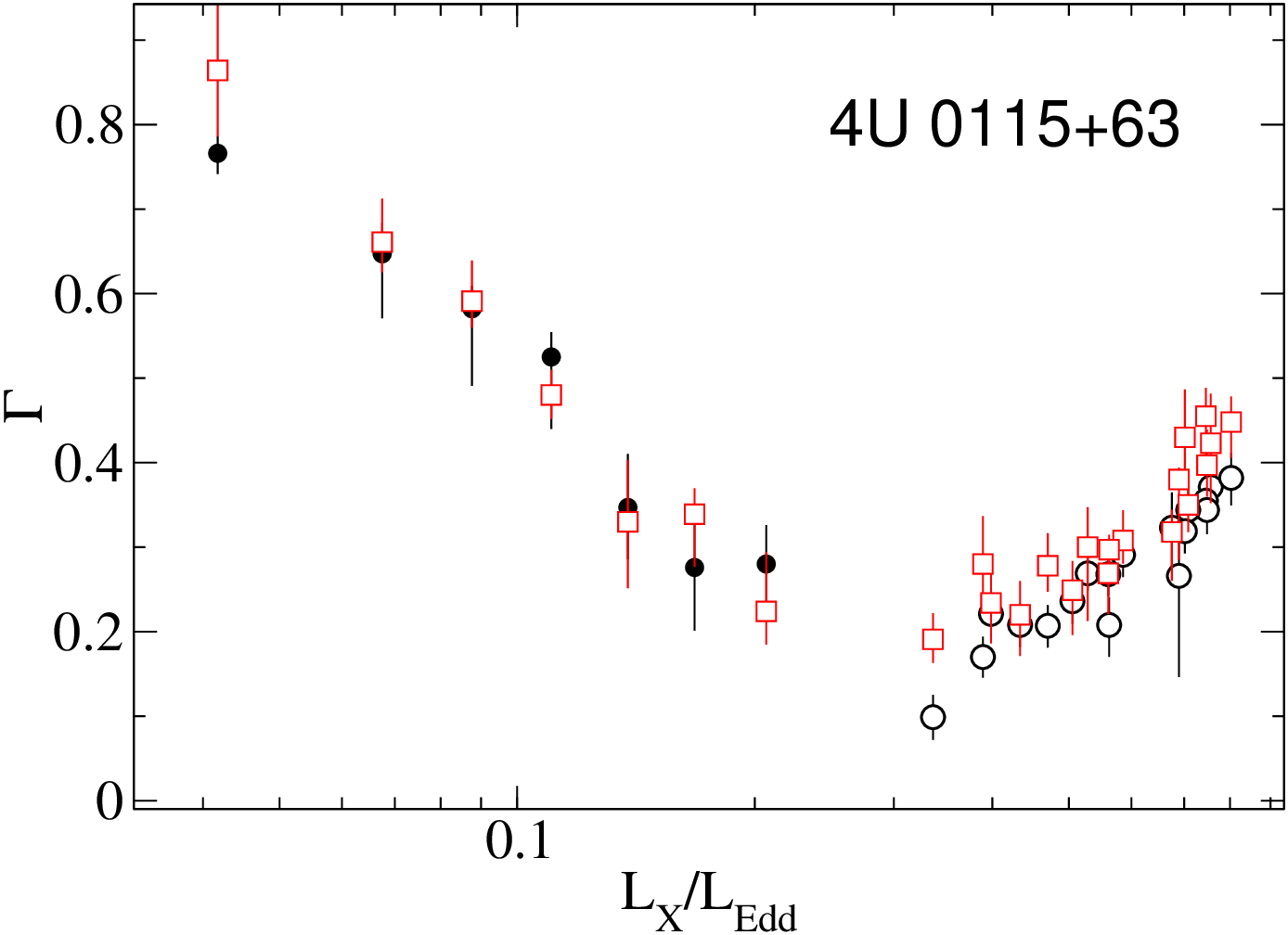}  &
   \includegraphics[width=5cm]{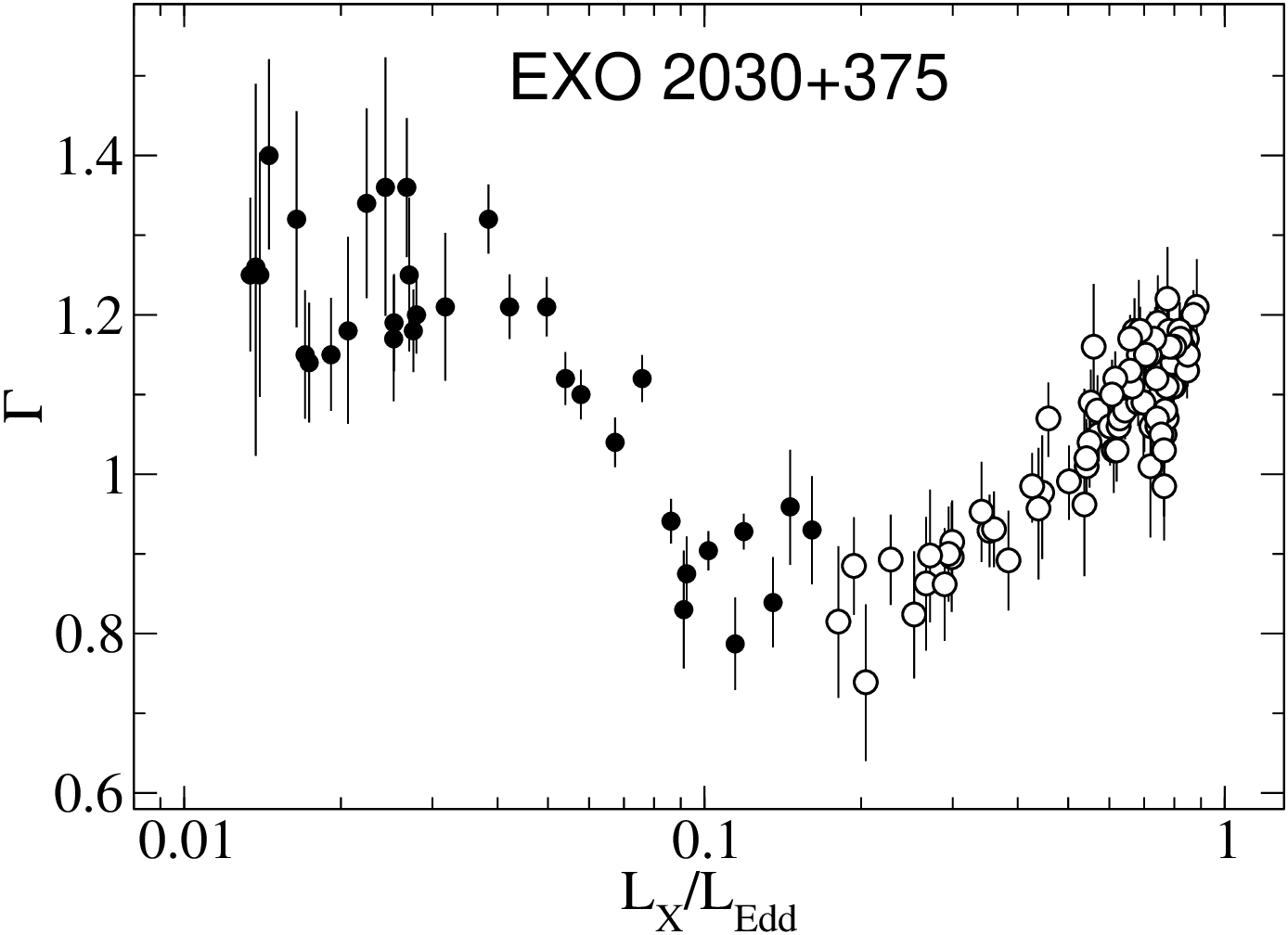} &
   \includegraphics[width=5cm]{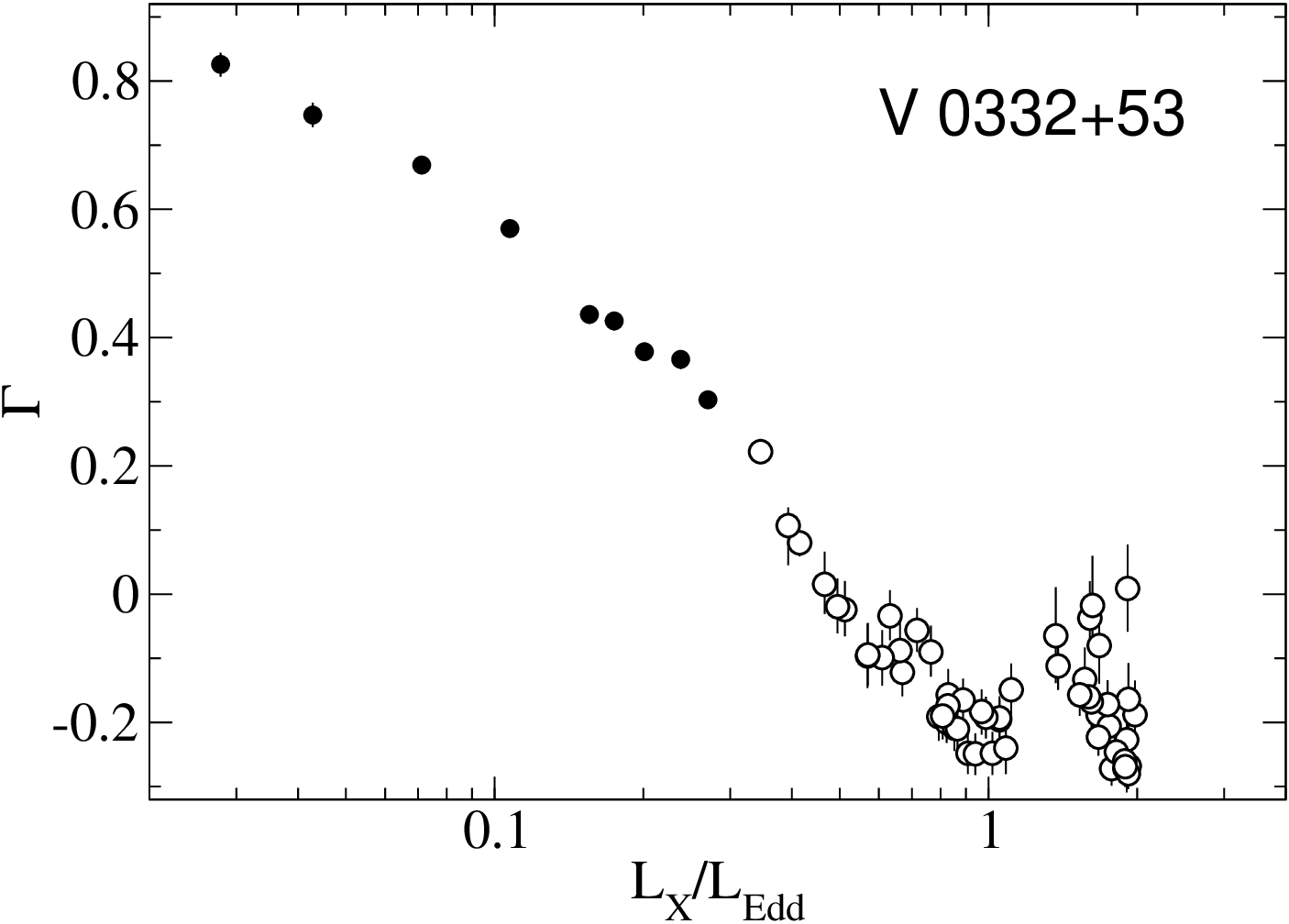}  \\
   \includegraphics[width=5cm]{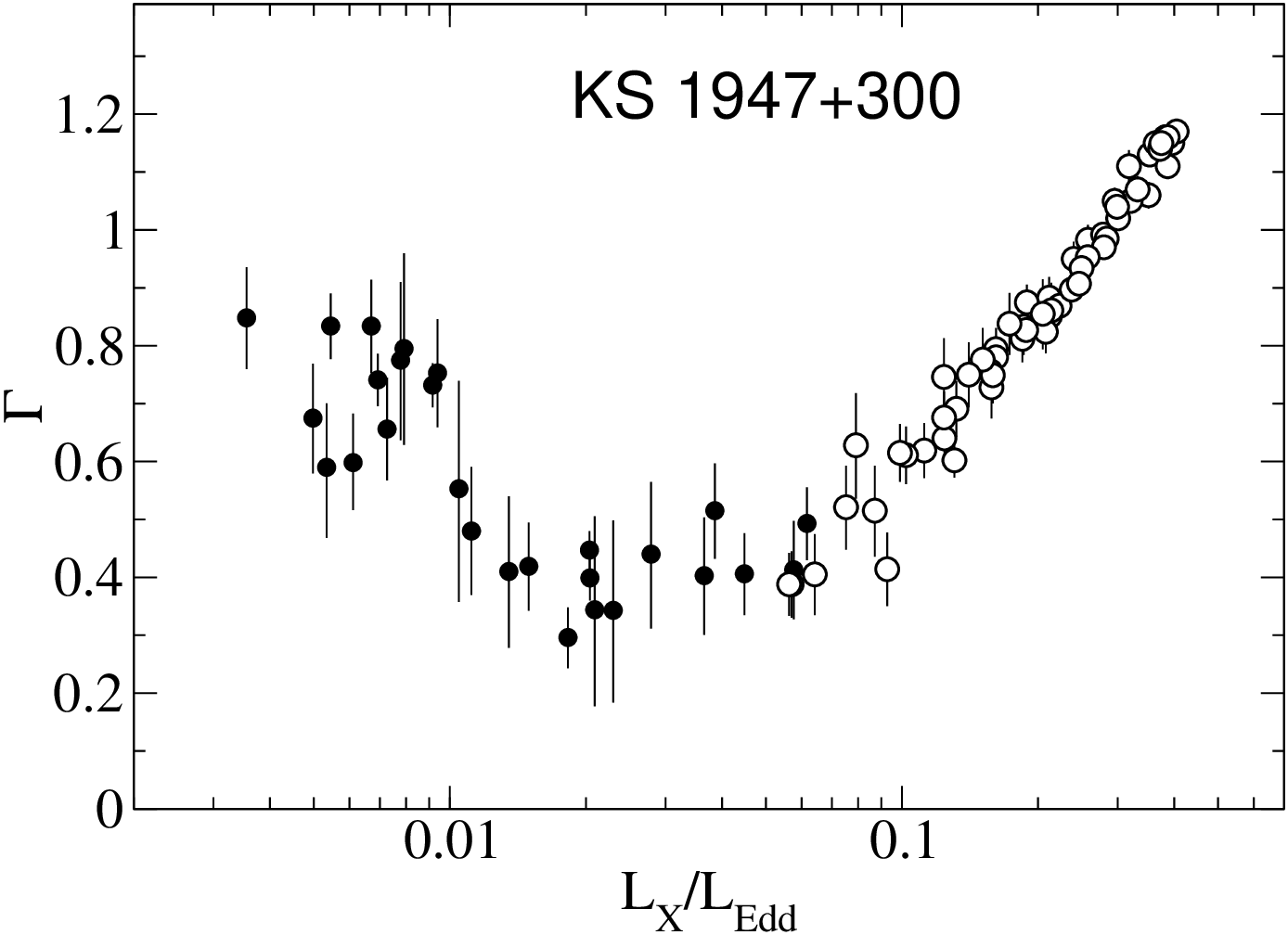}  &
   \includegraphics[width=5cm]{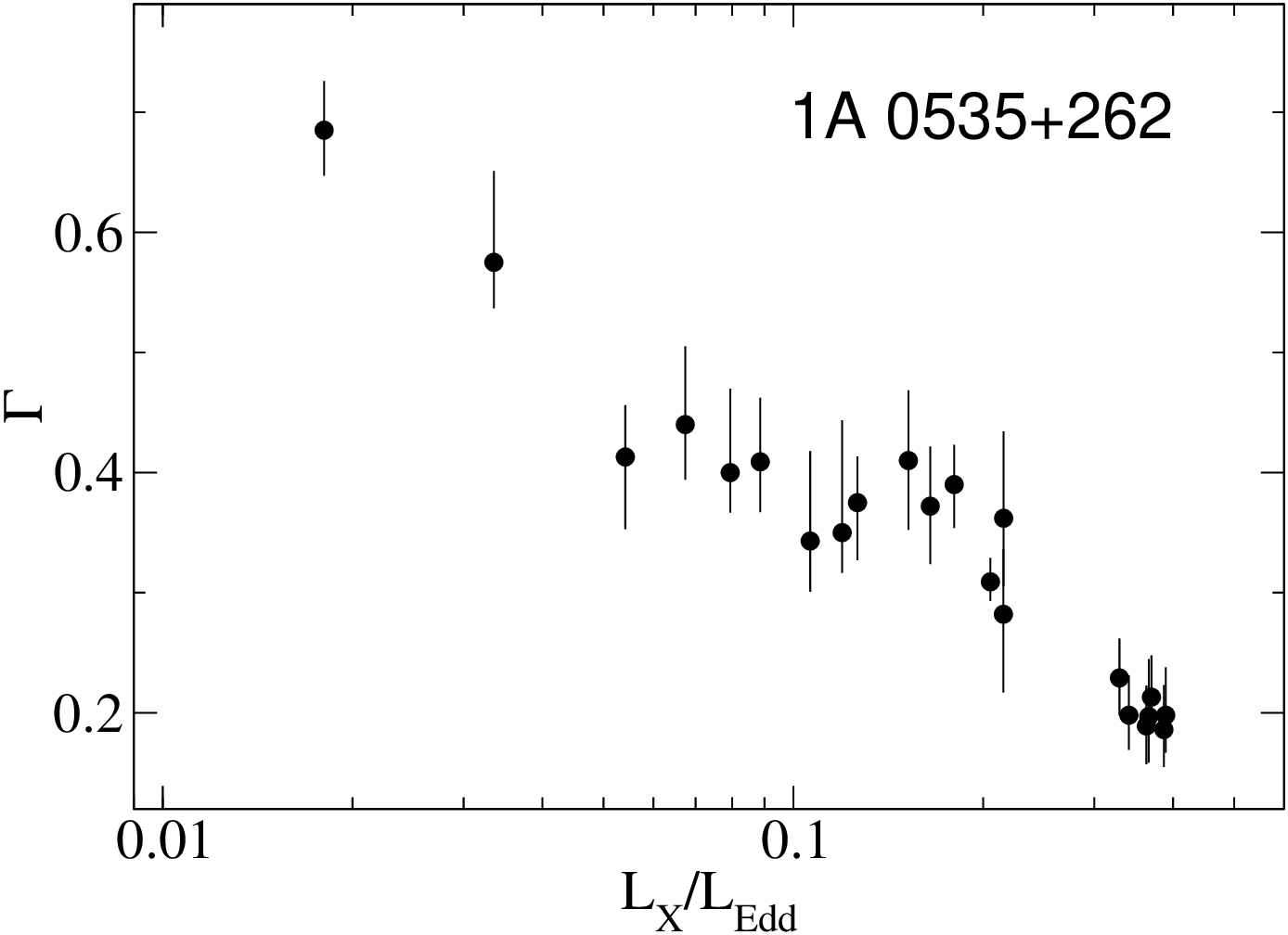}  &
   \includegraphics[width=5cm]{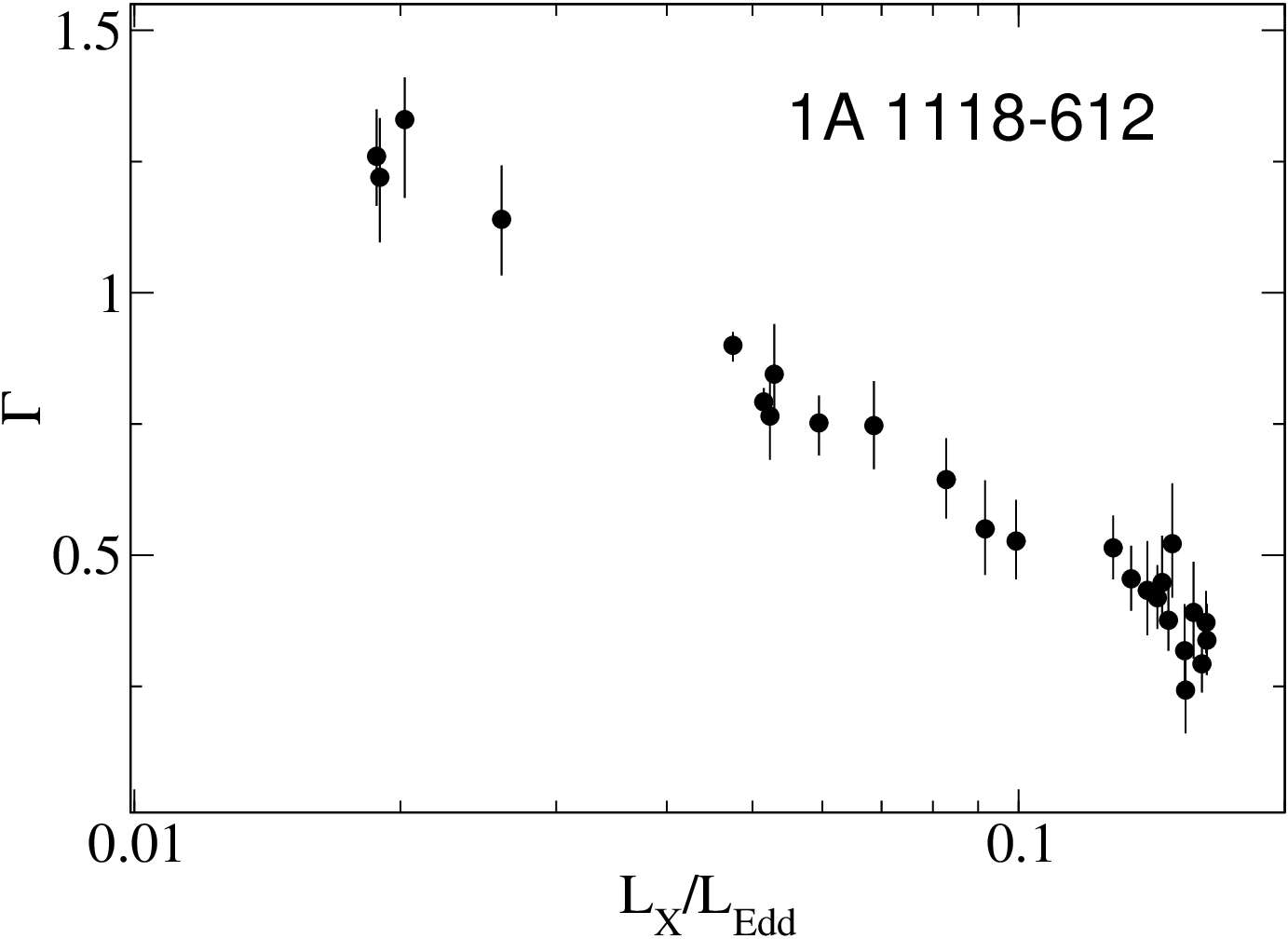}   \\
   \includegraphics[width=5cm]{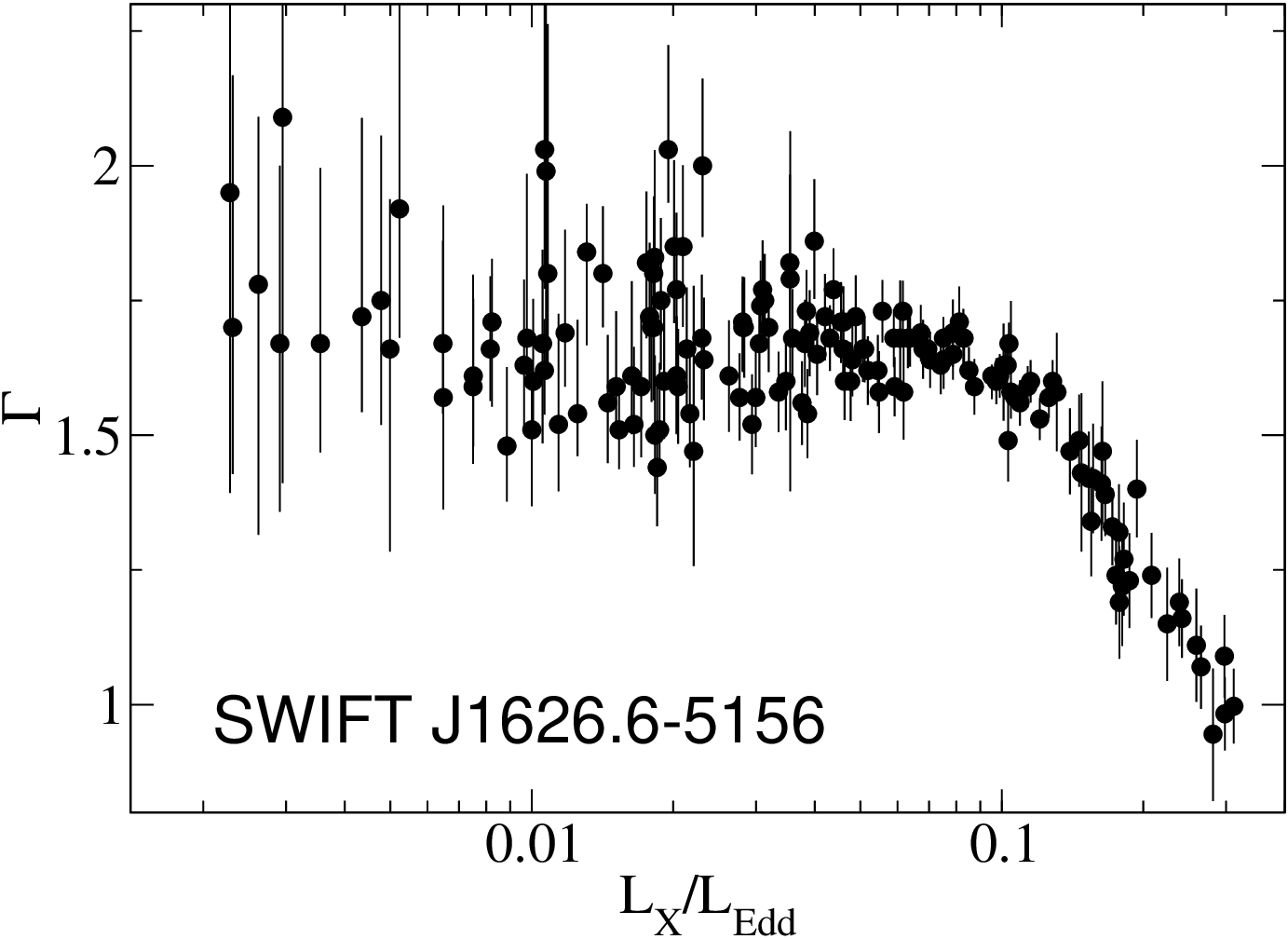}   &  
   \includegraphics[width=5cm]{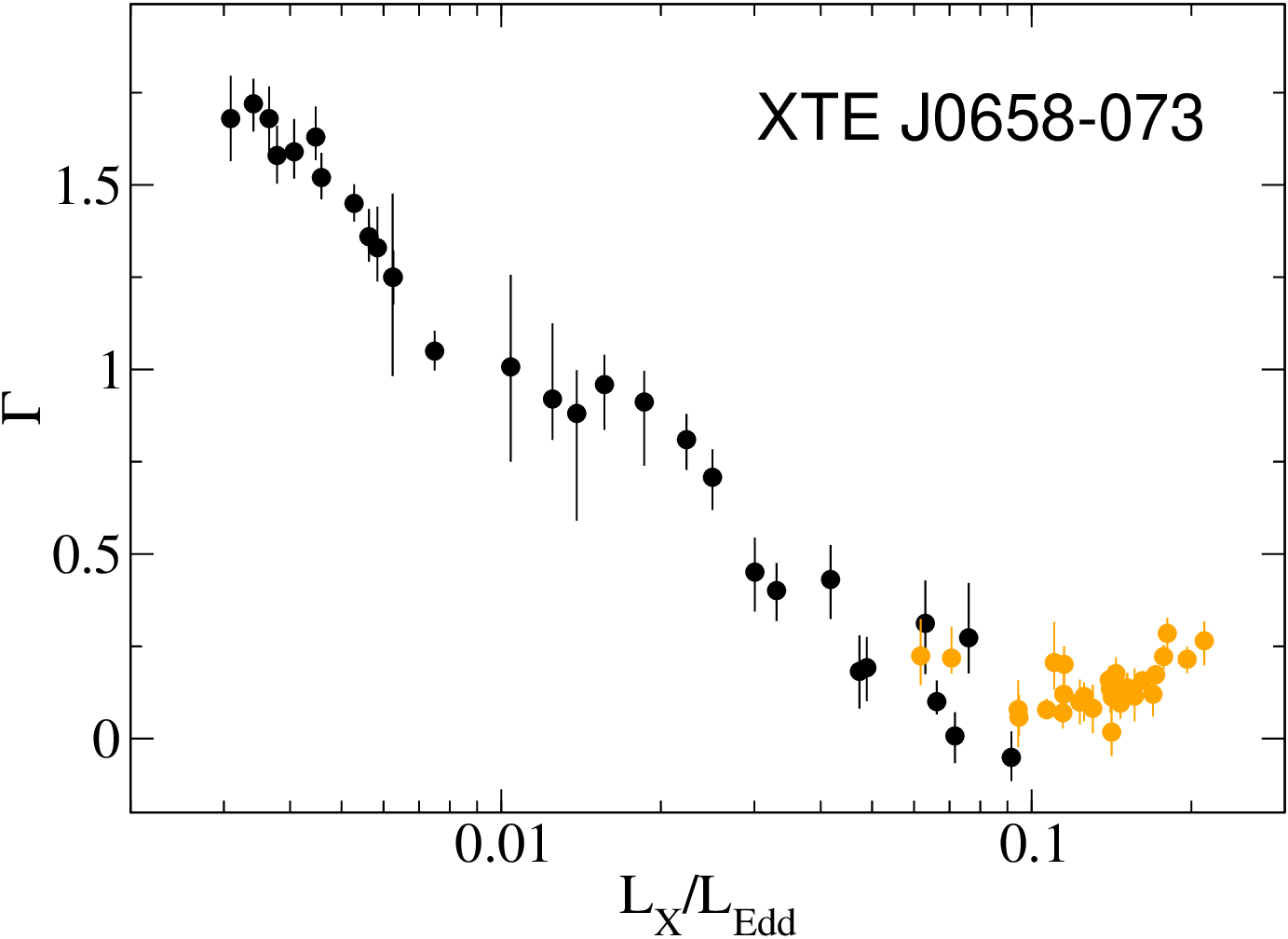}  &
   \includegraphics[width=5cm]{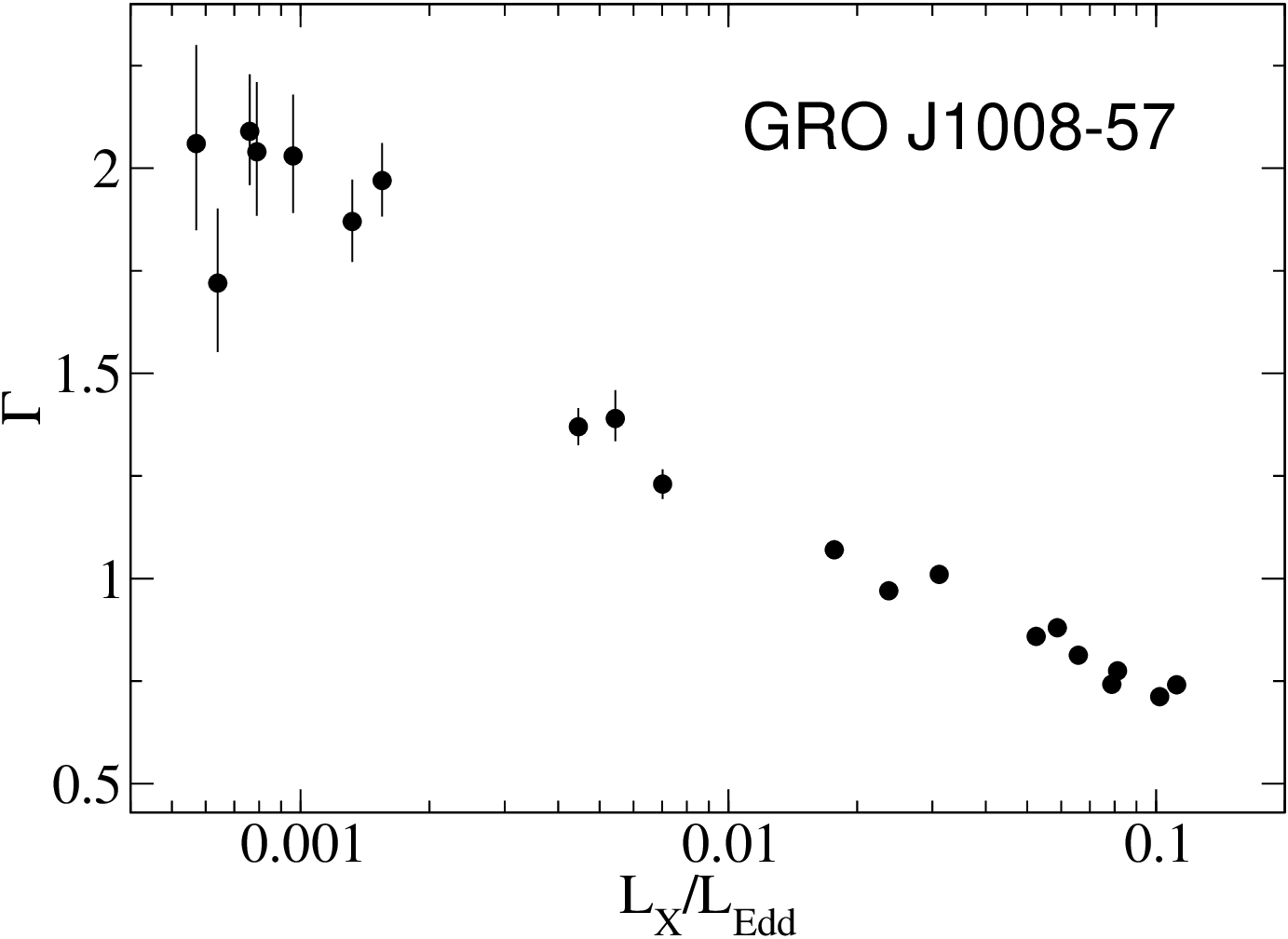}   \\   
   \end{tabular}
      \caption{Power-law photon index as a function of X-ray luminosity. Open
      circles represent the DB (supercritical regime), filled circles the HB 
      (subcritical regime). Orange data points
       in \xte\ correspond to the flaring episode during the peak of the
       outburst (see Fig.~\ref{rate-time}). In \vcas,
  red squares represent the results from the "bump" model (see text). The
       Eddington luminosity for a neutron star was taken to be $1.7\times
       10^{38}$ erg s$^{-1}$.}
  \label{gamma-flux}
  \end{figure*}
   \begin{figure*}
   \centering
   \begin{tabular}{ccc}
   \includegraphics[width=5cm]{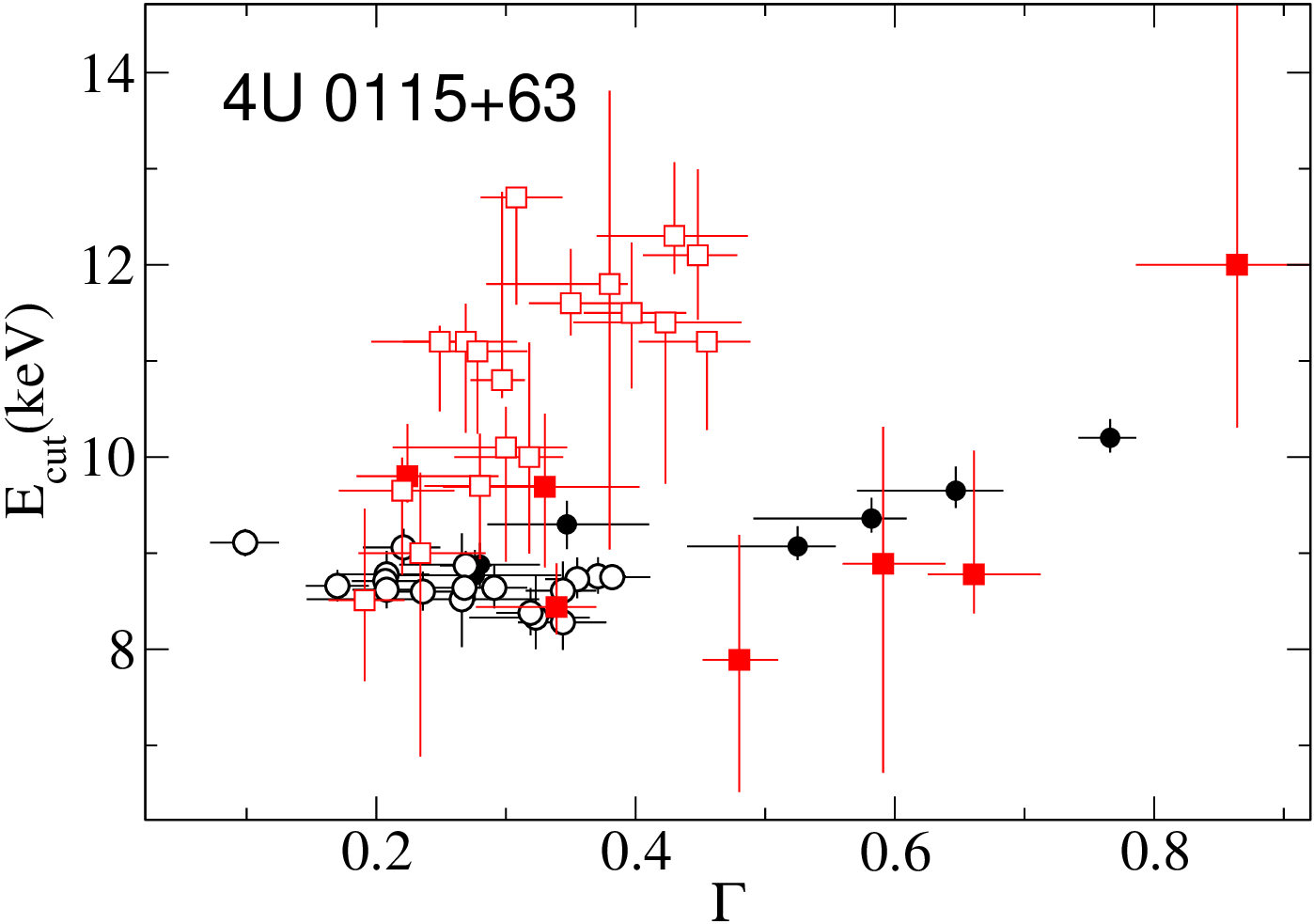}  &
   \includegraphics[width=5cm]{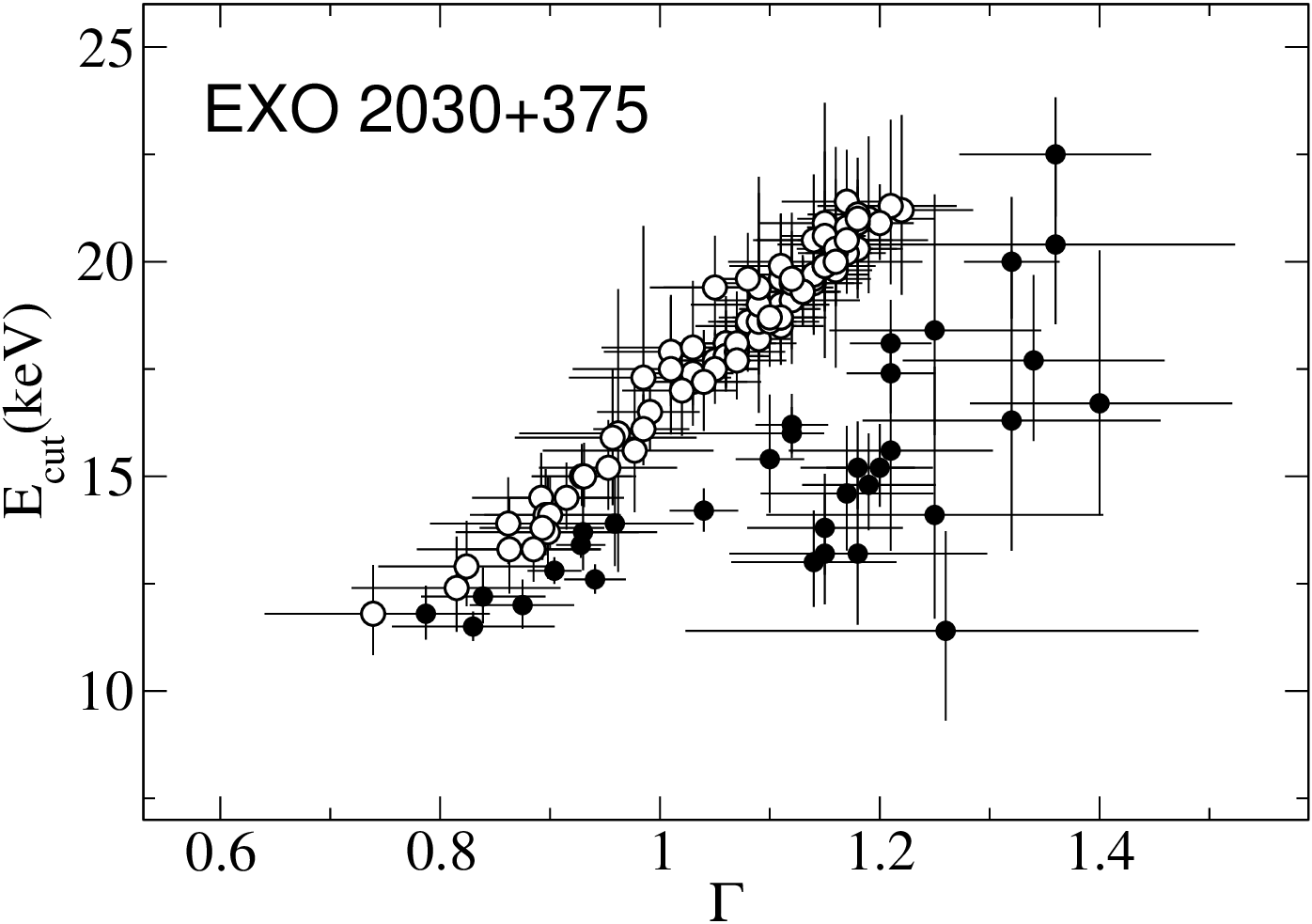} &
   \includegraphics[width=5cm]{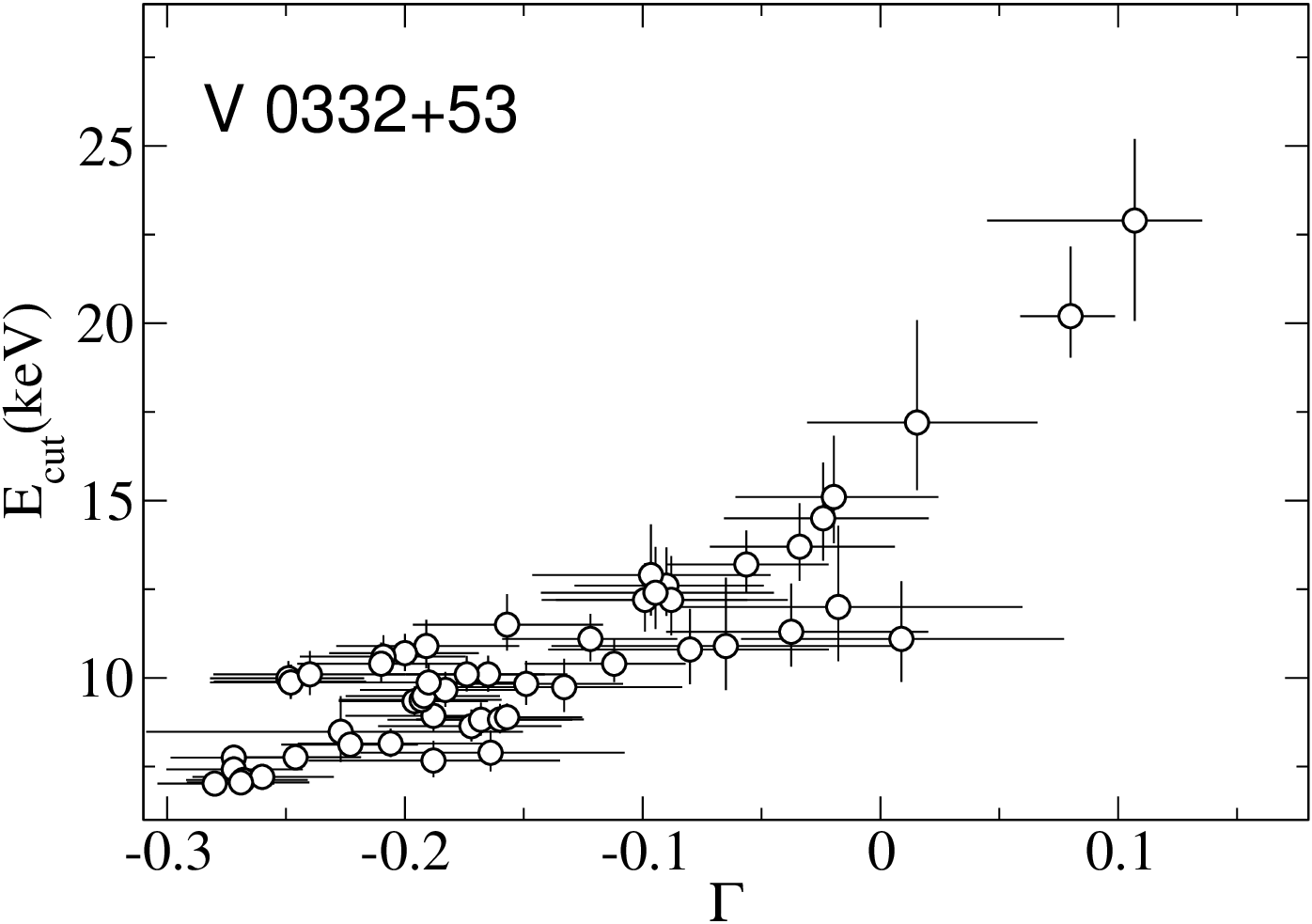}  \\
   \includegraphics[width=5cm]{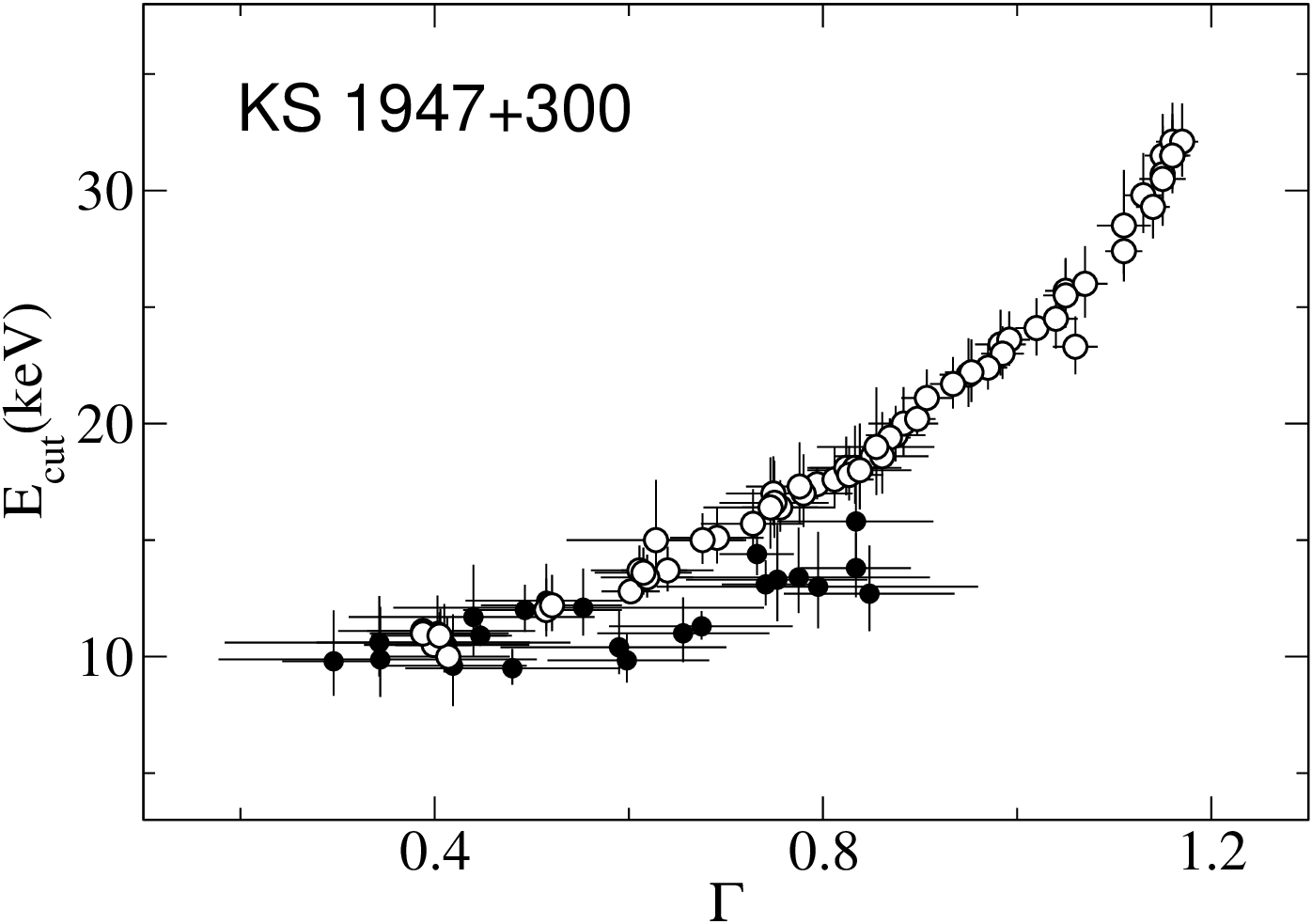}  &
   \includegraphics[width=5cm]{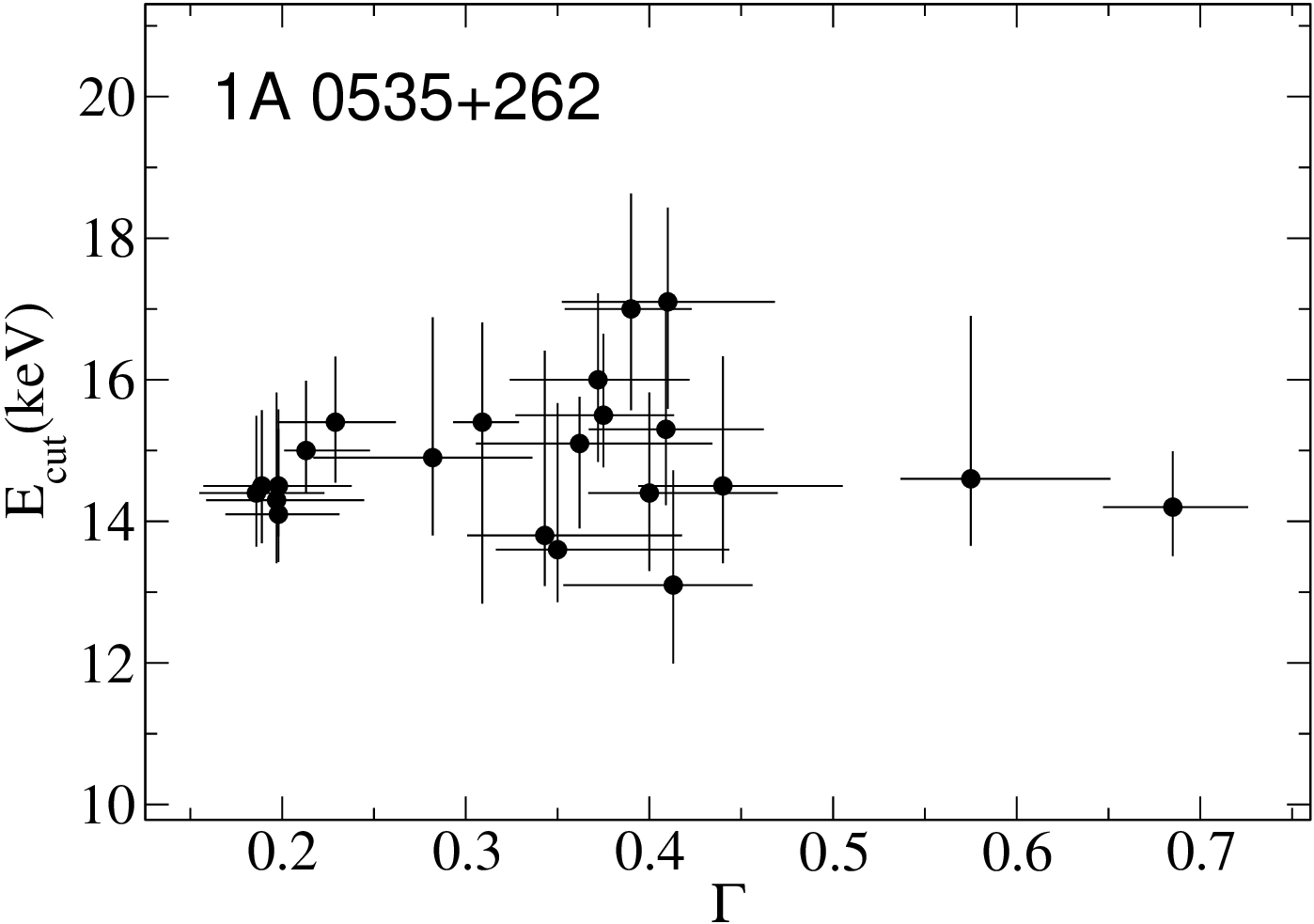}  &
   \includegraphics[width=5cm]{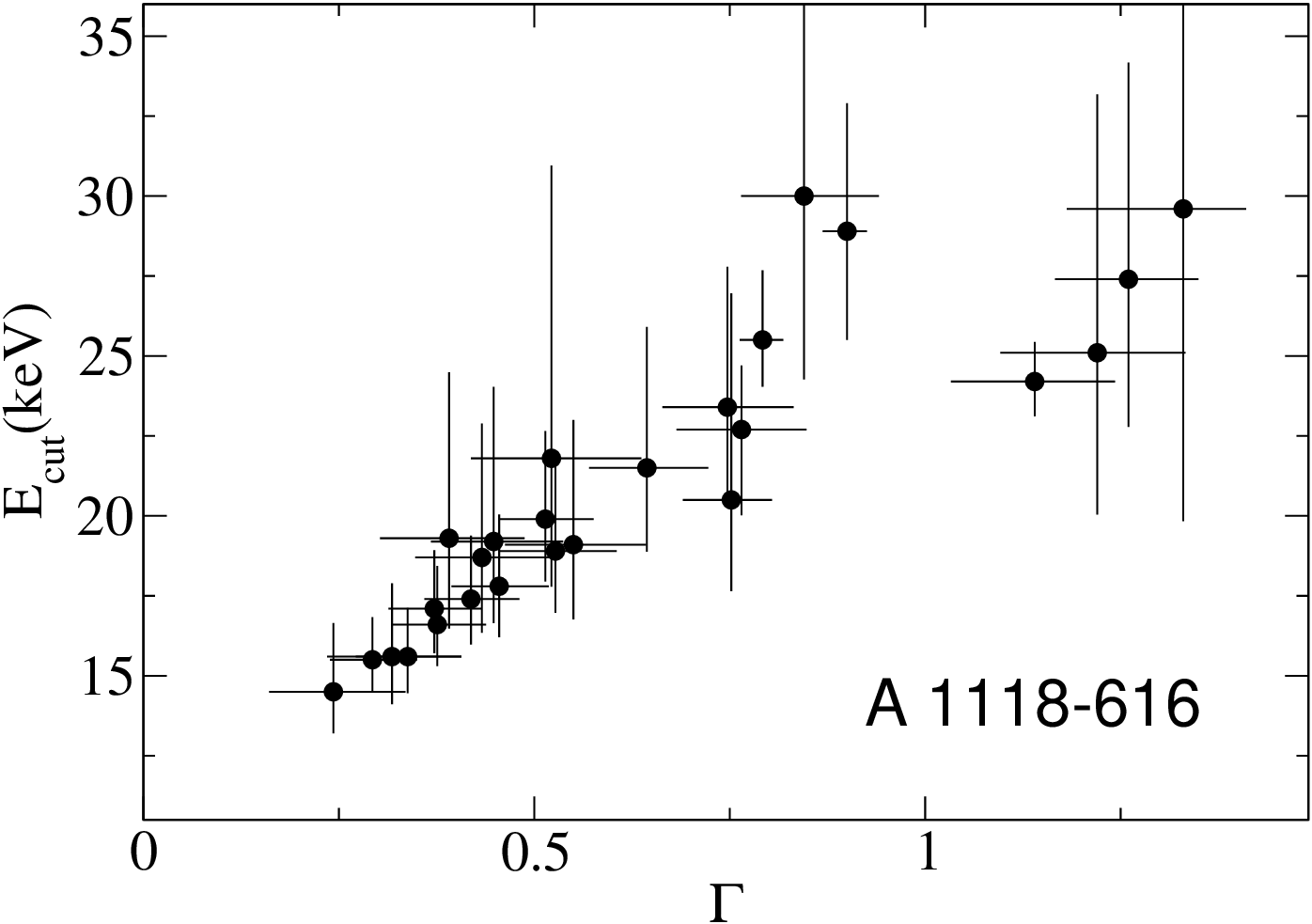}   \\
   \includegraphics[width=5cm]{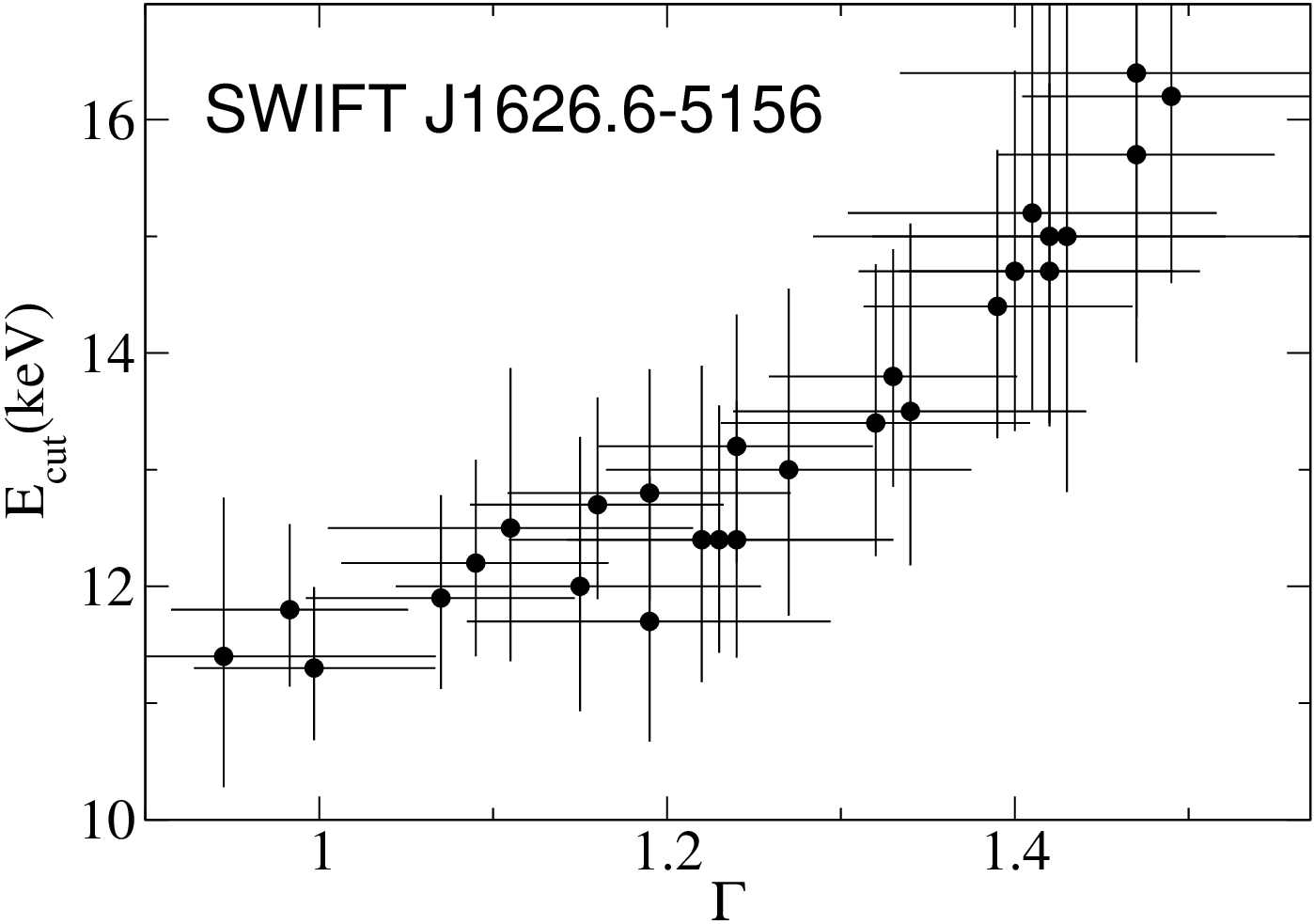}   &  
   \includegraphics[width=5cm]{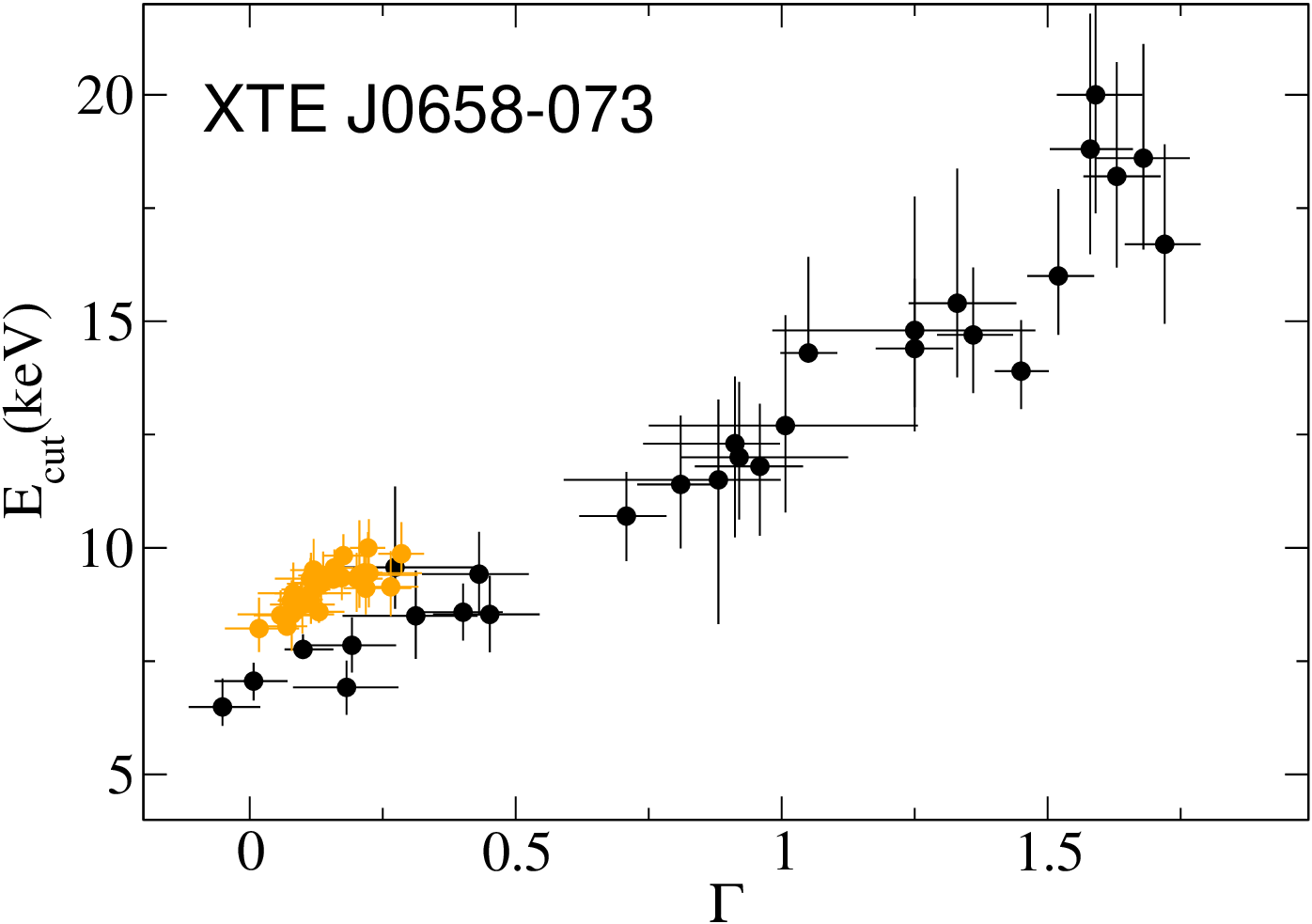}  &
   \includegraphics[width=5cm]{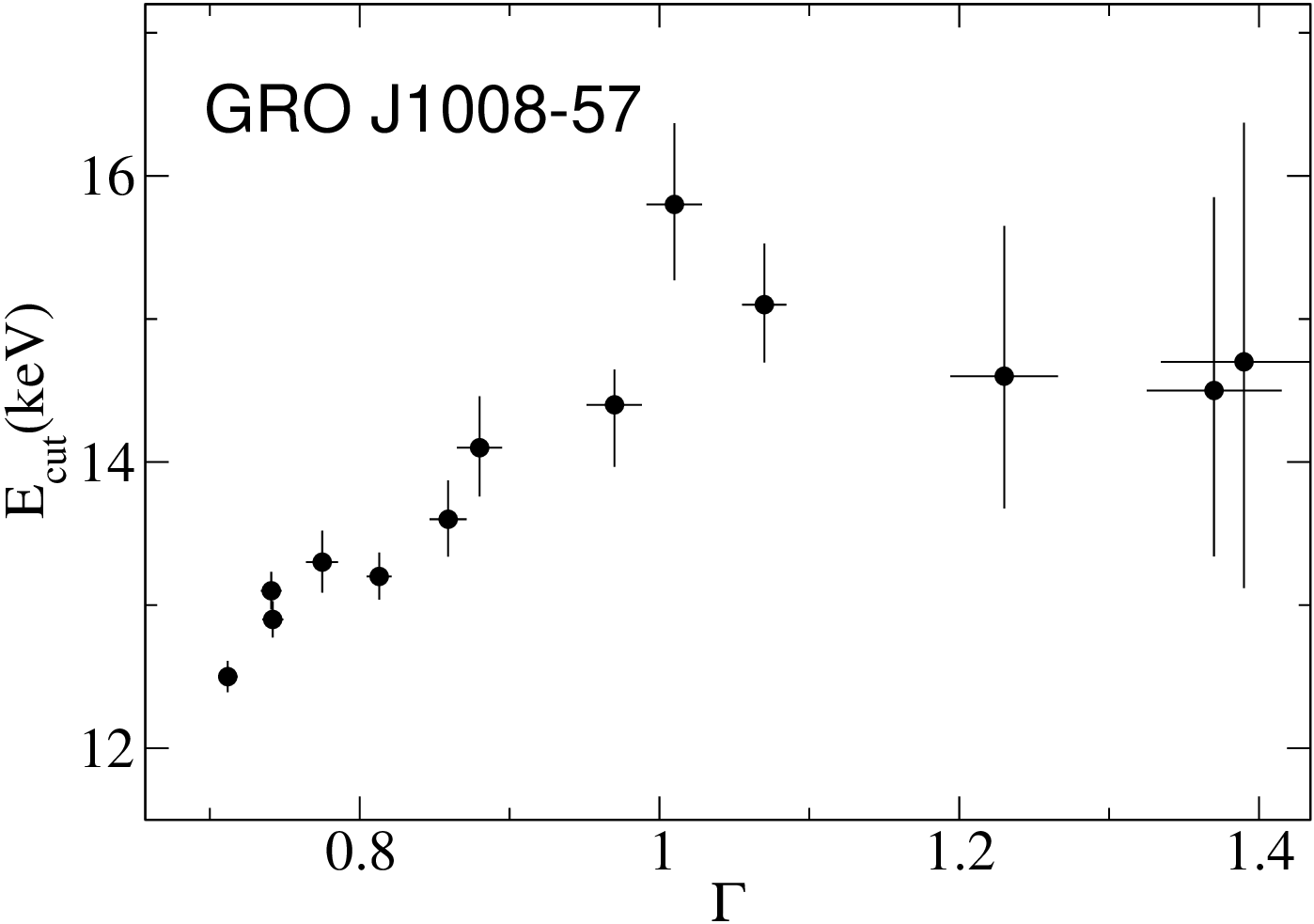}   \\   
   \end{tabular}
      \caption{Cutoff energy as a function of the photon index. Open
      circles represent the DB, filled circles 
      the HB. Orange data points
       in \xte\ correspond to the peak of the outburst. In \vcas,
  red squares represent the results from the "bump" model.}
  \label{ecut-gamma}
  \end{figure*}

   \begin{figure*}
   \centering
   \begin{tabular}{ccc}
   \includegraphics[width=5cm]{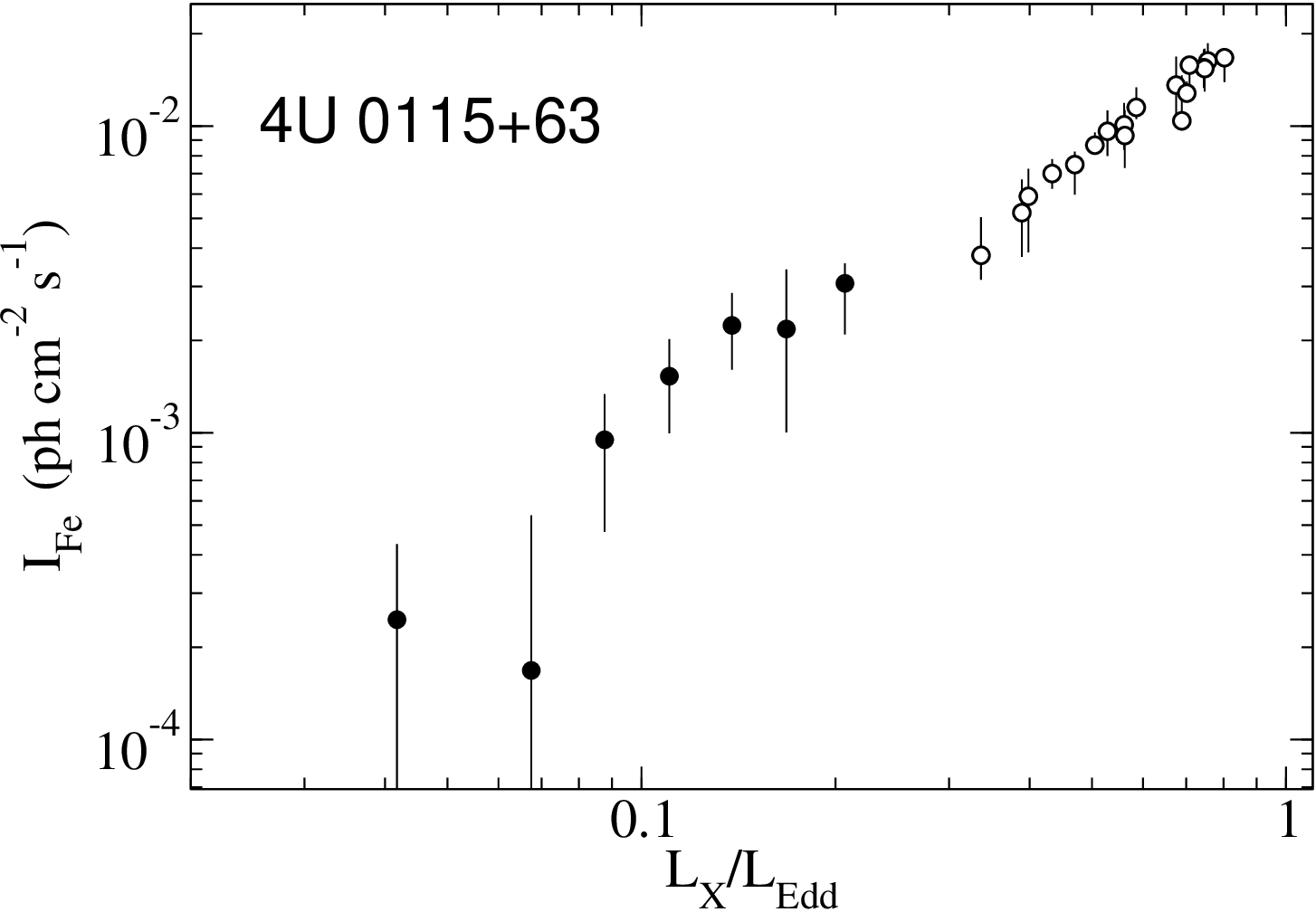}  &
   \includegraphics[width=5cm]{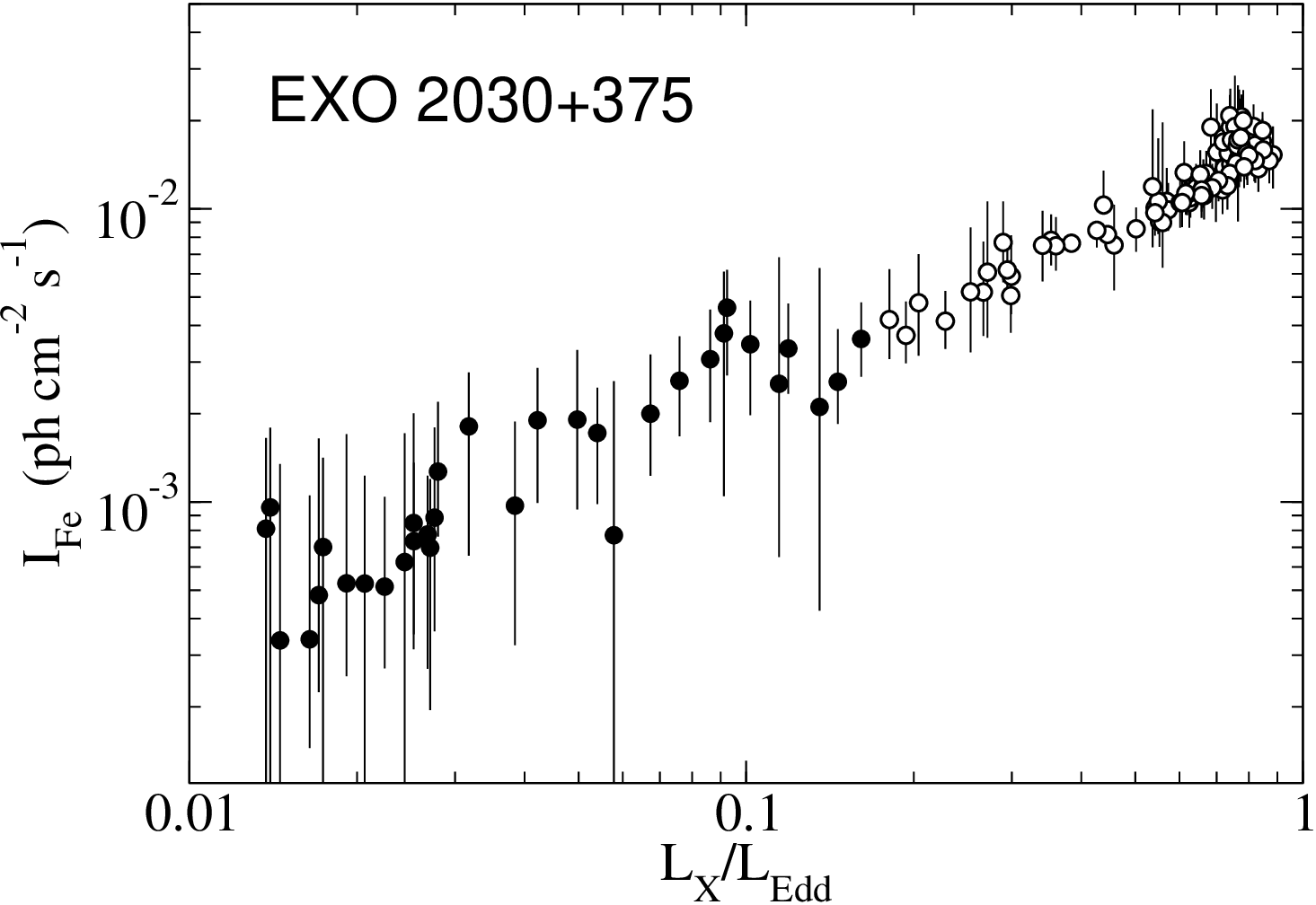} &
   \includegraphics[width=5cm]{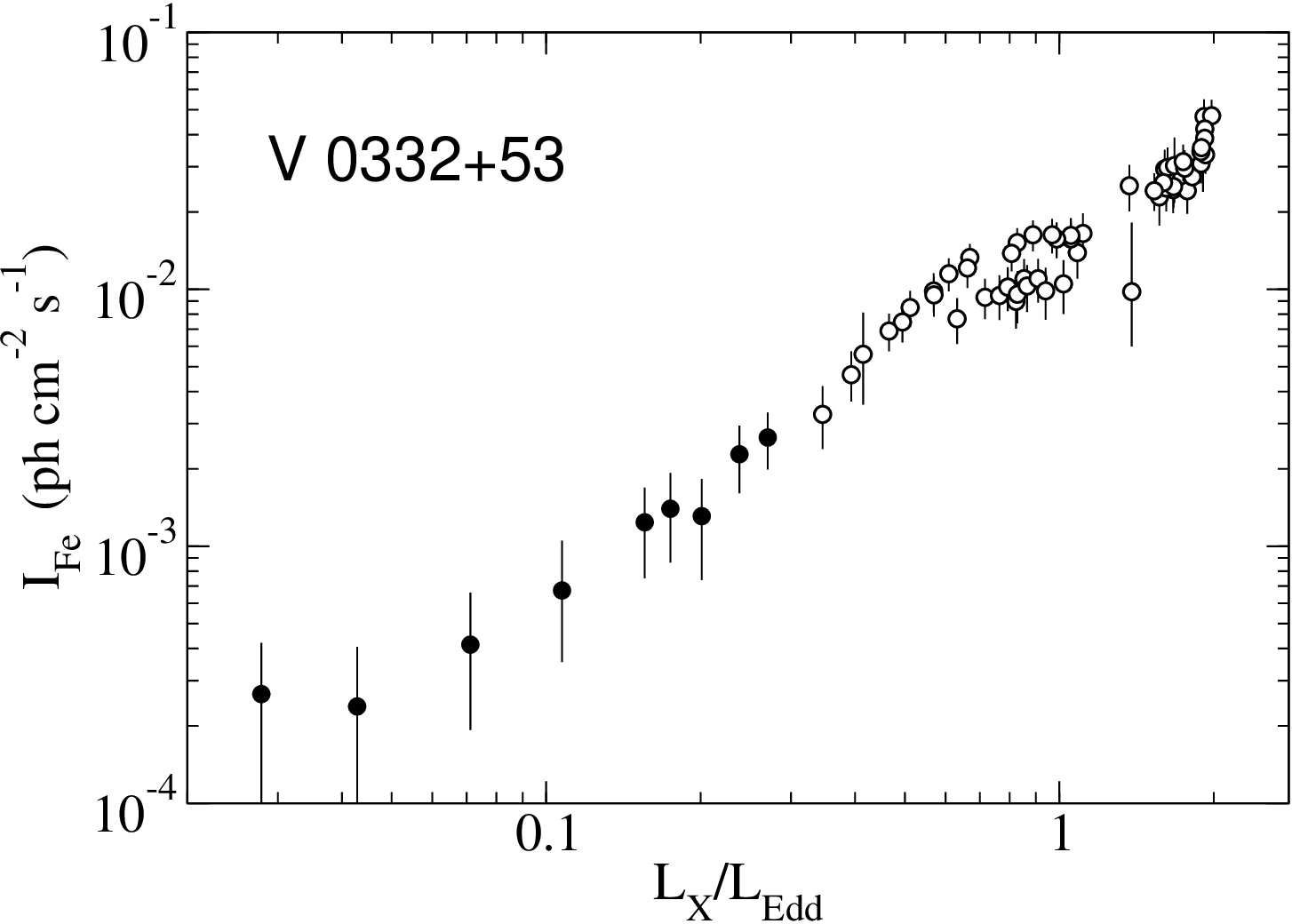}  \\
   \includegraphics[width=5cm]{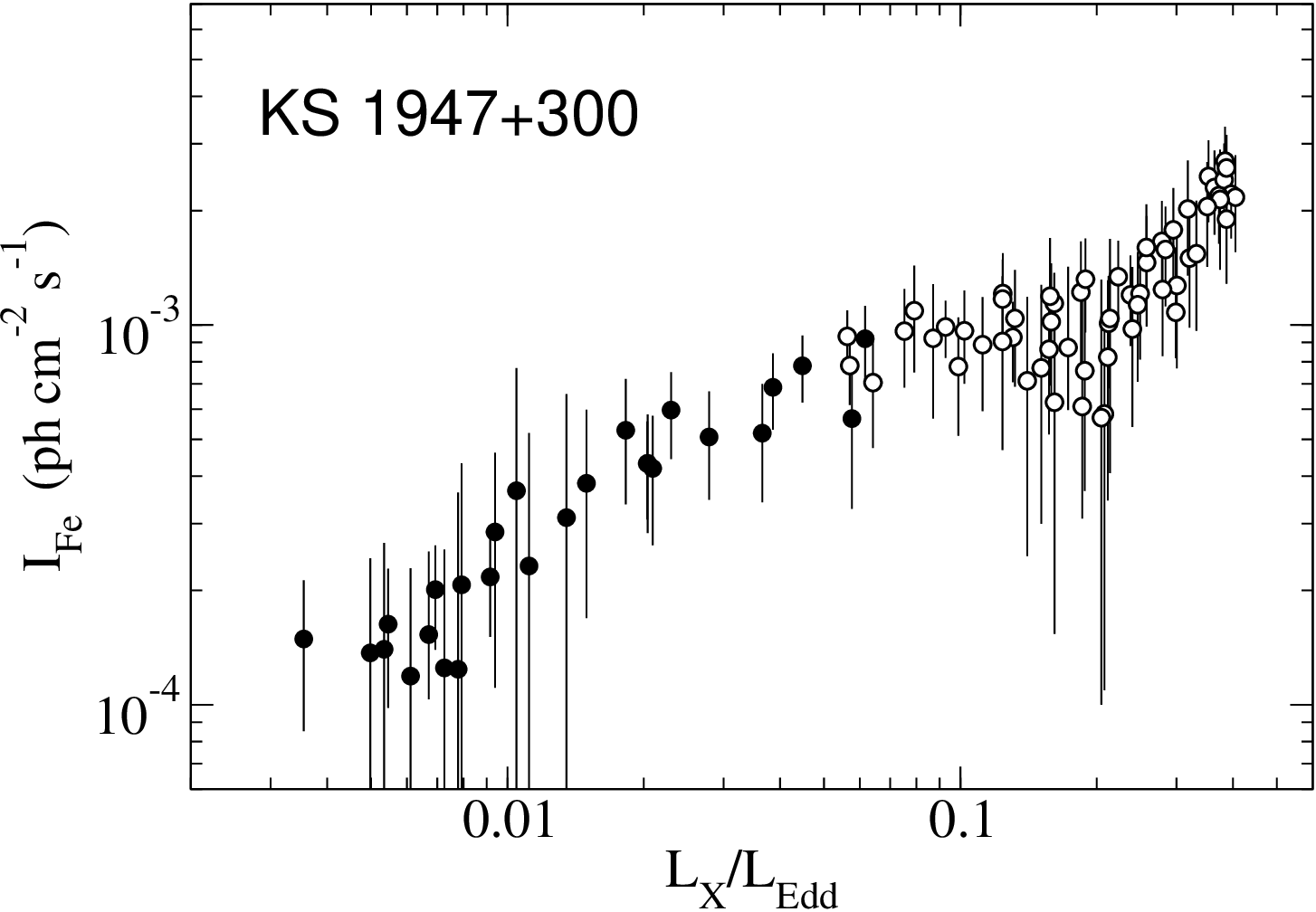}  &
   \includegraphics[width=5cm]{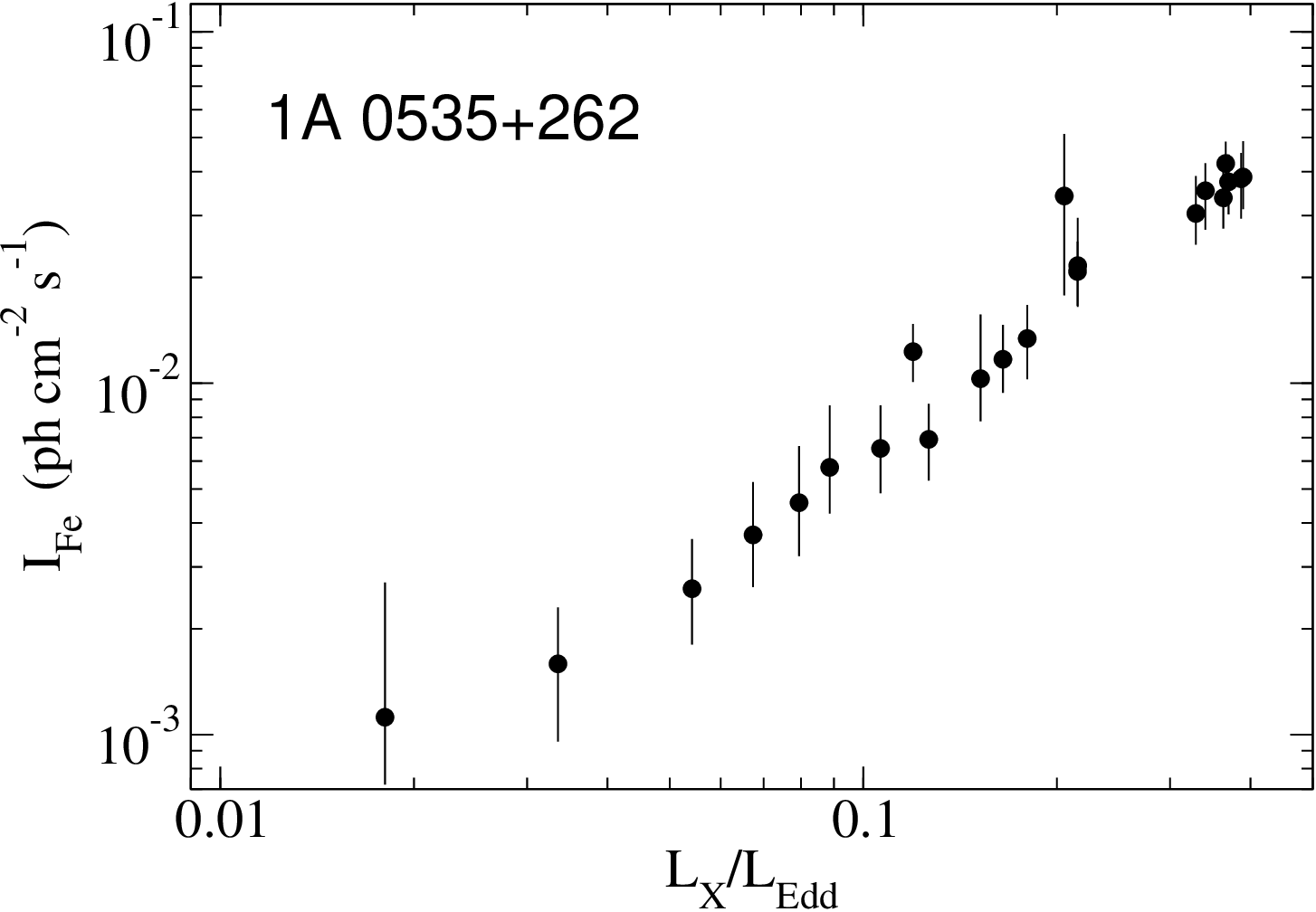}  &
   \includegraphics[width=5cm]{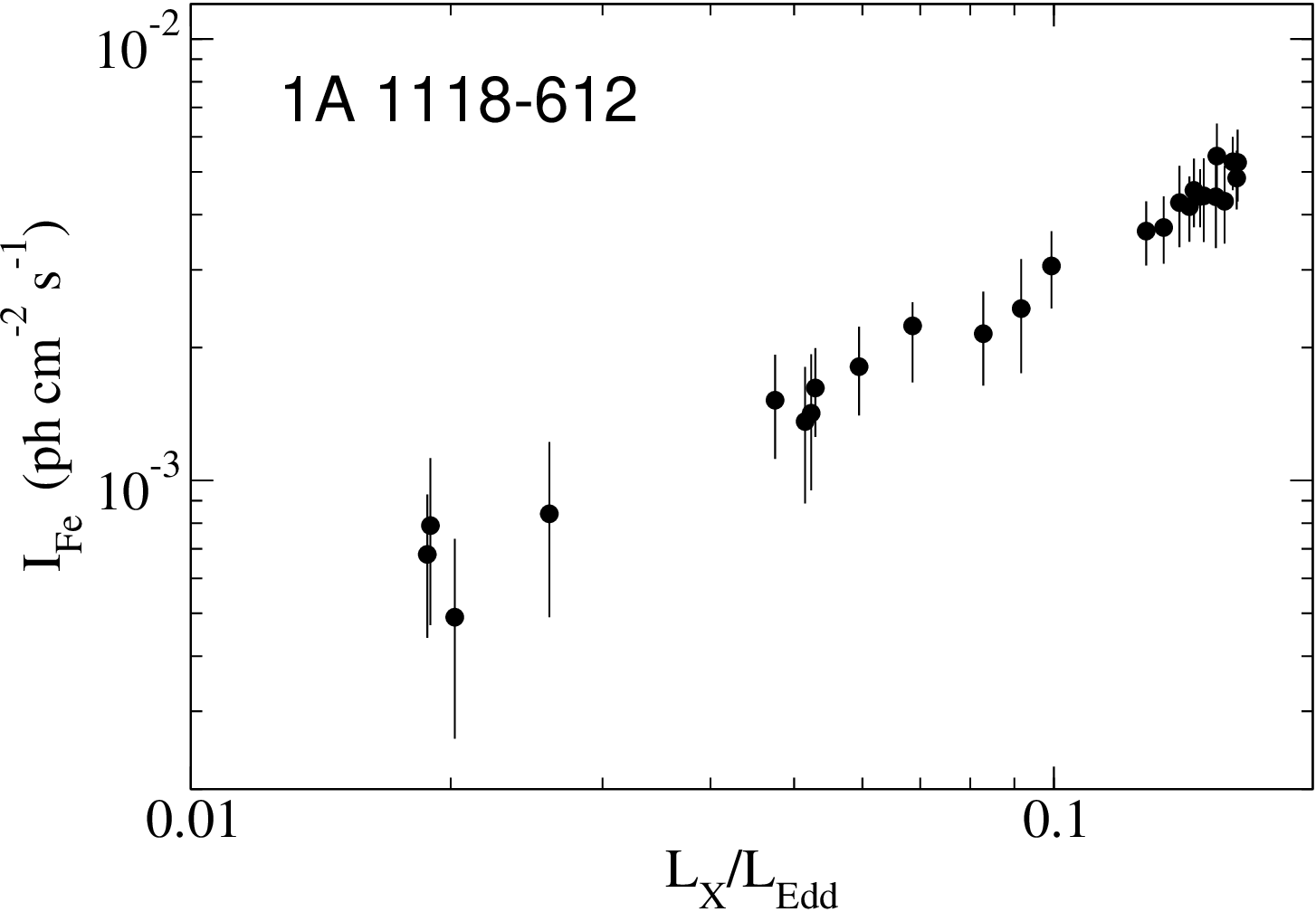}   \\
   \includegraphics[width=5cm]{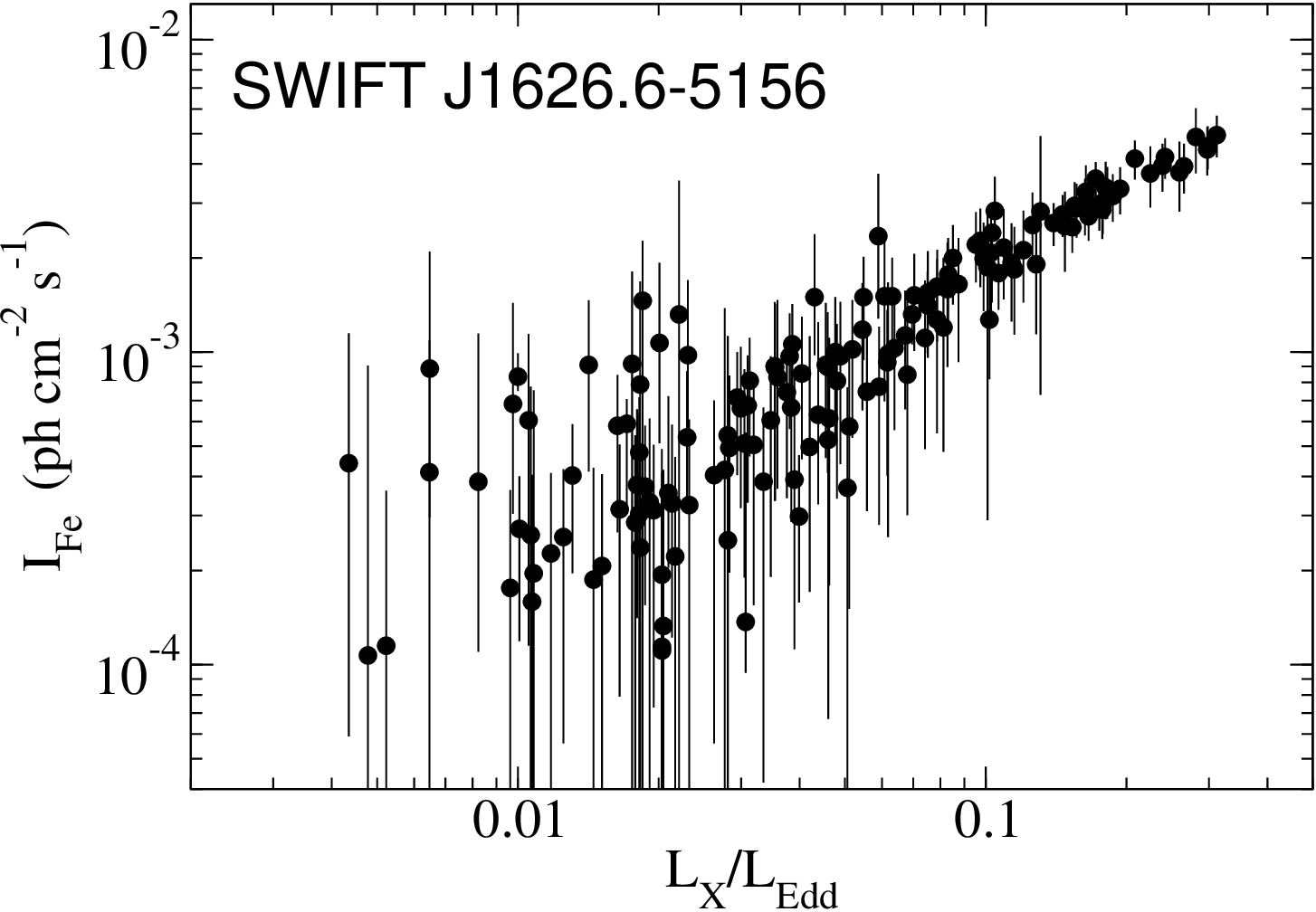}   &  
   \includegraphics[width=5cm]{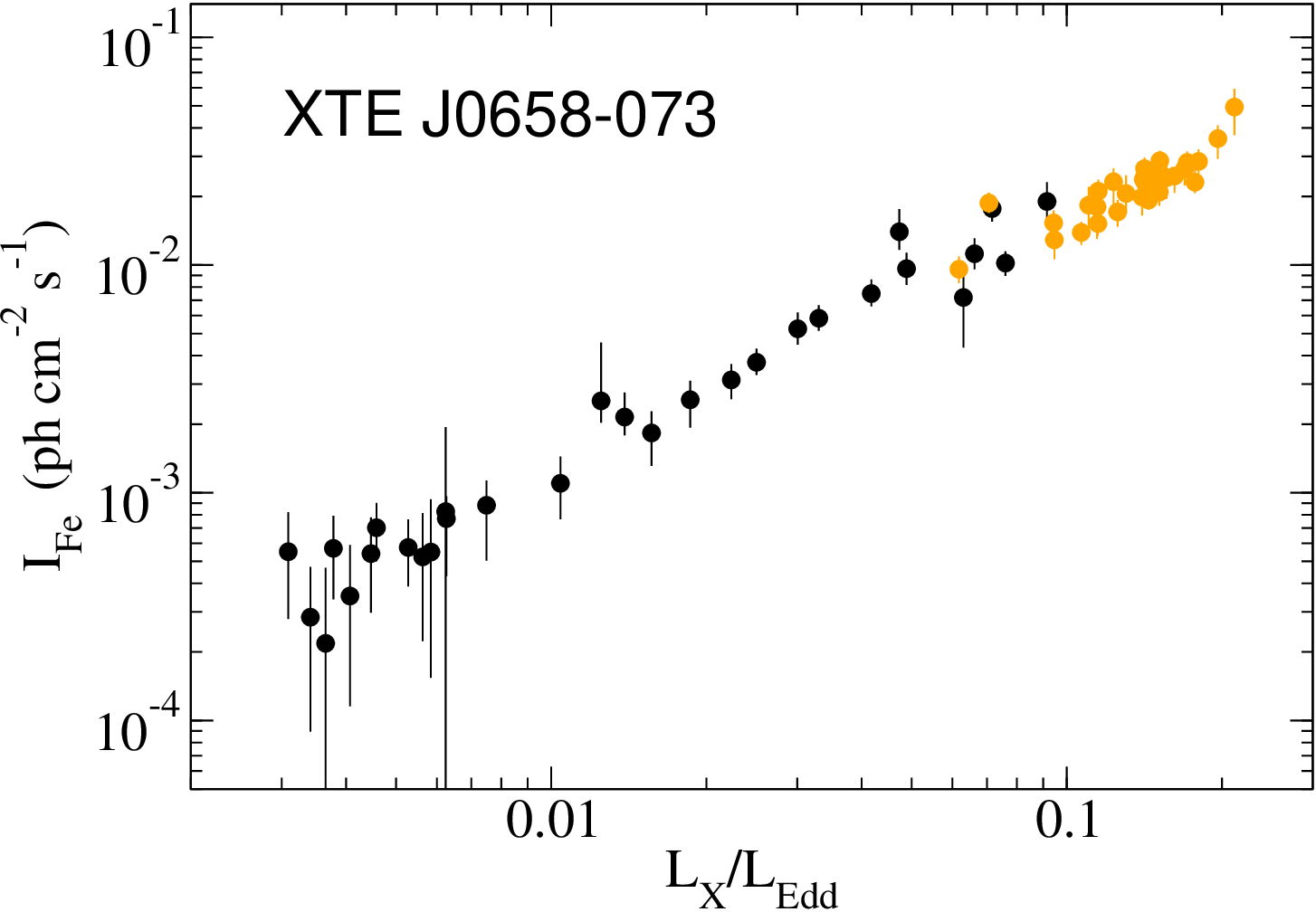}  &
   \includegraphics[width=5cm]{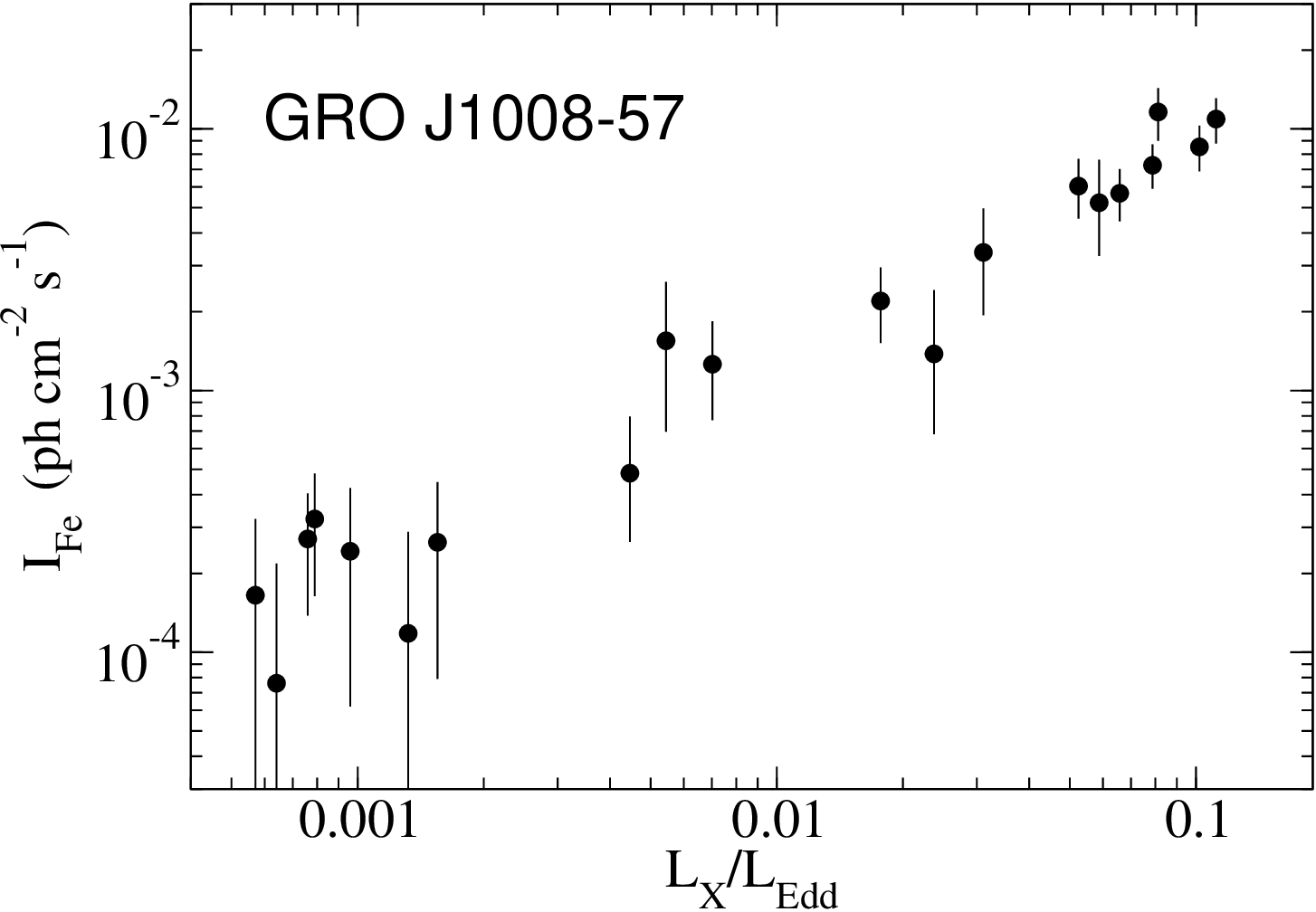}   \\   
   \end{tabular}
      \caption{Flux of the iron line as a function of the X-ray continuum flux.
      We assume $L_{\rm Edd}=1.7\times 10^{38}$ erg s$^{-1}$. Open
      circles represent the DB and filled circles the HB.}
  \label{iron-flux}
  \end{figure*}

\section{Results}
\label{res}

In this section we present the results of our analysis. First, we performed
an independent analysis with each one of the three techniques described
above, namely, colour-intensity diagrams, energy spectra and power
spectra.  We searched for correlations of the colour, spectral and timing
parameters as a function of X-ray flux and between various spectral
parameters. Then, we investigated whether it is possible to establish
relationship between parameters across different techniques. The parameters
involved in these correlations are the soft (SC) and hard (HC) colours, the
photon  index ($\Gamma$), the cutoff energy ($E_{\rm cut}$), the  energy
($E_{\rm Fe}$) and intensity ($I_{\rm Fe}$) of the iron line and the
characteristic frequency ($\nu_i$) and fractional amplitude of variability
($rms_i$) of the broad-band noise components.

To quantify the strength of the relationships between different parameters
we used the Pearson's correlation coefficient $r$, whereas the probability
$p$ was derived to assess the significance of the correlation. $p$
expresses the probability that the observed correlation happened by chance.
The smaller the $p$, the more significant the relationship. If $p <
\alpha$, the probability  that the relationship happened by chance is small.
$(1-\alpha)\times100$ is the confidence level (typically 95\% or 99\%). $p$
was computed from the $t-value$ defined as 

\begin{equation}
t=r \,\sqrt{\frac{N-2}{1-r^2}} 
\end{equation}

\noindent where N is the number of data points. $N-2$ gives the number of
degrees of freedom and $p=T(N-2,t)$, where $T$ is the Student's t
distribution (two-tailed).

Table~\ref{pearson} presents a compilation  of the Pearson's correlation
coefficients and their significance for most of the samples discussed in
this paper.  To have an approximate perception of the quality of the fits,
Columns 7 and 8 in this table give the mean and standard deviation of the
reduced $\chi^2$ values obtained from the fits to the energy and power
spectra, respectively. In a few cases, most notably in \vtau, the
best spectral fits gave reduced $\chi_{\nu}^2$ slightly lower than 1. In
those cases, however, the residuals confirmed that all the components
employed by the model were necessary to describe the data. The exclusion of
any of them, resulted in an overall not acceptable fit due to marked
residuals that could be easily fitted with known components. For example,
the removal of the cyclotron line in the higher-flux spectra of \vtau\
increased the value of $\chi^2$ from 50--70 for 77 degrees of freedom to
above 200 for 81 degrees of freedom. Likewise, removing the broad Gaussian
component reulted in an increase of $\chi^2$ to 90--100 for 80 degrees of
freedom.  

\begin{table*}
\caption{Pearson's correlation coefficients for the various samples. All
correlations are significant at $>99$\%, unless stated otherwise. 
The $L_X-\nu_1$ and $L_X-I_{\rm Fe}$ relationships were fitted with a power law 
function 
(linear correlation in $\log-\log$ scale). A "---" indicates that no data 
points exist. Also shown (columns 7 and 8) is the mean and standard
deviation of the reduced $\chi^2$ of the fits. In \vcas, quoted value 
corresponds to the POWER$\times$HIGHECUT model. Using the "bump" model the
$\chi^2_{\nu}$ reduces to $1.0\pm0.2$ for this source.}
\label{pearson}      
\centering          
\begin{tabular}{lccccccc}
\hline\hline
Source name	&$L_X-\Gamma$	&$\Gamma$-$E_{\rm cut}$	&$L_X-I_{\rm Fe}$	&$L_X-\nu_1$ 	&$\Gamma-\nu_1$	&$\chi^2_{\nu}$	&$\chi^2_{\nu}$ \\
		&HB/DB		&HB/DB			&all			&all		&HB/DB		&spectra 	&power spectra\\
\hline
4U 0115+63	&--0.96/0.94	&0.89/--0.49$^f$&0.98	&0.85	&--0.80$^f$/0.46$^f$	&1.5$\pm$0.4	&1.4$\pm$0.3	\\
EXO 2030+375	&--0.85/0.89	&0.80/0.98	&0.96	&0.88	&--0.82/0.37$^f$	&1.1$\pm$0.1	&2.0$\pm$0.8	\\
V 0332+53	&--0.97/--0.52	&---/0.86	&0.95	&0.83	&0.63$^f$/--0.48	&1.6$\pm$0.5	&1.3$\pm$0.4	\\	
KS 1947+300	&--0.80/0.97	&0.67/0.97	&0.90	&0.95	&--0.48$^f$/0.77	&1.1$\pm$0.4	&1.8$\pm$0.7	\\
1A 0535+262	&--0.90/---	&0.05$^g$/---	&0.98	&0.92	&--0.91/---		&0.7$\pm$0.2	&1.3$\pm$0.3	\\
1A 1118--616	&--0.96/---	&0.88/---	&0.99	&0.96	&--0.86/---		&1.2$\pm$0.2	&1.2$\pm$0.3	\\	
Swift J1626.6--5156&--0.95$^a$/---&0.93$^b$/---	&0.94	&0.45$^d$&0.02$^g$/---		&0.9$\pm$0.2	&1.6$\pm$0.6	\\
XTE J0658--073	&--0.89$^e$/---	&0.88$^b$/---	&0.99	&0.96$^e$&--0.85$^e$/---	&1.1$\pm$0.3	&1.9$\pm$0.8	\\
GRO J1008--57	&--0.84/---	&0.94$^c$/---	&0.96	&0.80	&--0.29$^g$/---		&0.8$\pm$0.2	&1.3$\pm$0.3	\\	
\hline
\multicolumn{1}{l}{$^a$: For $L_X/L_{\rm Edd} > 0.05$} &
\multicolumn{2}{l}{$^b$: For $L_X/L_{\rm Edd} > 0.1$} &
\multicolumn{2}{l}{$^c$: For $L_X/L_{\rm Edd} > 0.01$} &
\multicolumn{2}{l}{$^d$: For $L_X/L_{\rm Edd} < 0.2$}  \\
\multicolumn{1}{l}{$^e$: For $L_X/L_{\rm Edd} < 0.1$} &
\multicolumn{2}{l}{$^f$: 95\% confidence level} &
\multicolumn{2}{l}{$^g$: $<$80\% confidence level}  &\\
\end{tabular}
\end{table*}

\subsection{Hardness-intensity diagrams}
\label{col}

Paper I and Paper II showed that hard X-ray transients exhibit two distinct
spectral branches in the HID that were called the horizontal branch (HB)
and the diagonal branch (DB) in Paper II. As in LMXBs, these names were
adopted because of the pattern that the source traces in the HID.
Figure~\ref{hid} shows a characteristic example of HID, that of the source
\ks, where the two branches can be clearly distinguished. The average
intensity in the HB is always lower than that of the DB.

The HB appears horizontal in Fig.~\ref{hid} because a linear scale was used
for the Y-axis. Had we used a logarithmic scale, the HB would also have a
diagonal pattern, although with opposite slope (Fig.~\ref{rate-SC}). In
this work we shall use a logarithmic scale because it facilitates the
comparison {\em i)} with other fainter accreting X-ray pulsars, as the
behaviour of the points that populate the HB stands out clearer and {\em
ii)} with the HID of BHB and LMXB, as for this type of binaries a
logarithmic scale is the usual way to represent the count rate.
Nevertheless, for the sake of consistency with previous work, we shall keep
the term {\em horizontal branch} to designate the low-intensity state and 
{\em diagonal branch} for the high-intensity state.

The  hardness-intensity diagrams (HID) of all the sources analysed in this
work, where the count rate in the 4-30 keV band is plotted as a function of
the soft colour, are shown in Fig.~\ref{rate-SC}. An important result that
emerges from the colour analysis is that only the brightest sources display
the DB (columns 10 and 11 in Table~\ref{info} give an estimate of the peak
luminosity). However, the X-ray luminosity does not seem to be the only
parameter triggering the transition between branches. \ks\ and \vtau\
exhibit similar peak luminosity but only \ks\ displays the two branches. We
shall come back to this different behaviour in Sect.~\ref{discussion} and
show that the magnetic field strength plays a crucial role in determining
the critical luminosity at which the transition takes place.

Due to their unpredictable nature, the initial stages of the rise of the
outbursts lack a proper coverage. Normally, non-scheduled observations are
triggered once the source is relatively bright and the reaction timescales
are often similar to the time scale of the rise. Only one system, \ks, was
observed for the complete duration of the outburst. In all other systems,
the first data point is located, in the best cases, half way through the
rise. We shall take \ks\ as an example of the motion of the source in the
HID. As the X-ray flux increases the SC increases, i.e., the source moves
right and depicts the HB (Fig.~\ref{hid}, see also Fig.~\ref{rate-SC}).
When the luminosity reaches a critical value,
the source enters the DB by making a sudden turn in the HID: it moves left
and the SC starts to decrease. The peak of the outburst corresponds to the
softest state of the DB. Then, as the flux decreases, the source moves back
following the same track but in the opposite direction. Below the critical
luminosity, the source enters the HB and the SC begins to decrease again.
Arrows in Fig.~\ref{hid} indicate the flow of time. In the sources that
exhibit only the HB, the peak of the outburst corresponds to the hardest
spectrum (largest SC).

Unlike the X-ray spectra of BHBs and LMXBs where the hard emission ({\it
i.e.,} above 10 keV) is well characterised by a simple power law component,
the higher energy part of the X-ray spectra of \bex\ is affected by extra
components such as exponential cutoffs and cyclotron lines and their
harmonics. As a result, the X-ray colours  do not always represent a
reliable measurement of the spectral slope. The HC is expected to be more
affected by the distortion of the continuum because these extra components
mainly appear above 10 keV. If the spectrum could be well represented by a
single power law, then we would expect the X-ray colours to decrease as the
emission becomes softer, that is, we would expect an {\em anticorrelation}
between the photon index and the colours. The SC in both branches and the
HC in the horizontal branch display this anticorrelation, albeit with large
scattering in the case of the HC (Fig.~\ref{gamma-HC}).  However, the
evolution of the HC in the DB is the opposite to what it is expected, that
is,  the HC increases but the gamma also increases, {\it i.e.,} the
spectrum becomes softer. An intriguing case is \bq\ (pink squares),
whose behaviour differs from the other sources by displaying a {\em
positive} correlation in three out of the four panels of
Fig.~\ref{gamma-HC}. This peculiar behaviour of the colours of \bq\ was
already reported in Paper II. Clearly, the energy spectra of BeXBs above
10 keV are more complex than a simple power law.

   \begin{figure}
   \centering
   \includegraphics[width=8cm]{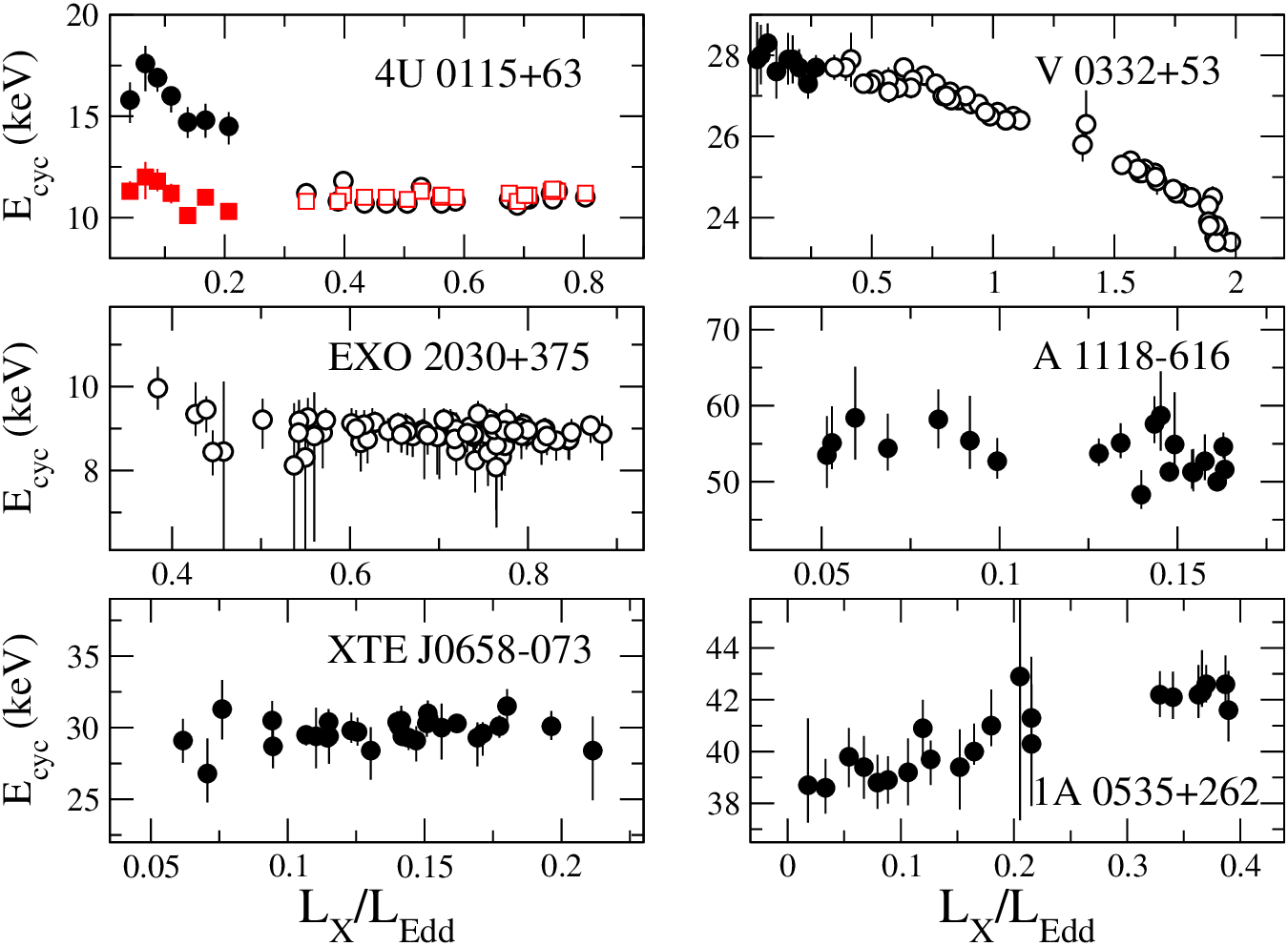}
  \caption{Variation of the energy of the cyclotron line with luminosity
  (3--30 keV). 
  Filled circles correspond to the HB and open circles to the DB. In \vcas,
  red squares represent the results from the "bump" model, whereas circles
  are the best-fit cyclotron energy with the POWERLAW$\times$HIGHECUT
  description of the spectral continuum. }
  \label{cyc-flux}
  \end{figure}

   \begin{figure}
   \centering
   \begin{tabular}{c}
   \includegraphics[width=8cm]{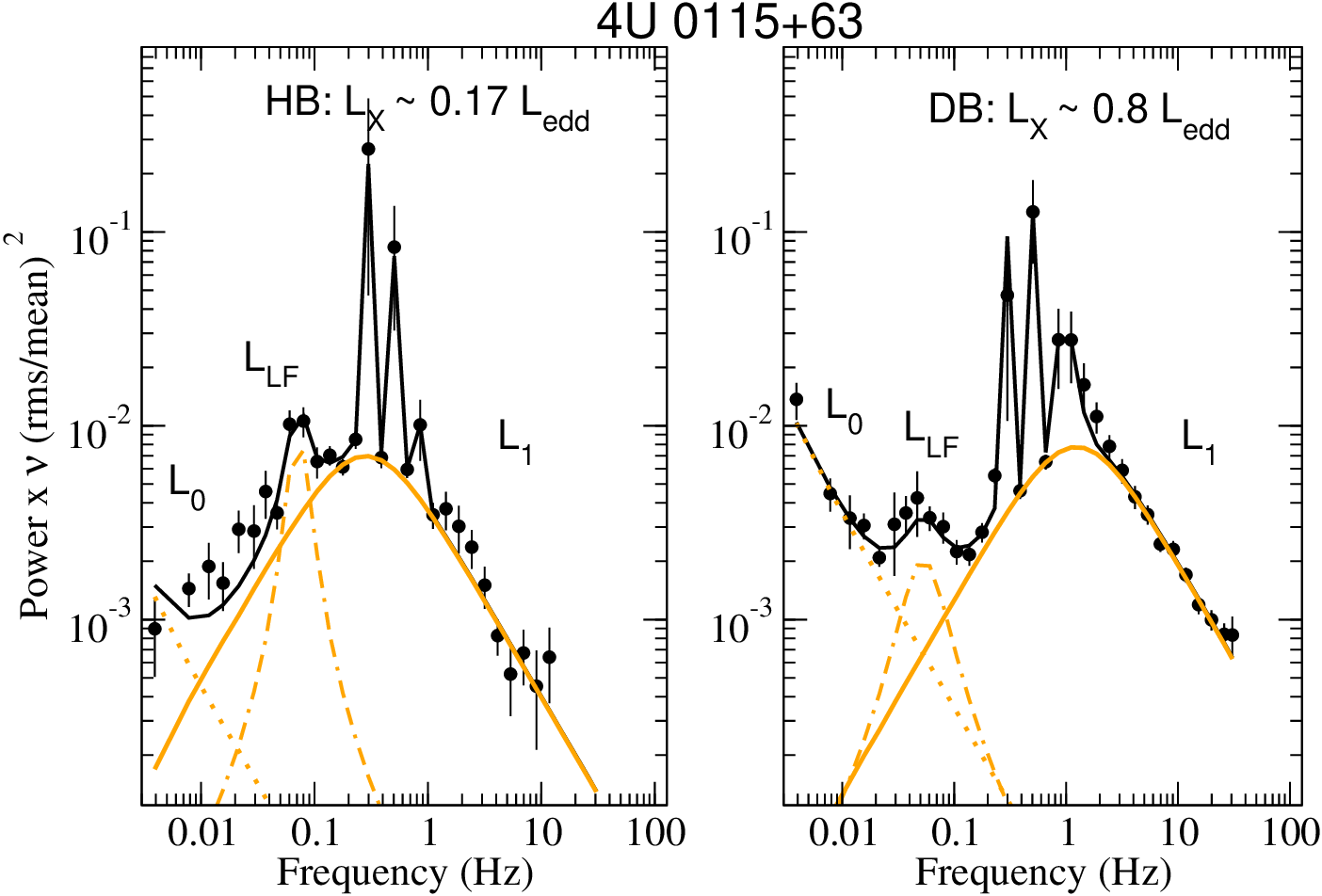} \\
   \includegraphics[width=8cm]{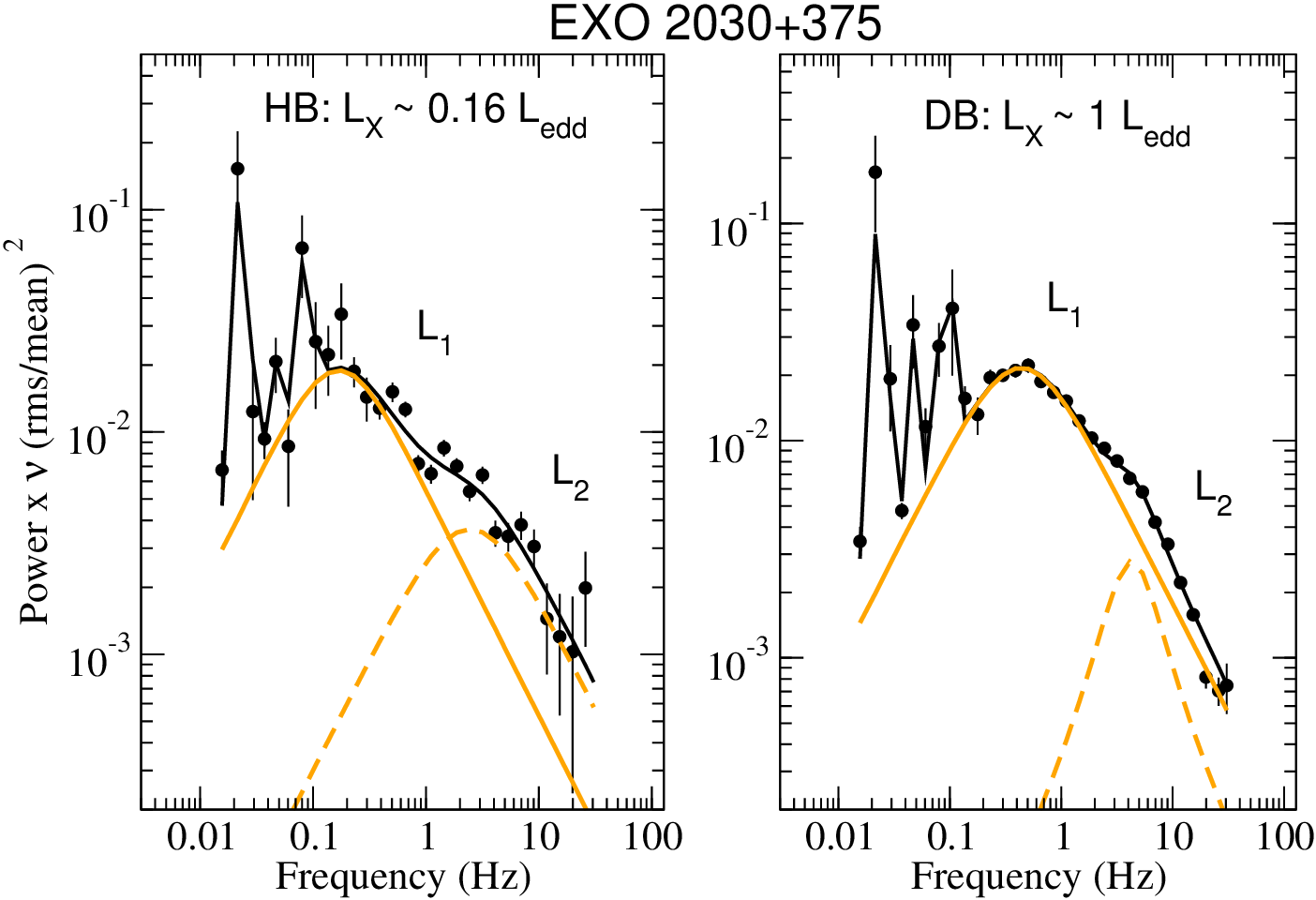} \\
   \includegraphics[width=8cm]{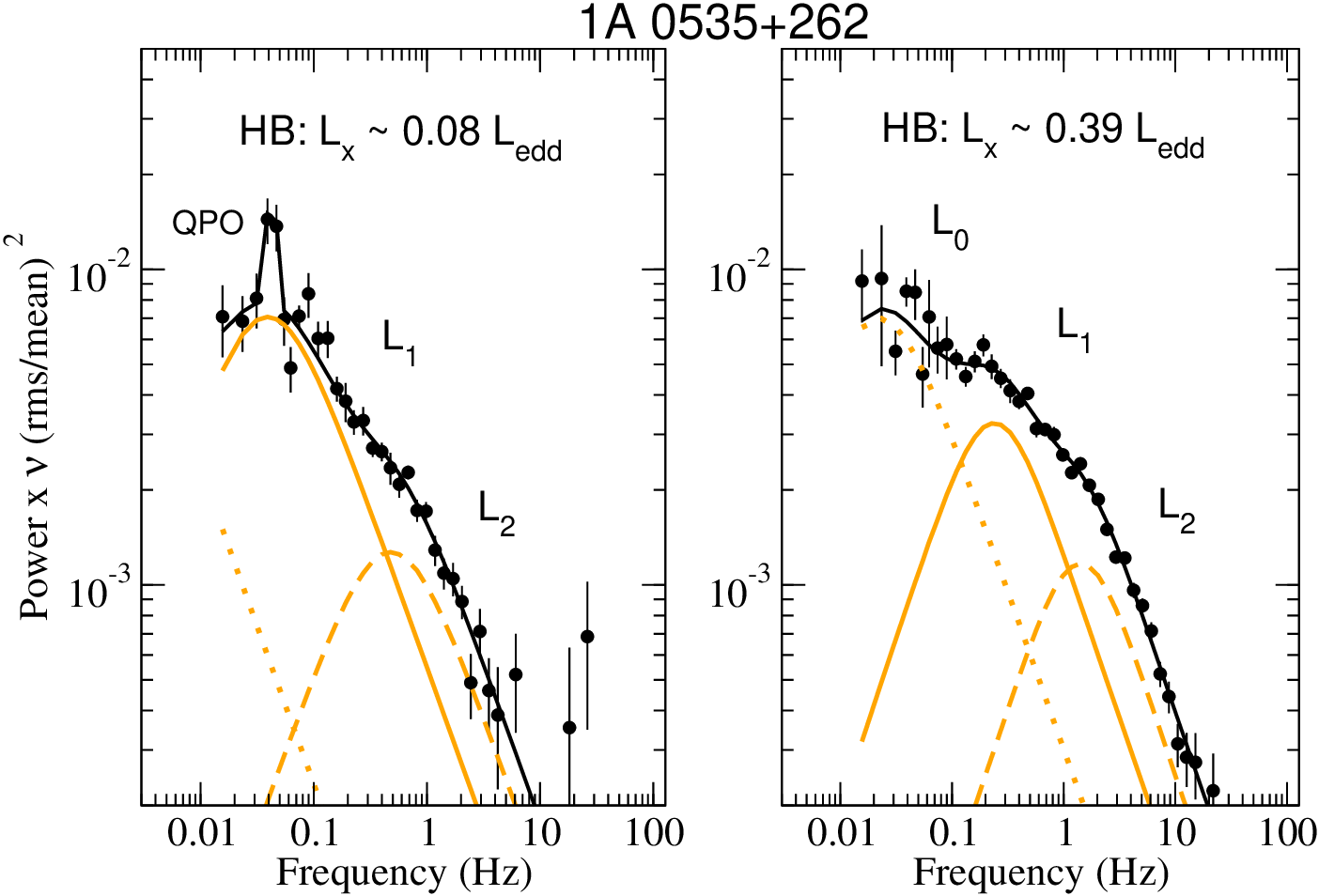} \\   
  \end{tabular}
  \caption{Representative power spectra of the horizontal and diagonal
  branches at various flux levels for the fast-rotating pulsar \vcas\ (up),
   \exo\ (middle) and the slow pulsator \vtau\ (down). Note the 
  shift toward higher frequencies at high flux. In \vtau, $L_0$ is not seen
  at low luminosities because it lies outside of the frequency interval 
  considered. Other features such as $L_{\rm LF}$ in \vcas\ or a QPO at 
  $\sim$40 mHz in \vtau\ are shown. The Lorentzian profiles of the pulse peaks 
  and the QPO were removed for clarity.}
  \label{psd}
  \end{figure}

\subsection{Spectral variability}

Although the energy spectra of all sources can be fitted with the models
outlined in Sect.~\ref{specana}, the different components (power-law,
cutoff,  cyclotron line) fit different parts of the spectrum in different
sources. Hence a direct comparison of the actual value of some of the
spectral parameters ({\it i.e.}, photon index) between different sources
may be misleading.  Nevertheless, because the spectral parameters vary
smoothly along the outbursts, the search for correlations among different
parameters and also with X-ray flux in each individual source is
meaningful. Figure~\ref{spectra} shows some representative spectra for
various flux levels of the HB and DB. The quoted X-ray luminosity in this
figure was derived for the 2--100 keV range and normalised to the Eddington
luminosity for a 1.4$\msun$ neutron star.  See Table~\ref{info} for the
assumed distance estimation.

The following general results are common to all or most of the sources:

\begin{itemize}

\item The photon index anticorrelates with X-ray flux in the HB and
correlates with it in the DB (Fig.~\ref{gamma-flux}). This means that as
the flux increases, the spectrum becomes harder in the HB and softer in the
DB.  In either branch, these correlations are strong and significant. The
Pearson's correlation coefficient for the variation of the photon index
with flux range is $\rho \simmore 0.9$ in most cases.  Although
correlations between the power-law photon index and the X-ray luminosity
have been reported in the past \citep{reynolds93,devasia11,klochkov11},
this is the first time that a change in the slope of the correlation is
seen, in correspondence with the position of the source in the HID. 
For \vcas, the results from the two different ways to describe the spectral
continuum are plotted in Fig.~\ref{gamma-flux}. Circles correspond to the
POWERLAW$\times$HIGHECUT model and squares to the CUTOFFPL+GAUSS (or bump)
model, where GAUSS is a broad emission Gaussian profile that accounts for
the "10-keV feature" \citep{muller12}.

\item The cutoff energy approximately follows the same trend as the photon
index, namely, it increases as the flux decreases in the HB and increases
as the flux increases in the DB. As a result there is a positive
correlation between the photon index and the high-energy cutoff: softer
spectra correspond to high values of the cutoff energy
(Fig.~\ref{ecut-gamma}). 


\item  The fluorescent iron K$\alpha$ line feature that results from
reprocessing of the hard X-ray continuum in relatively cool matter is
ubiquitous in all sources. Near neutral iron generates a line centered at
6.4 keV. As the ionisation stage increases so does the energy of the line.
However, the energy separation is so small that even ionised iron up to Fe
XVIII can be thought as part of the 6.4 keV blend 
\citep{ebisawa96,liedahl05}. The central energy of this component does not
vary significantly during the outbursts and it is consistent with
"near-neutral" material, that is, with the composite of lines from Fe II-Fe
XVIII near 6.4 keV. The flux of the line increases with the continuum
X-ray flux, indicating that as the illumination of the cool matter
responsible for the line emission increases, so does the strength of the
line (Fig.~\ref{iron-flux}). In general, the line equivalent width remained
insensitive to luminosity changes. The strong correlations of
Fig.~\ref{iron-flux} imply that the contribution from thermal hot plasma
located along the Galactic plane \citep{yamauchi09} is not significant. We
would expect the ridge emission to affect the data only for very faint
observations, {\it i.e.}, at the start and end of the outbursts
\citep{ebisawa08}. The small deviation from the linear trend with flux at
very low fluxes would be consistent with this statement
(Fig.~\ref{iron-flux}). 

\item There is no universal trend between the energy of the cyclotron line
and X-ray luminosity (Fig.~\ref{cyc-flux}). Previous studies have shown a
strong anticorrelation in \bq\ \citep{tsygankov10}, somehow weaker
anticorrelation in \vcas\ \citep{mihara04,nakajima06} and no correlation in
\vtau\ \citep{caballero07}. We confirm the decrease of the cyclotron line
energy with luminosity in \bq, but we do not find a smooth anticorrelation
in \vcas. When the POWERLAW$\times$HIGHECUT model is used, the energy
of the cyclotron line  remains fairly constant within each branch, but its
value is higher in the HB ($\sim16$ keV) than in the DB ($\sim11$ keV).
This result agrees with that of \citet{tsygankov07} and \citet{li12}, who
reported a discontinuity in the $E_{\rm cyc}-L_X$ relationship of this
source with larger values of the cyclotron line energy at low flux.
However, if the continuum is fitted with the CUTOFFPL model and a broad
emission Gaussian, then no correlation is seen, in agreement with 
\citet{muller12}. In Fig.~\ref{cyc-flux},  red squares represent the
best-fit values of the energy of the cyclotron line using the "bump" model,
while circles represent those obtained with the high-energy exponential
cutoff model. 

Our results also show no clear trend between the energy of the
cyclotron line and luminosity in \hen, \exo, and \xte.  However, we report,
for the first time, a positive correlation between the cyclotron energy and
X-ray luminosity in \vtau\ on long timescales (Fig.~\ref{cyc-flux}). A
correlation had been reported  only in the pulse-to-pulse analysis
\citep{klochkov11}. Previous studies did not find any evidence for
variability \citep{terada06,caballero07}.  However, note that these works
show observations of \vtau\ when the X-ray luminosity was $\simless 1
\times 10^{37}$ erg s$^{-1}$ ($L_X/L_{\rm Edd} < 0.1$). As can be seen in
Fig.~\ref{cyc-flux}, the energy of the cyclotron line begins to increase
above this value\footnote{In fact, the two data points above $\sim 1 \times
10^{37}$ erg s$^{-1}$ in Fig. 4 of \citet{terada06} and in Fig. 5 of
\citet{caballero07} do not rule out an increase of the cyclotron line
energy at higher luminosity. }. We verified that this relationship cannot
be due to any artificial correlation between the line energy and the other
parameters of the model by plotting the contours of the confidence regions
of $E_{\rm cyc}$ and any other parameter. Additionally, a linear
correlation analysis gives a linear correlation coefficient of 0.89 and a
probability that the points are not correlated of $6\times 10^{-8}$. 

To ensure that the Lorentzian profile was fitting the CRSFs
and not the continuum, we carefully verified that the corresponding width
had reasonably narrow values. In all the cases, we obtained width values
compatible with previous works. For \exo\ and \vcas, which displays the
lowest energy CRSF, the obtained width is of the order of 3 keV
\citep{wilson08,muller12}; for \bq\ and \vtau, the retrieved width was of
the order of 6--8 keV \citep{tsygankov10,caballero07};
\hen\ and \xte, the systems that display the highest energy
CRSF, also show the largest width, of 12--15 keV
\citep{doroshenko10,mcbride06}.


\item The X-ray spectrum of the accreting pulsars analysed in this work is
characterised by a power law ($\Gamma=0-1)$ with an exponential cutoff in
the range 10--20 keV. However, the spectral continuum may be very distorted
by the presence of cyclotron resonant scattering features. Unlike BHBs or
anomalous X-ray pulsars where the presence of different components (disc
blackbody and power law, or two power laws) causes a sudden break in the
continuum, the X-ray spectral continuum of accreting pulsars does not show
any evidence for an abrupt change below or above 10-20 keV
(Fig.~\ref{spectra}), despite the fact that physical models such as those
by \citet{becker07} and \citet{ferrigno09} show that several well-defined
physical components ({\it i.e.,} thermal and bulk Comptonisation, blackbody
emission at the base of the accretion column) are present.

\end{itemize}

   \begin{figure*}
   \centering
   \begin{tabular}{ccc}
   \includegraphics[width=5cm]{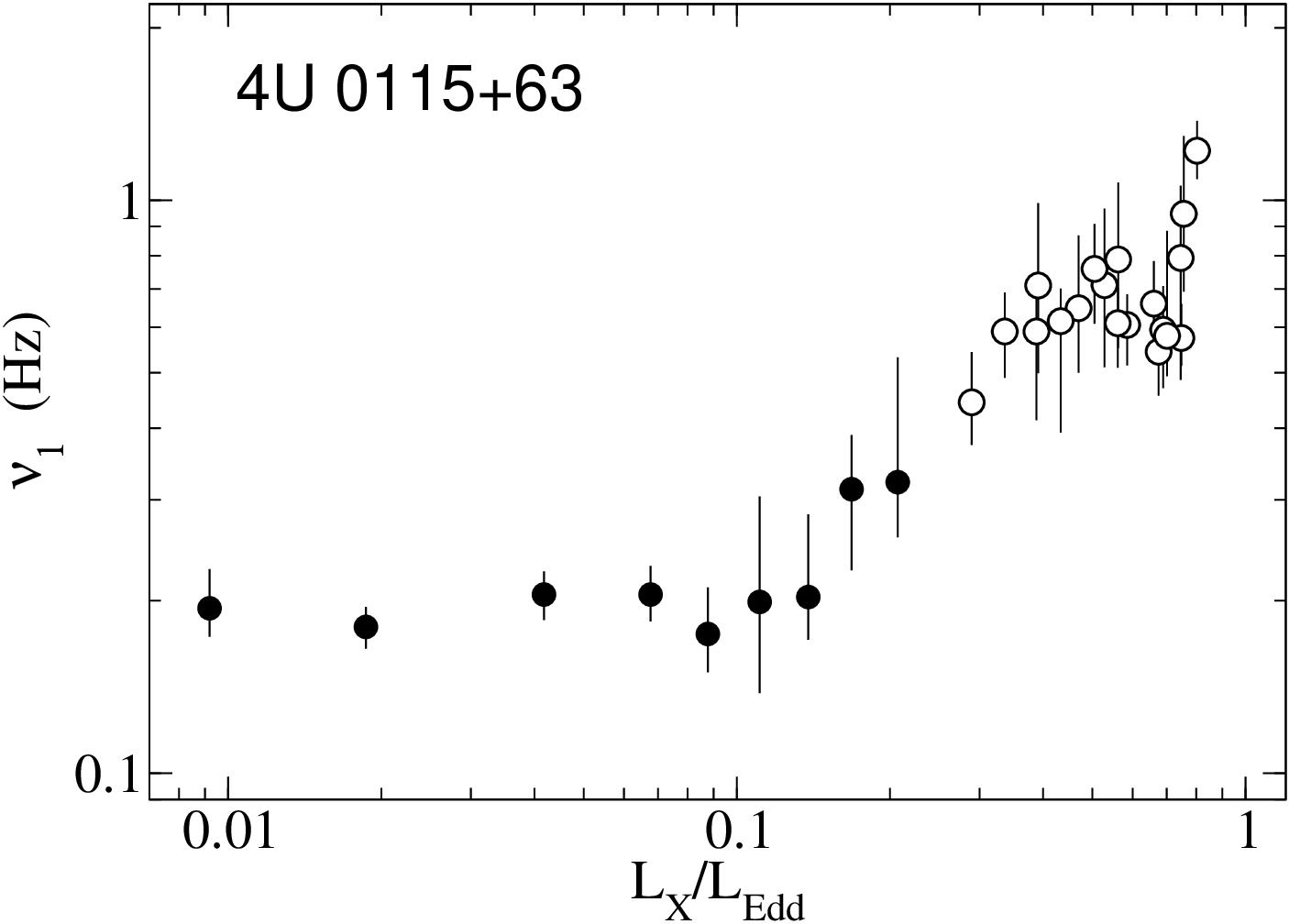}  &
   \includegraphics[width=5cm]{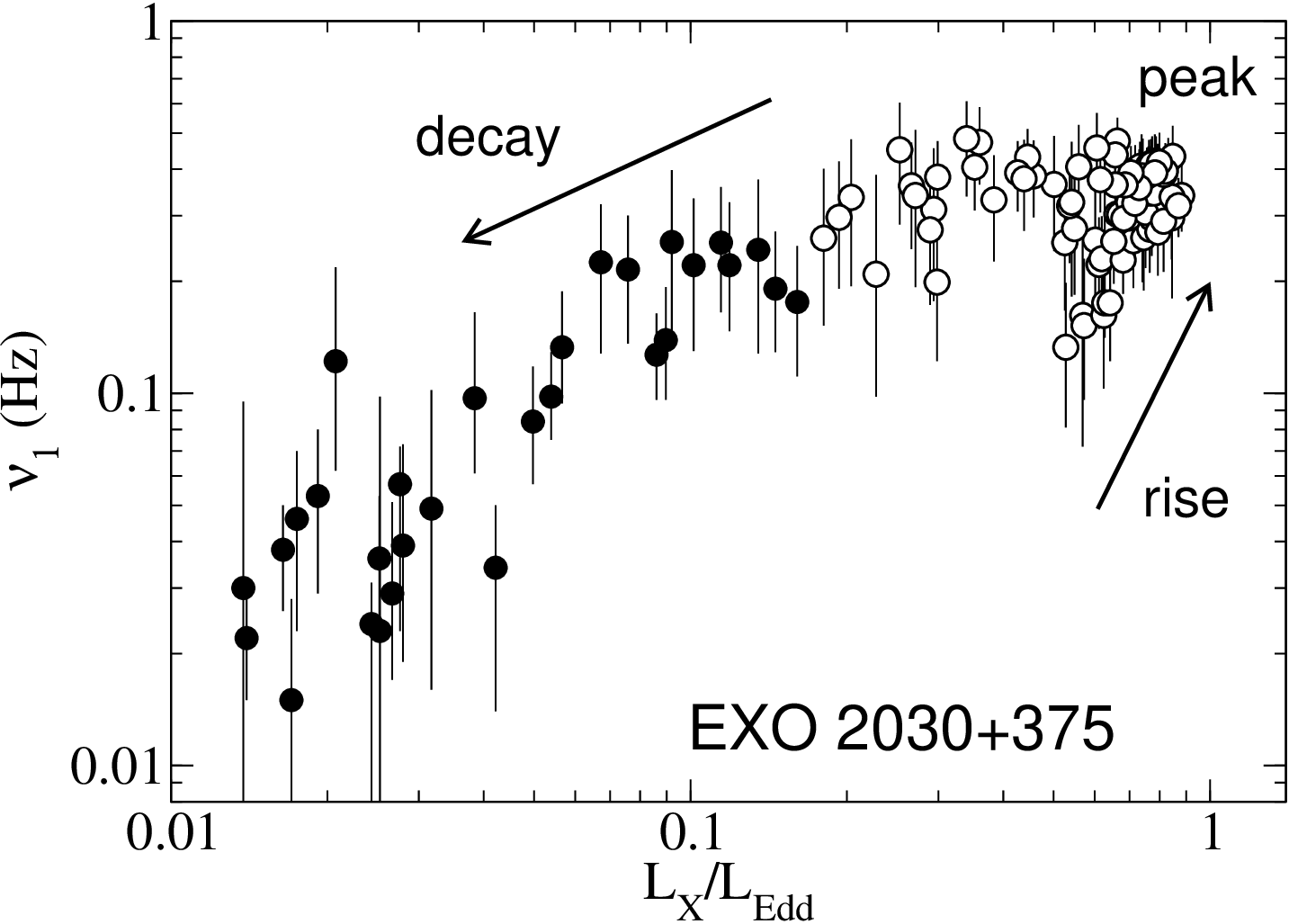} &
   \includegraphics[width=5cm]{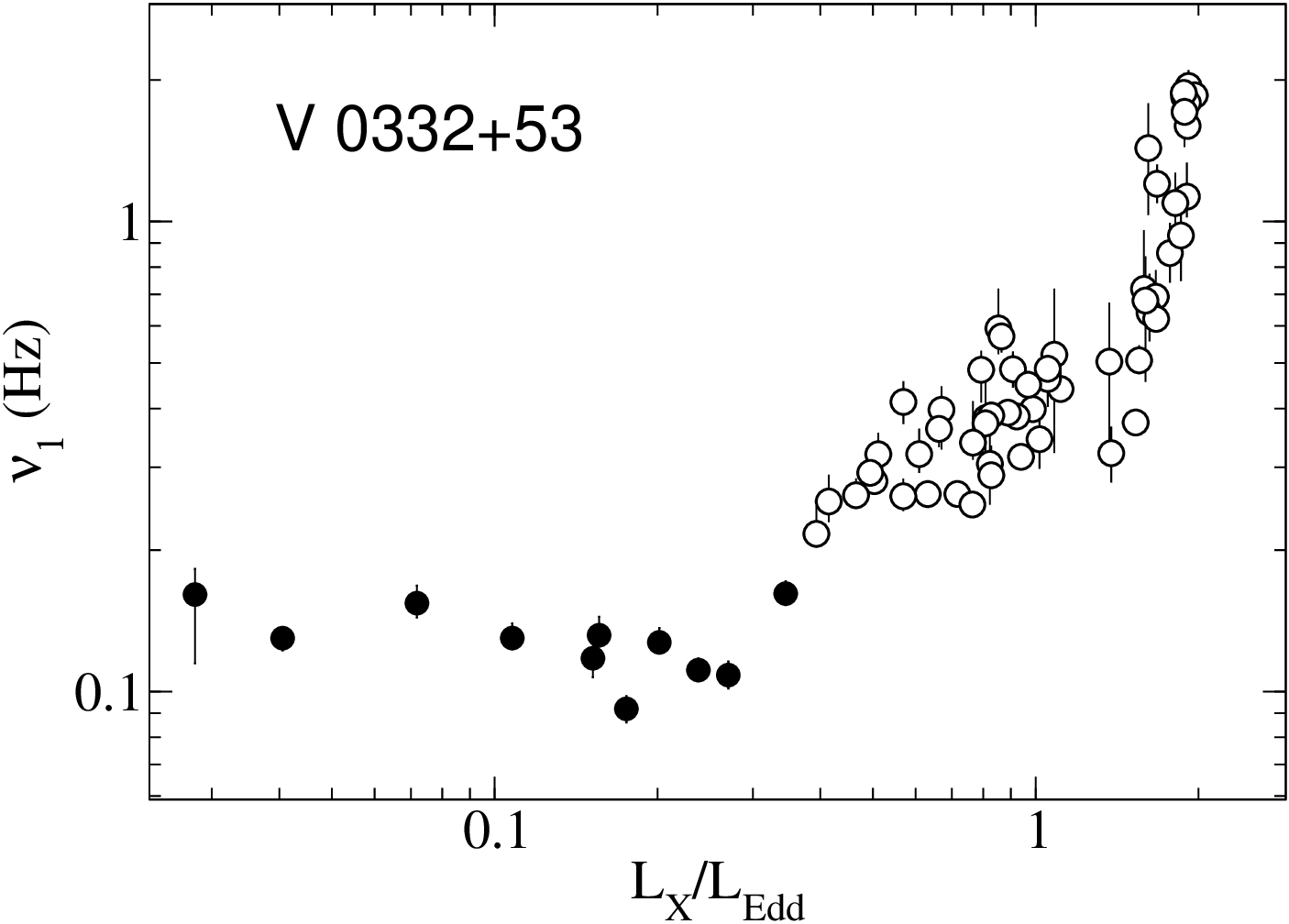}  \\
   \includegraphics[width=5cm]{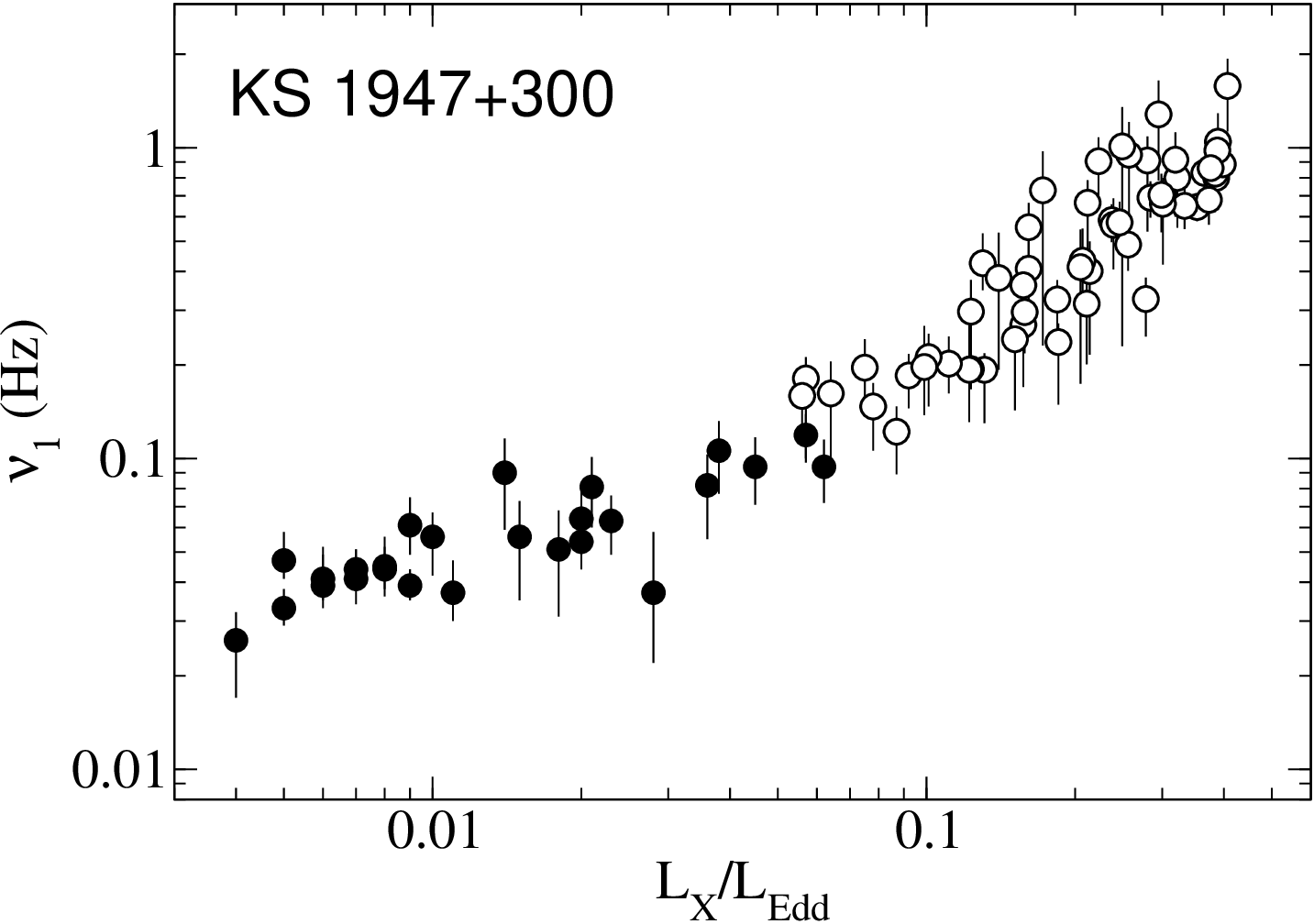}  &
   \includegraphics[width=5cm]{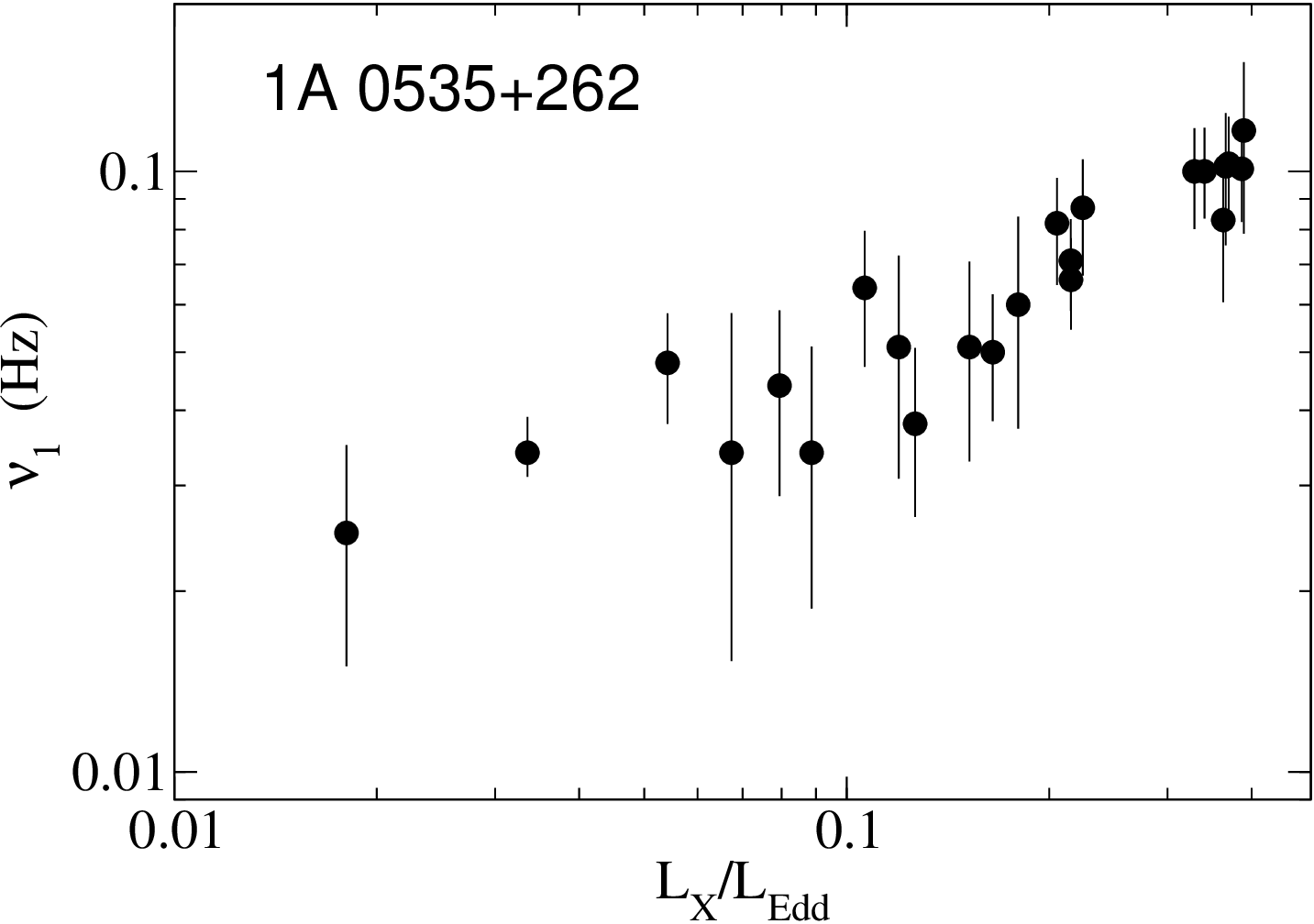}  &
   \includegraphics[width=5cm]{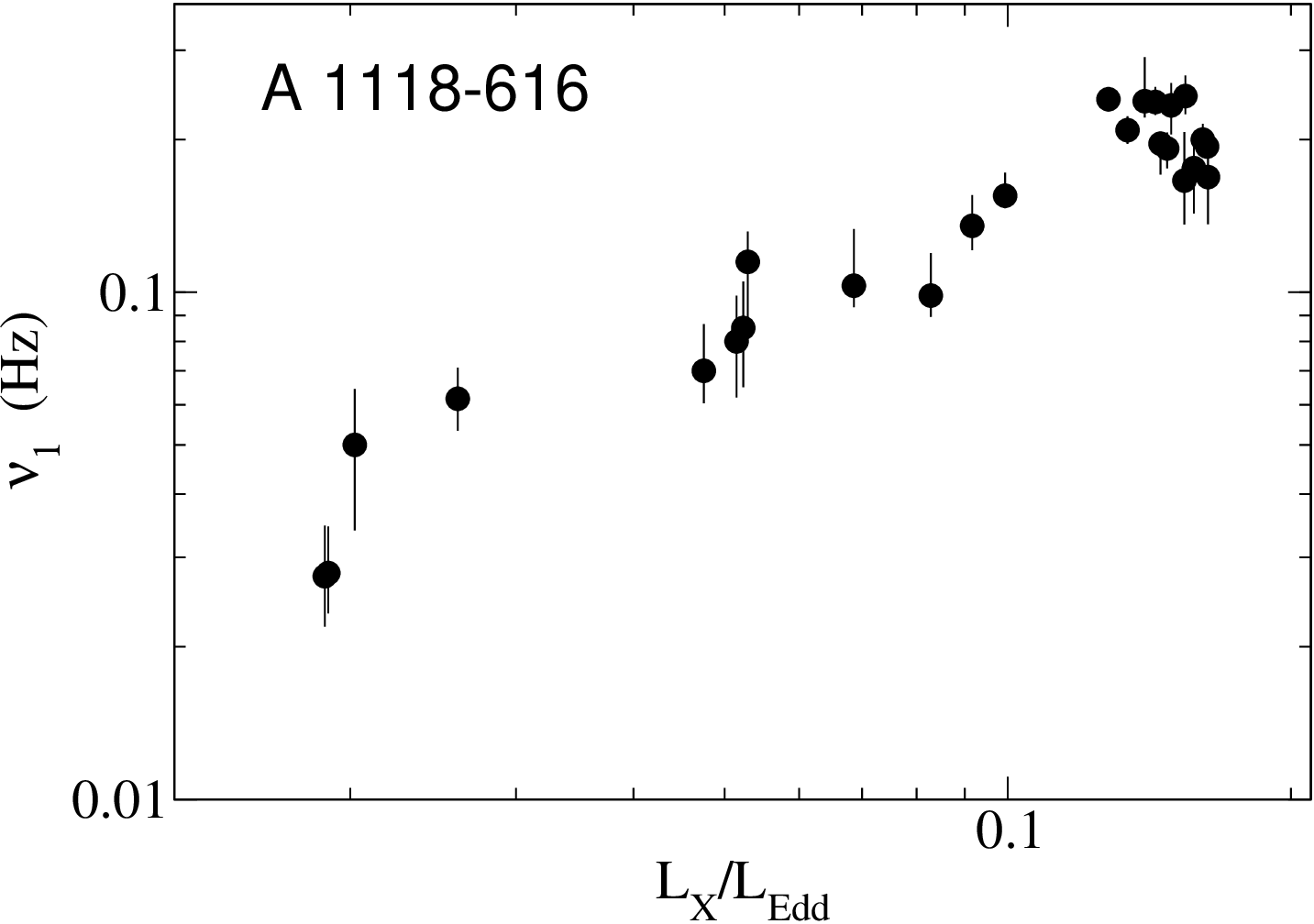}   \\
   \includegraphics[width=5cm]{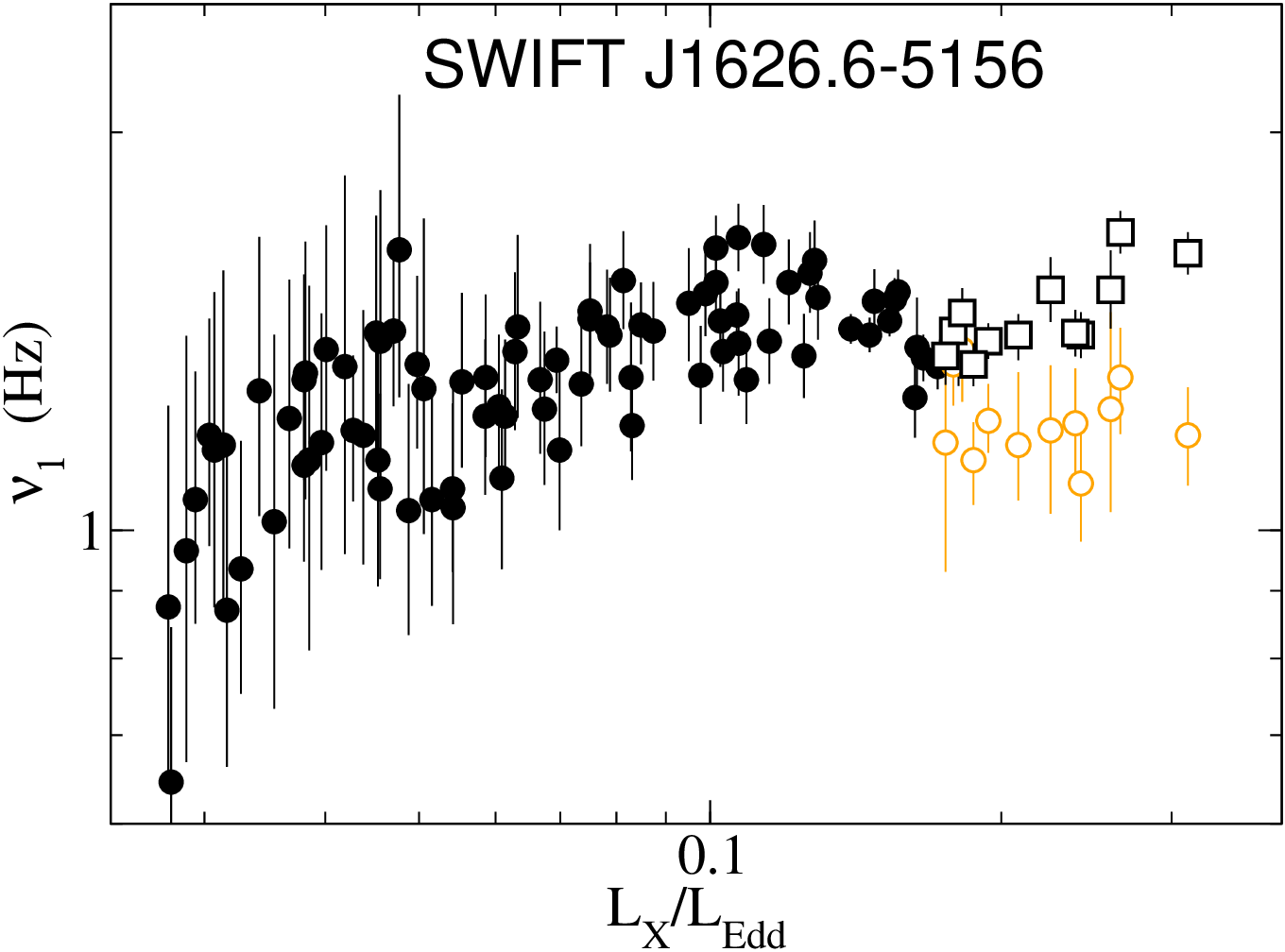}   &  
   \includegraphics[width=5cm]{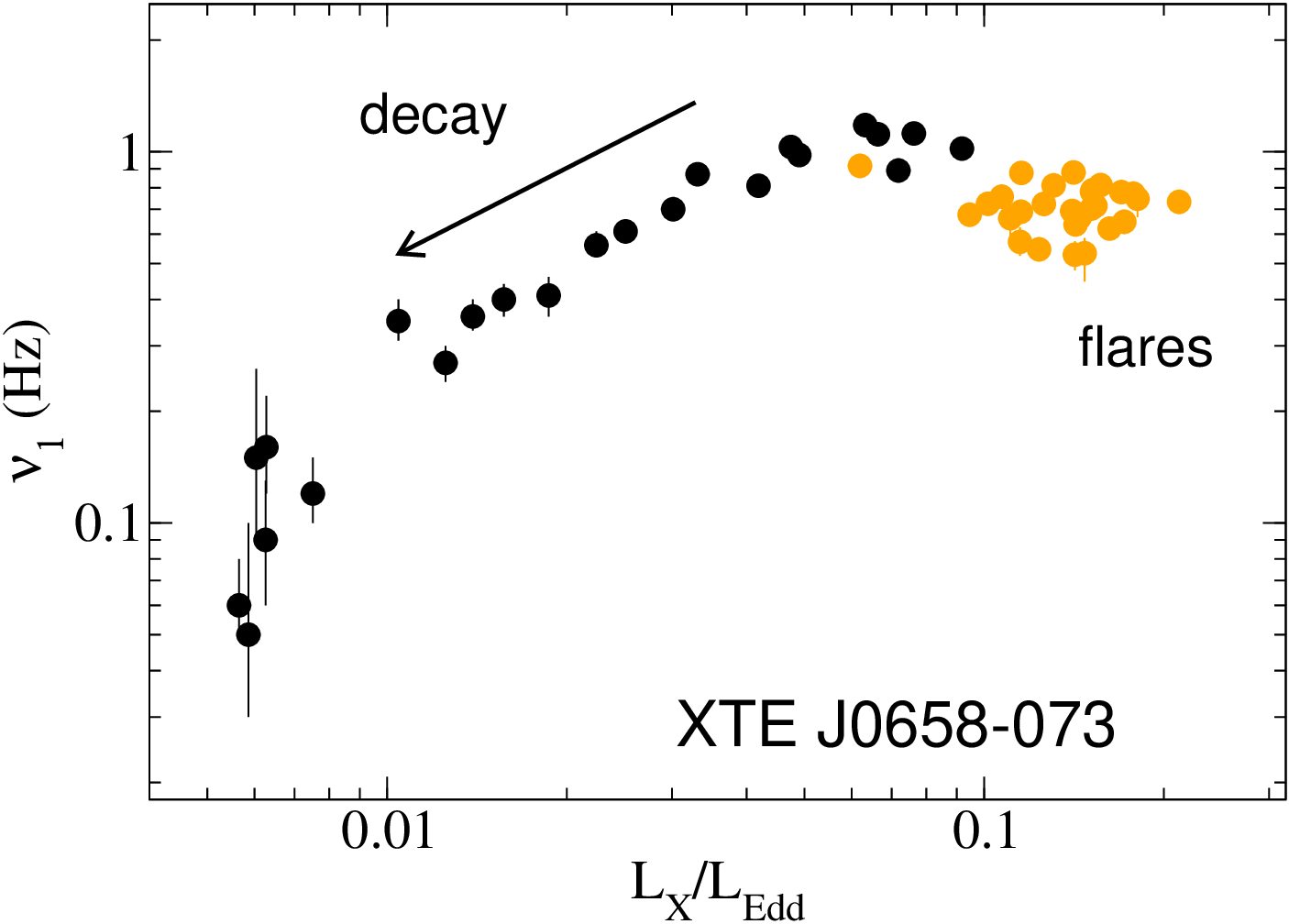}  &
   \includegraphics[width=5cm]{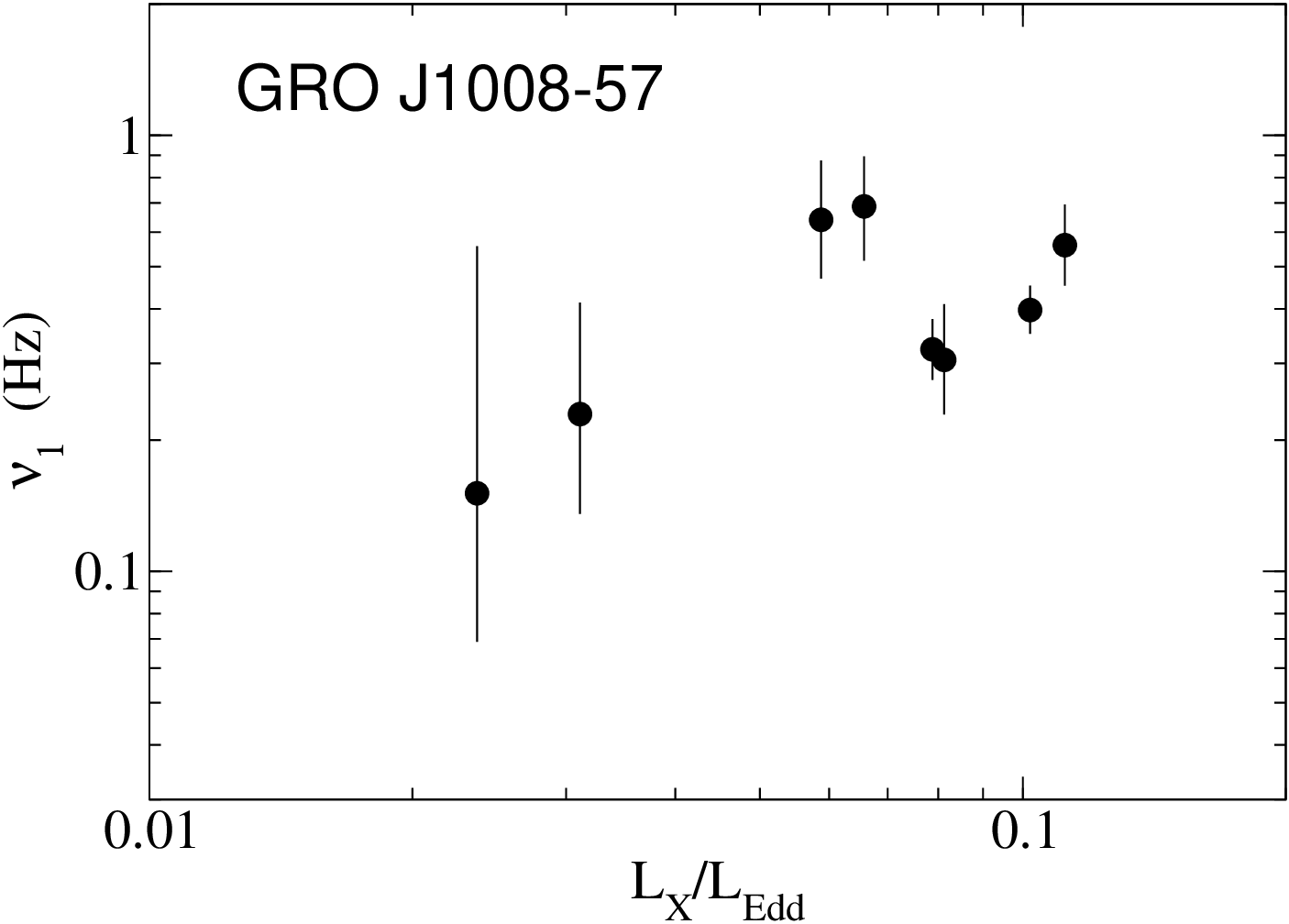}   \\   
   \end{tabular}
      \caption{Characteristic frequency of the main component of the 
      broad-band noise as a function of X-ray luminosity. Open
      circles represent the DB (supercritical regime), filled circles 
      the HB (subcritical regime). In \xte, orange
      circles correspond to the flares. In \sw, open circles (in orange)
      and open squares give the frequency of $L_1$ with and without the
      $L_2$ component, respectively. We assume 
      $L_{\rm Edd}=1.7\times 10^{38}$ erg s$^{-1}$.
      }
  \label{freq-flux}
  \end{figure*}

\subsection{Timing aperiodic variability}
\label{aperiodic}

We conducted a systematic analysis of the rapid aperiodic
variability of the sources. Our aim is to provide a consistent
description of the power spectra of all sources,  investigate whether the
changes in spectral state are also evident in the power spectra, and study
the evolution of the noise parameters throughout the outburst. 

We  consistently used  the same model to fit the power spectra of all
sources.  We found that the majority of power spectra are well represented
by the sum of up to three Lorentzian profiles that were termed as $L_i$,
where $i=0,1,2$.  Some power spectra displayed more complex
substructure, which mainly consisted of the presence of  wiggles without
any preferred frequency range or narrow, generally weak, features confined
to only one bin. Although these features may lead to a formally unacceptable
fit ($\chi^2 > 2$) in some cases, no attempt to correct for this effect was
carried out because it does not affect our results significantly.  Here we
focus on the broad-band noise components, that is, power covering 
relatively large frequency intervals.  The use of Lorentzian profiles is
particularly suitable for this kind of studies as it has been demonstrated
in black-hole binaries \citep{pottschmidt03, axelsson05}. Although it
cannot model very narrow features, it provides a simple but consistent way
to fit the main characteristics of the power spectra across different
states and to track changes occurring on short timescales.   In
addition to the broad-band noise represented by $L_i$, the fast rotating
pulsars \vcas\ and \bq\ show other narrower components, whose
characteristic frequency does not vary with luminosity. For a more detailed
analysis of these components the reader is referred to Paper II.   The
peaks that correspond to the spin period and its harmonics were also fitted
with Lorentzian profiles but the centroid frequency and width were fixed. 
Figure~\ref{psd} shows some representative power spectra at various flux
levels.

$L_0$ is a zero-centred Lorentzian that accounts for the noise below 0.05
Hz. $L_1$ is the main noise component and accounts for the noise in the
range 0.1-1 Hz. It is also a zero-centred Lorentzian whose characteristic
frequency increases as the flux increases. Its $rms$ is always larger than
15\%. $L_2$ accounts for the broad-band noise at higher frequencies. Its
characteristic frequency peaks typically in the range 1--5 Hz, while the
$rms$ is normally below 15\%. $L_2$ appears at high flux, {\it i.e.}, near
the peak of the outbursts. This component is normally not statistically
significant at the end of the outbursts. Although in most cases it shows up
as a zero-centred Lorenztian, it may turn into peaked noise, especially at
the highest flux. Because of the narrower frequency range covered by
the $L_0$ component and typically larger error bars of low-frequency points,
the addition of $L_0$, especially for the first appearance of this
component, did not generally improved $\chi^2$ significantly. Hence, the
addition of $L_0$ was based on visual inspection of the residuals. In
contrast, the introduction or removal of $L_2$ was based on whether its
presence or absence resulted in a significant improvement of the best fit
$\chi^2$.

The following general results are common to all or most of the sources:

\begin{itemize}

\item The characteristic frequency increases as the X-ray luminosity
increases (Fig.~\ref{freq-flux}).

\item Whereas the overall 0.01-10 Hz $rms$ does not change significantly
within a branch, the fractional amplitude of variability of the main
component ($L_1$)  tends to decrease as the X-ray luminosity increases
(Fig.~\ref{rms1-flux}). With the exception of \exo, this trend is more
distinct for sources exhibiting two branches. That is, the source tends to
be more variable in the HB. 

\item $L_0$ and $L_2$ are usually not present in the HB. The
disappearance of $L_0$ in the HB is probably due to the fact that this
component falls below the frequency interval considered ($\nu_0 < 0.01$ Hz)
at very low fluxes in agreement with the observed frequency shift  
(Fig.~\ref{freq-flux}).

\item Whenever $L_1$ and $L_2$ appear simultaneously in the power spectrum,
their frequencies correlate (Fig.~\ref{freq-freq}).

\end{itemize}


The saturation at high luminosity seen in the luminosity-frequency
diagram of some sources (most notably in \mxb, \sw, and \exo) is due to
the appearance of the $L_2$ component. The correlation between the
characteristic  frequency and the flux is broken when this extra component
is added because it prevents $L_1$ from shifting toward higher
frequencies.  As an exercise, we removed the $L_2$ component from the power
spectra of \sw. The absence of $L_2$ gives worse fits, but $\nu_1$ shifts
up (squares in Fig.~\ref{freq-flux}) and the correlation holds even at the
highest luminosity.

\subsection{Correlation between spectral and timing parameters}

The spectral parameters correlate tightly with the X-ray flux. In some
sources, though, this correlation changes sign at very low fluxes (in the
HB). The characteristic frequency of the broad-band noise also shows a
smooth relationship with flux, albeit with more scattering. The question is
then whether the spectral and timing parameters correlate with each other.
Figure~\ref{freq-gamma} displays the $L_1$ maximum frequency as a function
of the photon index. These two parameters seem to correlate in most
sources, although the sign and strength of the correlation vary
significantly. We shall discuss the implication of this result in
Sect.~\ref{spec-tim}.

\section{Discussion}
\label{discussion}

The main goal of this study is to perform a detailed spectral and timing
analysis of accretion-powered pulsars with Be companions in an attempt to
characterise this type of systems as a group. More precisely, we wish to
investigate {\em i)}  whether accreting X-ray pulsars display spectral
states as black-hole and low-mass X-ray binaries do and characterise those
states, and {\em ii)} to search for correlations between spectral and
timing parameters during major outbursts in an attempt to constrain the
accretion models in X-ray pulsars.

 In this section we discuss the implications of our work. We interpret
our results in the context of the model that suggests the existence of two
different accretion regimes, defined by a critical luminosity
\citep{becker12}. First, we make a short introduction of the spectral
formation mechanism in accreting pulsars and then we summarise and discuss
our results.

\subsection{Spectral formation}

A detailed description of the emission properties of an accretion-powered
pulsar is a complex task. First, it requires an understanding of the
processes by which mass is transferred, captured, deposited on the neutron
star surface in the form of an accretion column and converted into
high-energy radiation.  Second, this radiation has to find its way out,
which involves a knowledge of the interactions of the X-rays produced close
to the neutron star surface with the highly magnetized plasma forming the
magnetosphere.  Third, the X-rays emanating from the
polar caps are subjected to absorption and reflection processes due to the
ambient matter.


This complexity translates into a lack of a fundamental physical model that
yields results that agree with the observations. The common practice is
then to fit phenomenological multicomponent models to the energy spectra of
accretion-powered pulsars. The parameters of these model components are
difficult to interpret physically. Some attempts to alleviate this
situation have been made by \citet[][see also
\citealt{ferrigno09}]{becker05,becker07}, who developed a new model for the
spectral formation process in X-ray pulsars based on the bulk and thermal
Comptonisation of photons due to collisions with the shocked gas in the
accretion column. The accretion flow is channelled by the strong magnetic
field into the polar caps, creating an accretion column. Most of the
photons are produced at the base of the column, just above the neutron star
surface (thermal mound). The low-energy (blackbody) photons created in the
mound are upscattered  due to collisions with electrons that are infalling
at high speed (bulk Comptonisation). Thermal Comptonisation (in this case
high-energy photons lose energy) also plays a significant role in shaping
the spectrum and it is responsible for the formation of the exponential
cutoff. 

\subsubsection{Two accretion regimes}

There exists a critical luminosity, $L_{\rm crit}$ , at which the
deceleration of the accreting flow to rest at the neutron star surface
changes from being dominated by radiation pressure to occur via Coulomb
interactions \citep{basko76}.
\citet{becker12}
suggested that these two different regimes can explain the bimodal
behaviour of the variability of the cyclotron energy with luminosity. Some
sources, such as  \bq\ display a negative correlation between the
cyclotron energy and the source luminosity, others, such as Her
X--1 \citep{staubert07}, show a positive correlation.  

Two accretion regimes have also been invoked by \citet{klochkov11} to
explain the pulse-to-pulse variability in some X-ray pulsars. It is
illustrative to compare our results with those of \citet{klochkov11}
because of the different timescales involved. We have studied the long-term
variability patterns of accreting pulsars by sampling timescales of X-ray
variability of the order of 1000 s. \citet{klochkov11} studied the
pulse-to-pulse spectral variability, {\it i.e.}, timescales of the order of
the pulse period ($P_{\rm spin} \sim 1-100$ s). Their sample included four
X-ray pulsars, three of which appear in our list: \bq, \vcas, and \vtau. To
achieve enough photon statistics they restricted their analysis to
high-flux states. In the context of the present work, this means that they
analysed data of the DB in \bq\ and \vcas\ and of the high-flux part of the
HB in \vtau. Despite the order of magnitude difference in the variability
range, the same correlations are observed. They found that the photon index
of the power-law component correlates with X-ray intensity in \bq\ and
\vcas, but anticorrelates in \vtau, which is the same result as our 
Fig.\ref{gamma-flux}. Likewise, they reported a correlation between the
energy of the fundamental cyclotron line with count rate in \vtau, as we do
with our long-term pulse-average spectral analysis (Fig.~\ref{cyc-flux}).

   \begin{figure}
   \centering
   \includegraphics[width=8cm]{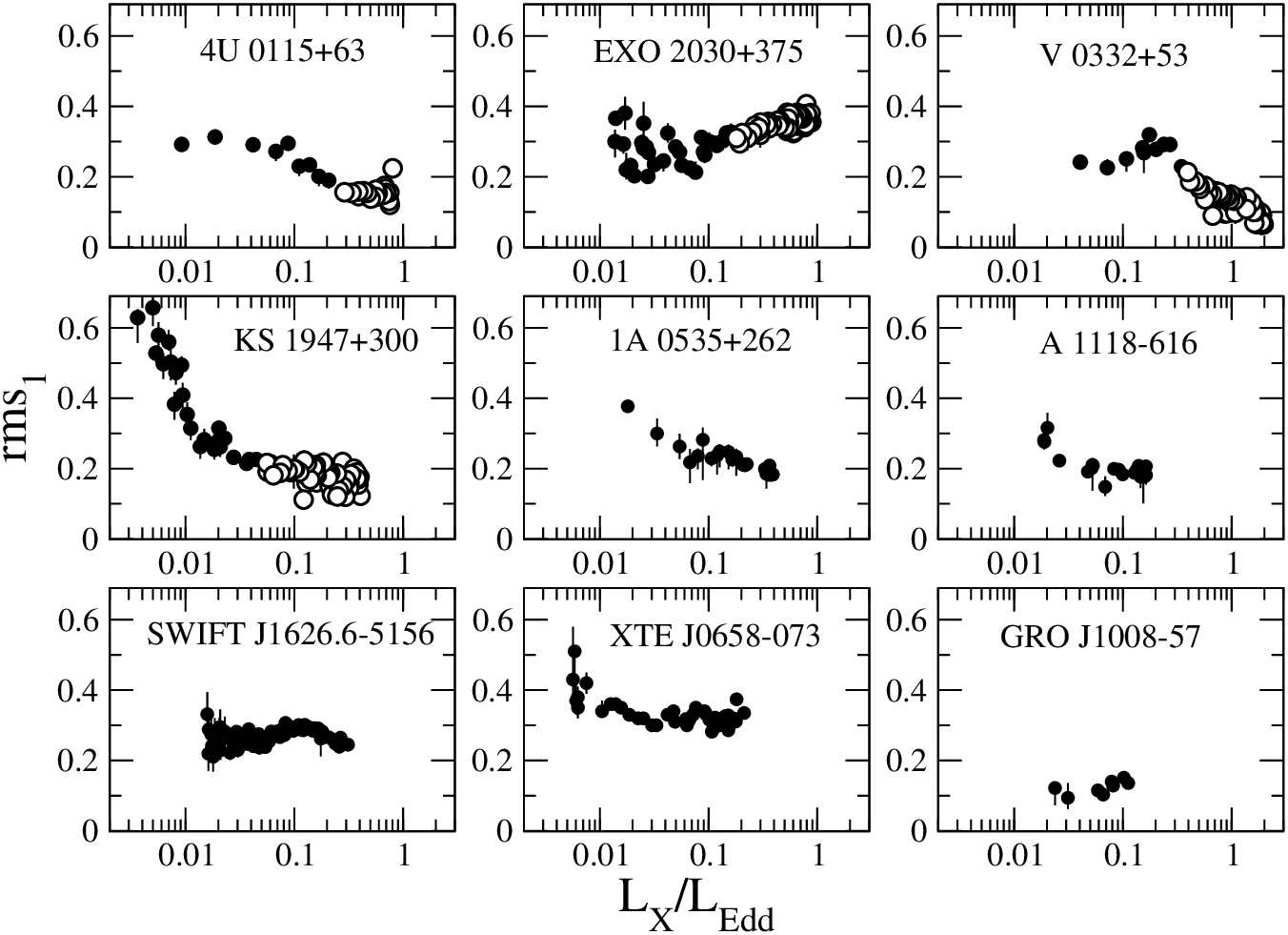}
  \caption{Variation of the fractional amplitude of variability of
  component $L_1$ with luminosity. Filled black circles correspond to the
  HB and open circles to the DB.}
  \label{rms1-flux}
  \end{figure}
   \begin{figure}
   \centering
   \includegraphics[width=8cm]{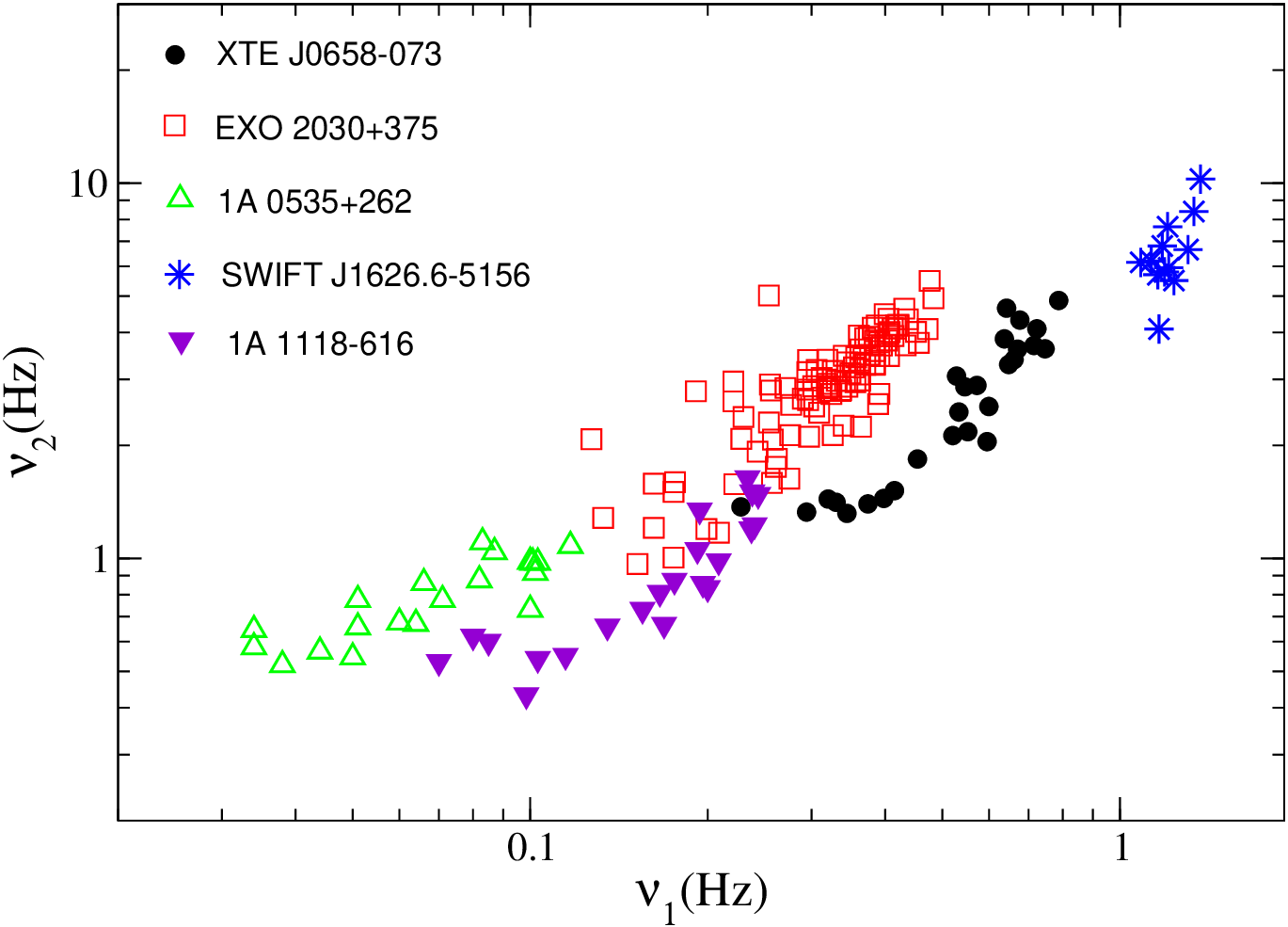}
  \caption{Relationship between the peak frequencies of $L_1$ and
  $L_2$. }
  \label{freq-freq}
  \end{figure}
   \begin{figure*}
   \centering
   \begin{tabular}{ccc}
   \includegraphics[width=5cm]{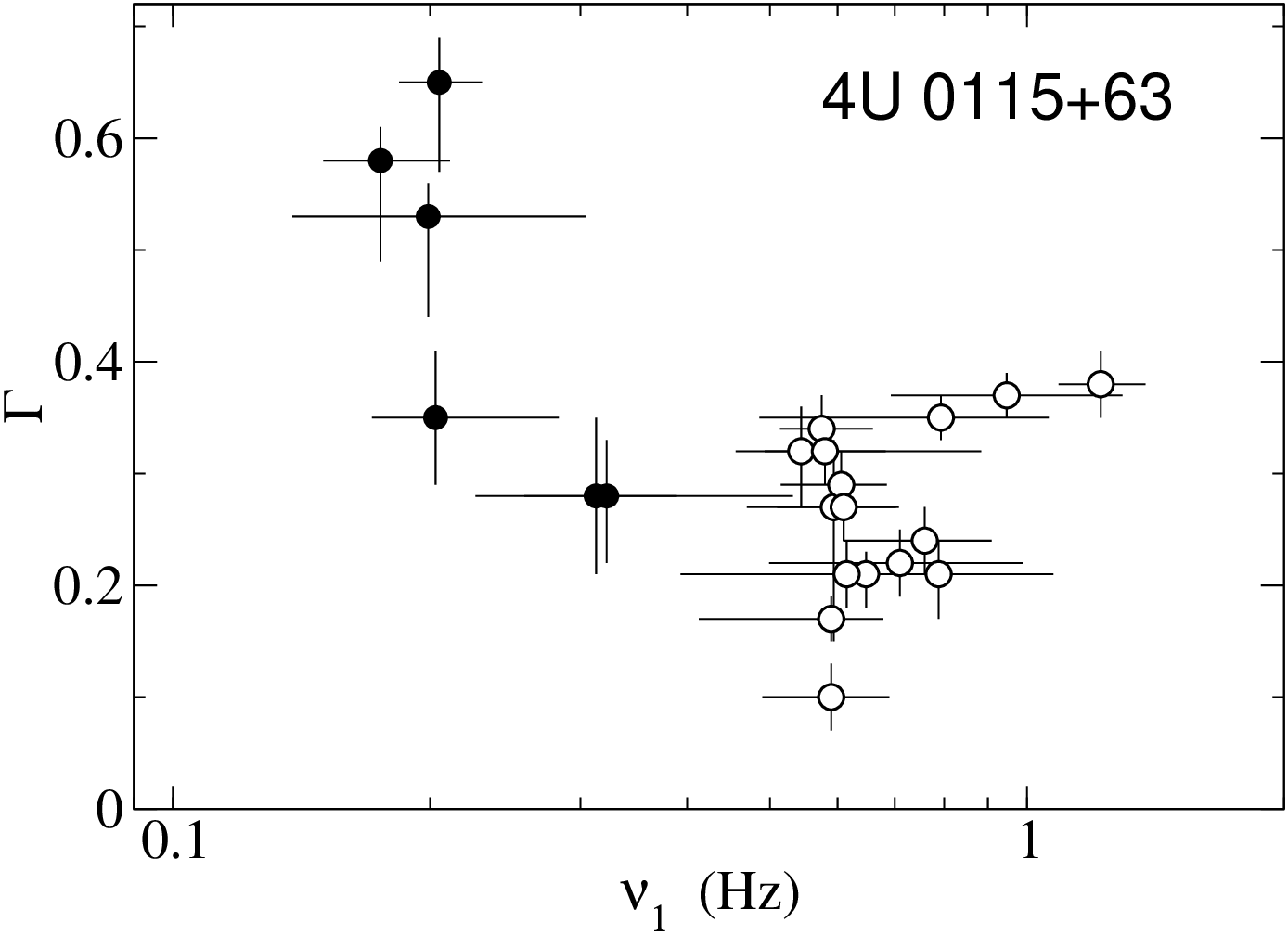}  &
   \includegraphics[width=5cm]{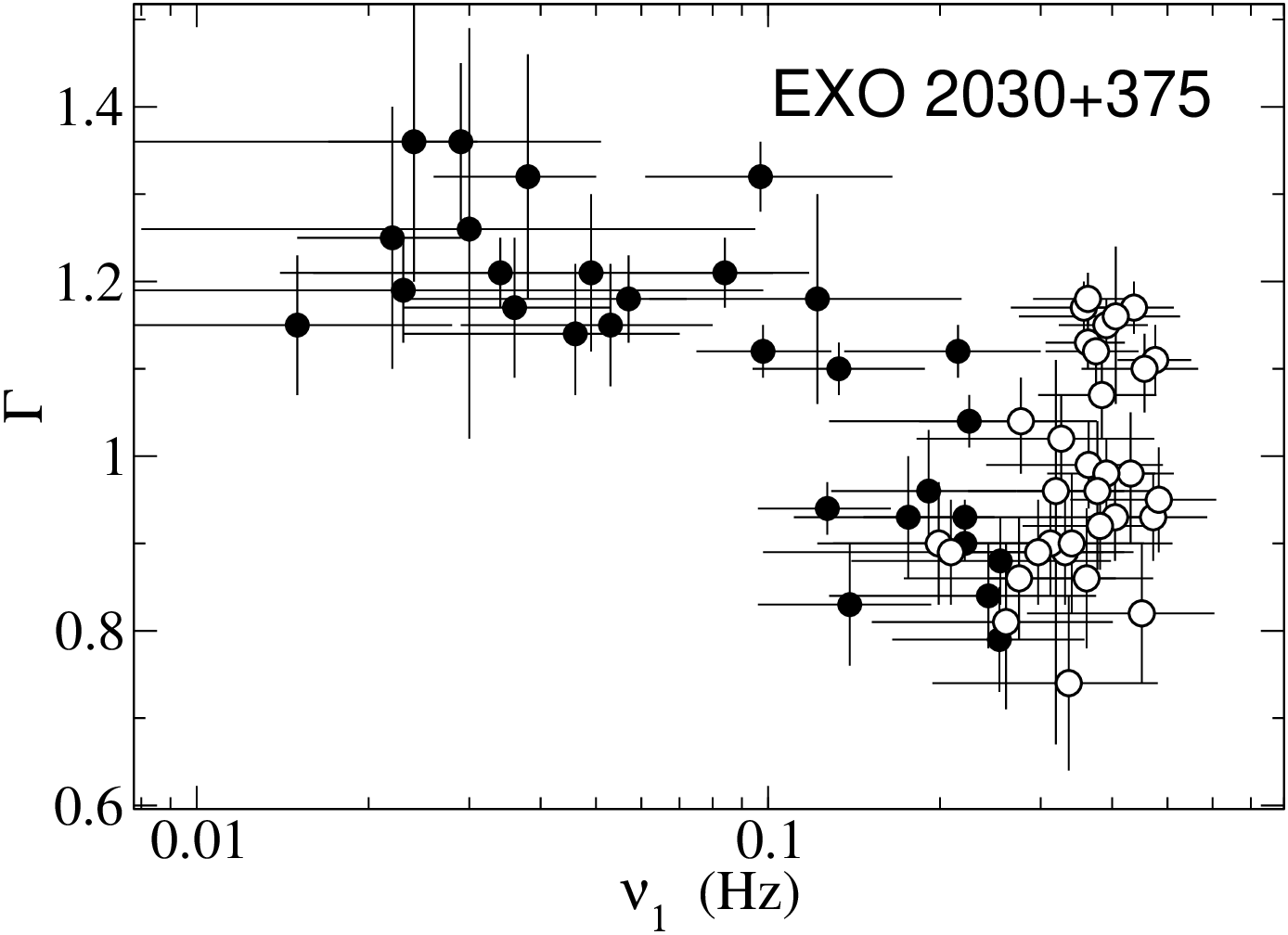} &
   \includegraphics[width=5cm]{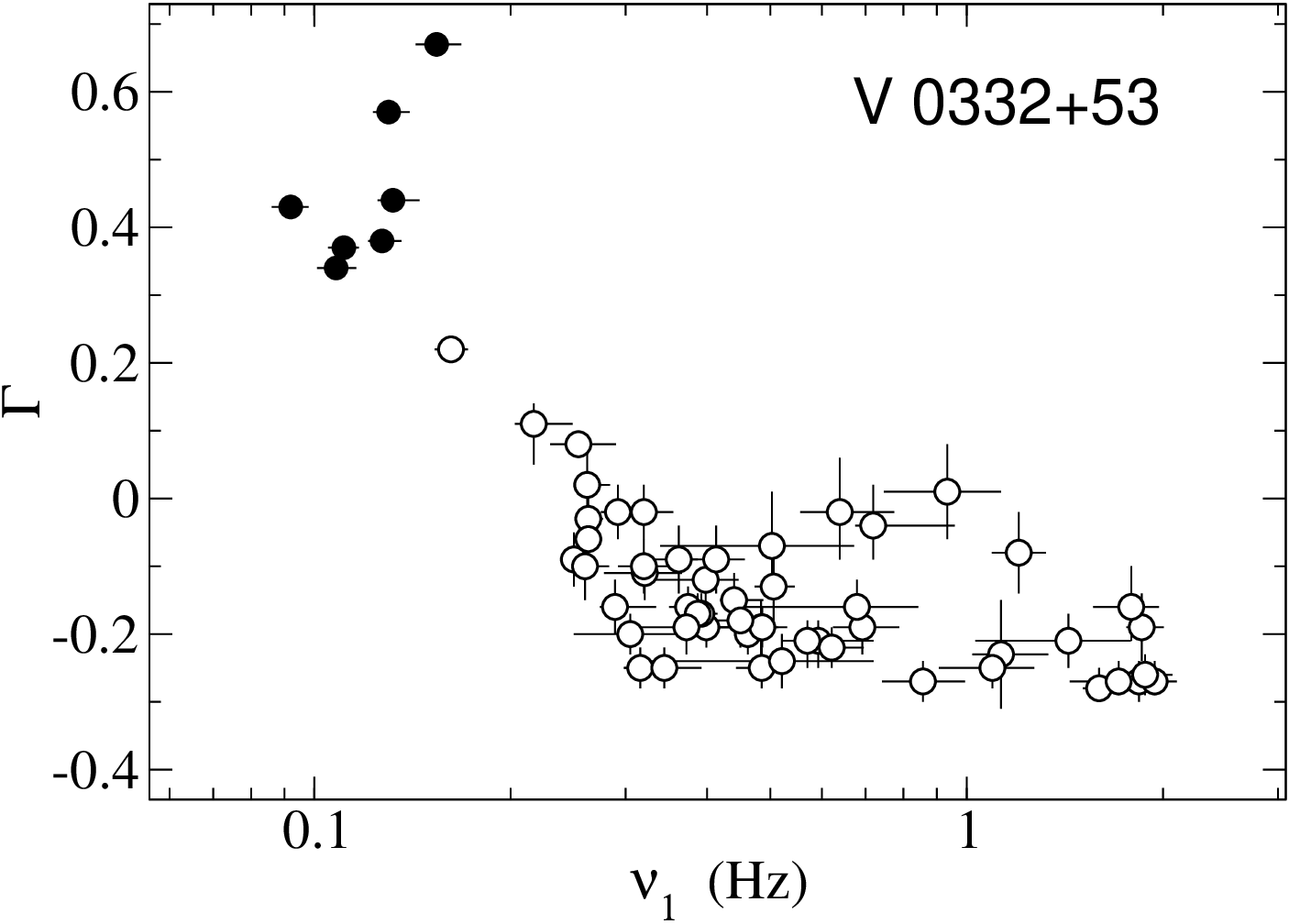}  \\
   \includegraphics[width=5cm]{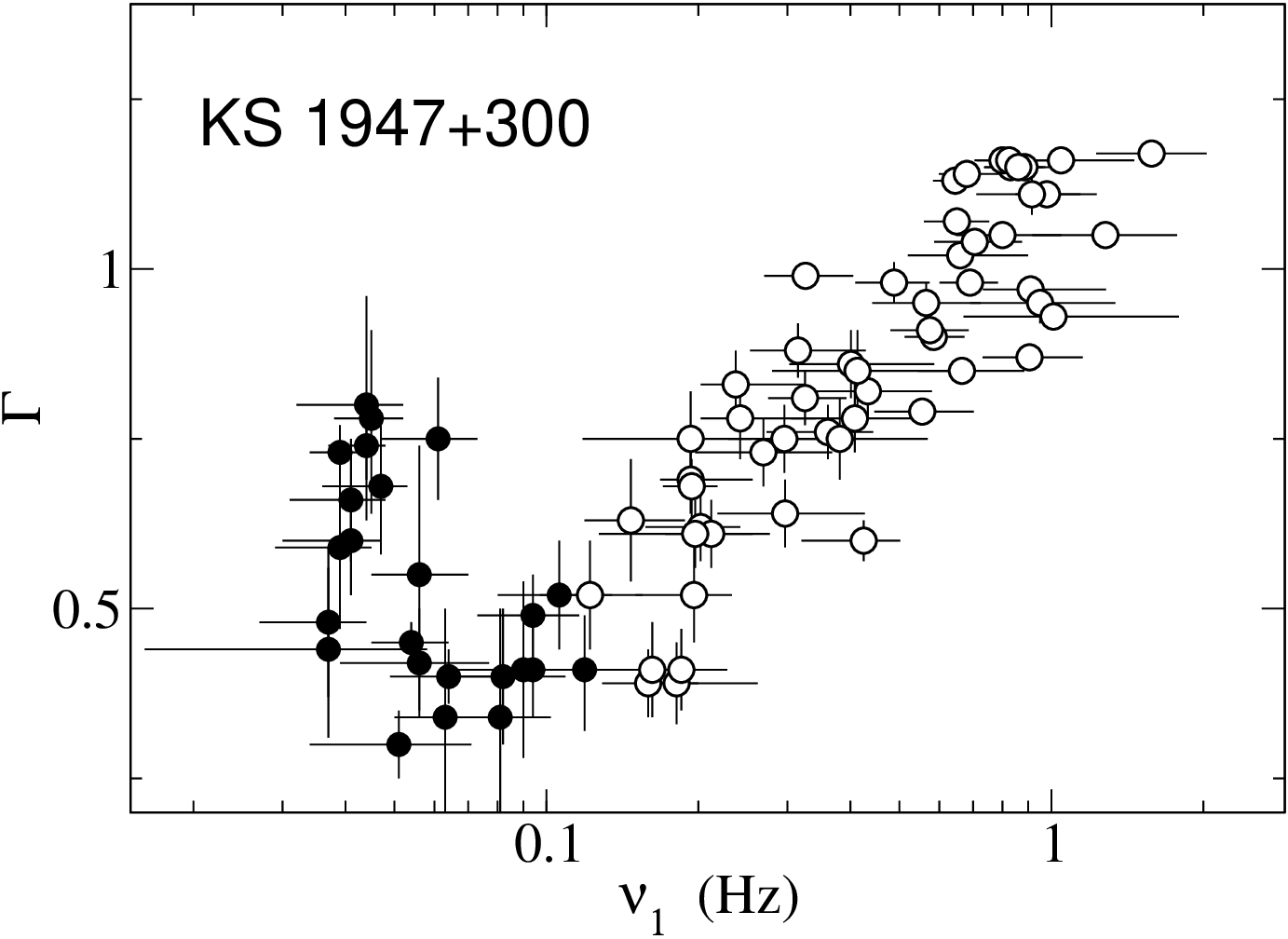}  &
   \includegraphics[width=5cm]{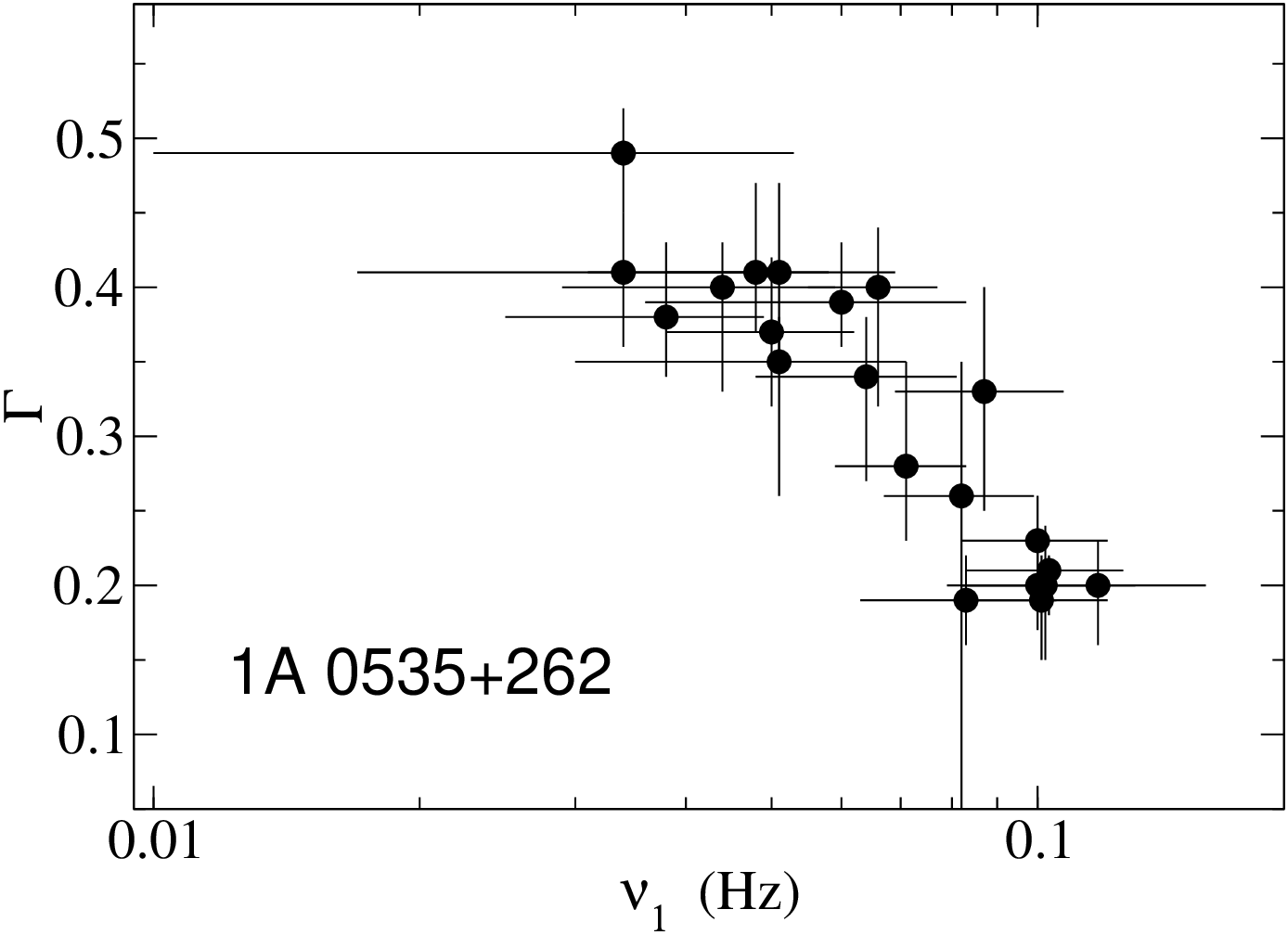}  &
   \includegraphics[width=5cm]{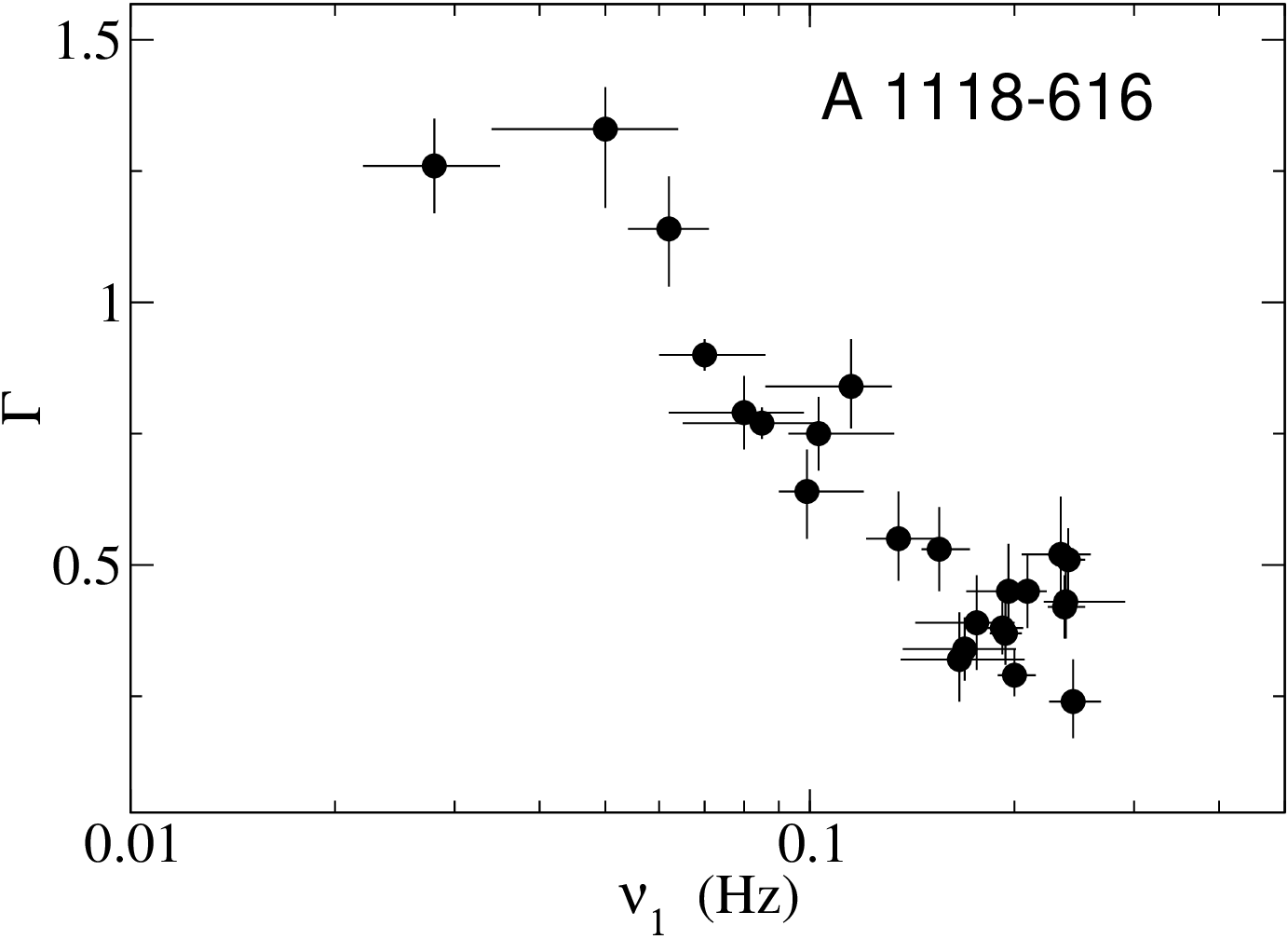}   \\
   \includegraphics[width=5cm]{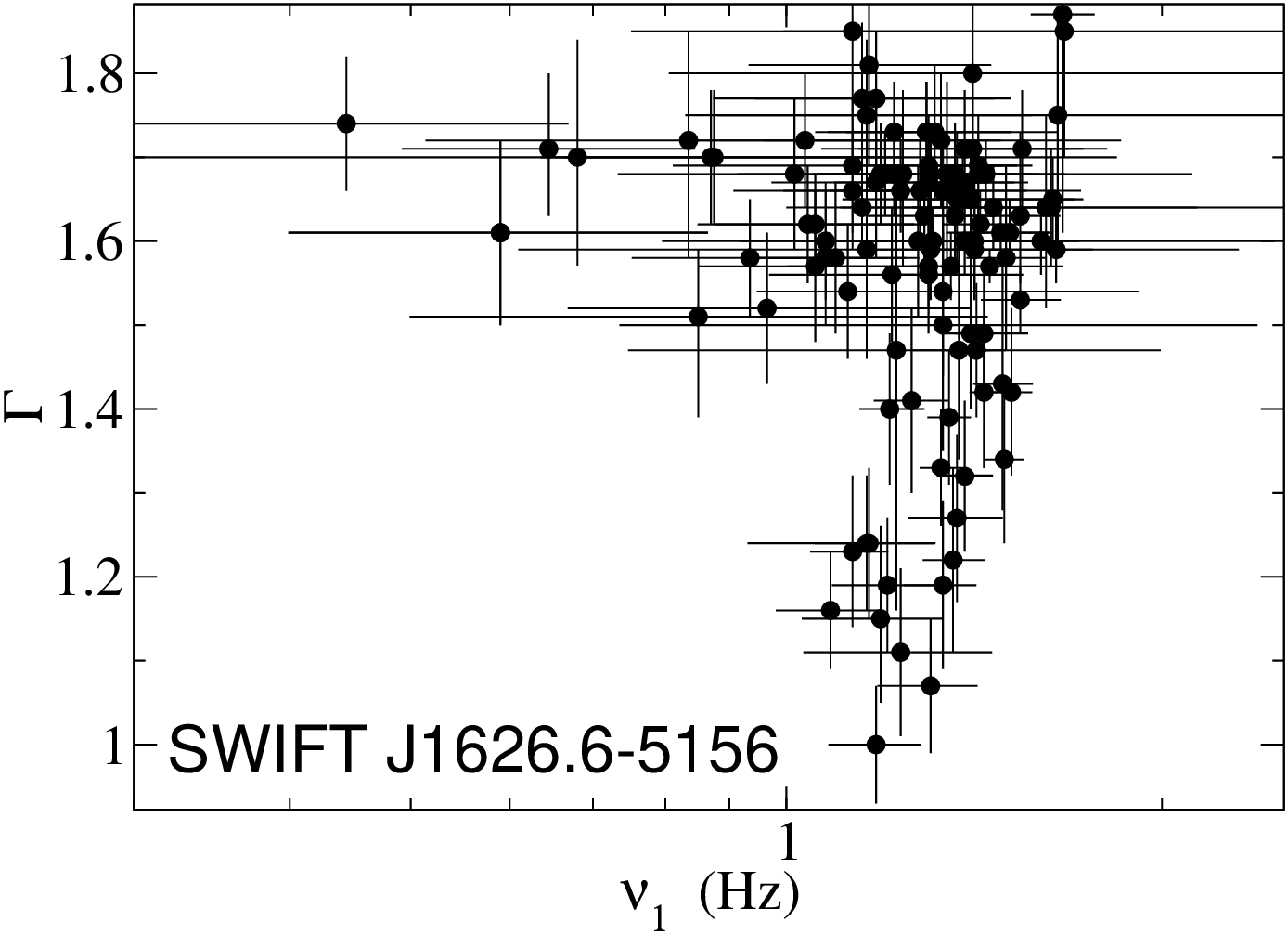}   &  
   \includegraphics[width=5cm]{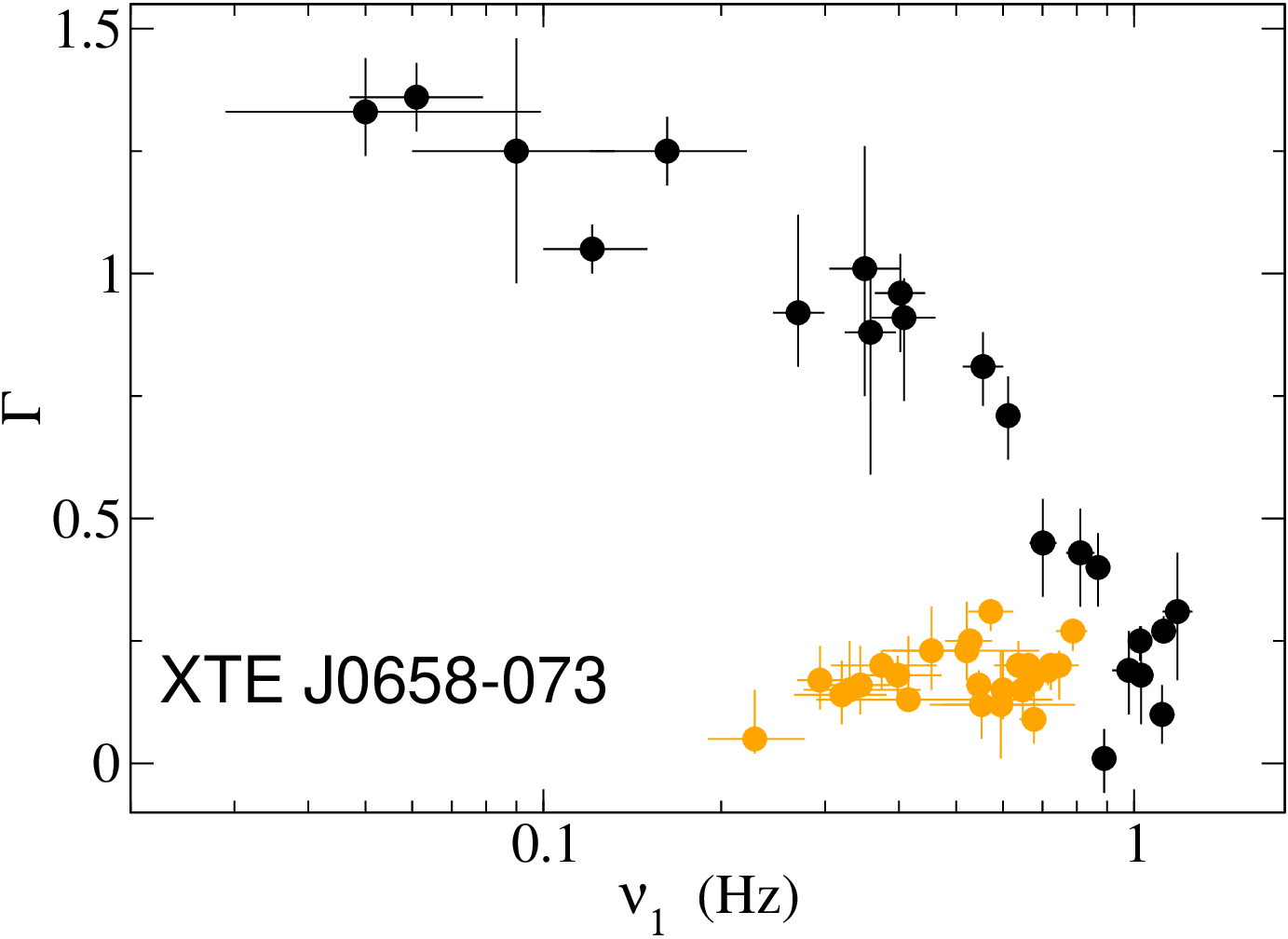}  &
   \includegraphics[width=5cm]{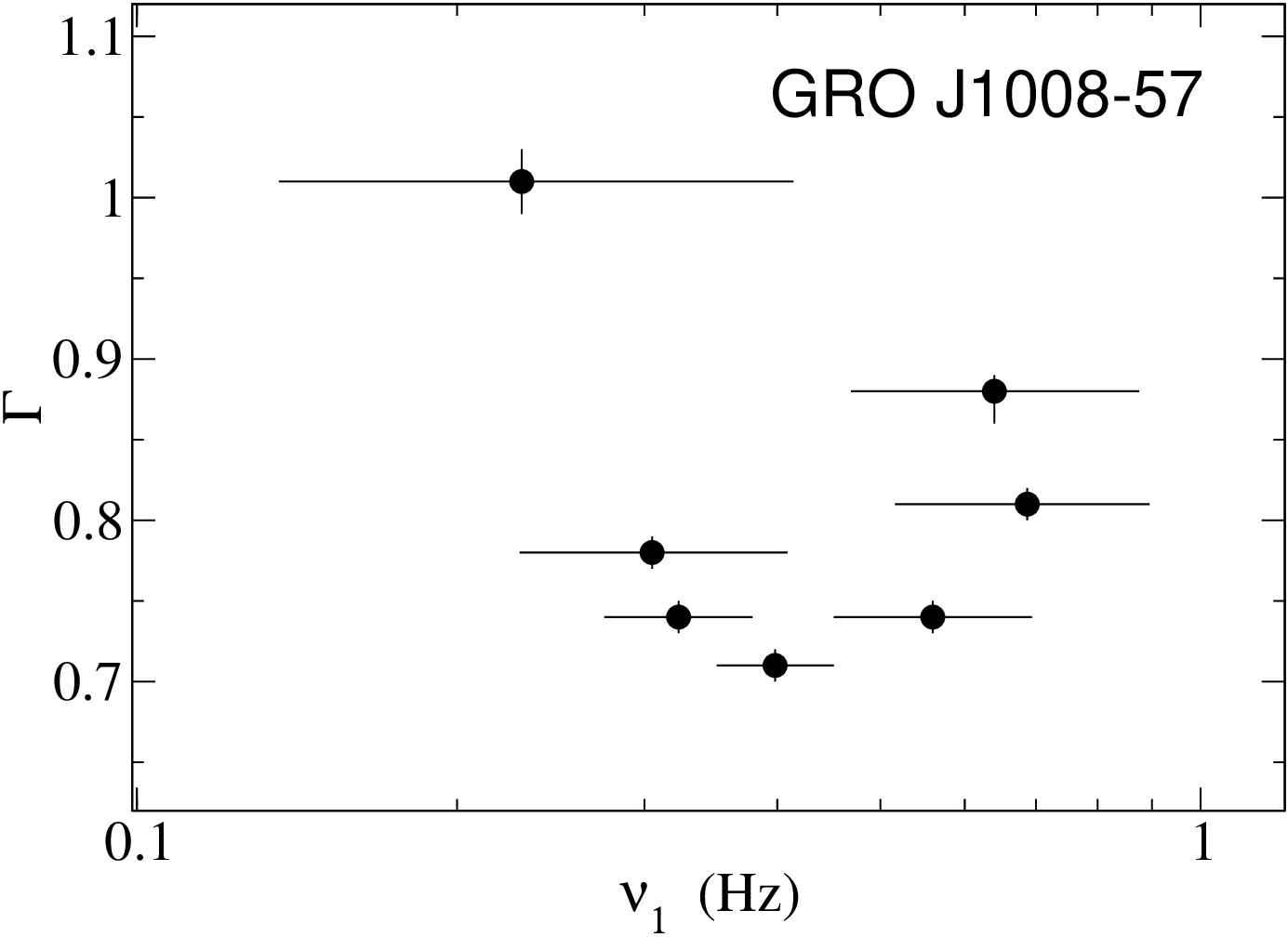}   \\   
   \end{tabular}
      \caption{Relationship between the characteristic frequency of the 
main component of the broad-band noise and the power-law photon index. 
In \exo, due to the hysteresis observed in the frequency-luminosity 
diagram (Fig.~\ref{freq-flux}) only points that correspond to the decay 
(after MJD 53980) have been plotted for clarity.}
  \label{freq-gamma}
  \end{figure*}

\subsection{Source states}
\label{soustat}

This work shows that, as BHBs and LMXBs, accreting X-ray binaries with
high-mass companions also exhibit spectral states in the hardness-intensity
diagrams. At low and intermediate X-ray luminosities the sources populate
the horizontal branch (HB), whereas at higher luminosities the sources
trace a diagonal branch (DB). However, unlike BHBs and LMXBs, where the
hardness of the source is a distinguishing property of the accretion states
(the so-called {\em hard} and {\em soft} states), the spectral branches in
accretion-powered pulsars roughly sample the same colour interval.  If a
source displays the two branches, then  the timescales of the colour and
flux changes are faster in the HB than in the DB. The source covers the HB
in hours to few days and the DB in days to weeks. In contrast, systems that
show only the HB can remain in that branch for weeks. 

With the exception of \vcas\ and \bq\ (see Paper II), there is no
clear evidence for hysteresis in the HID, {\it i.e.}, the same colour
corresponds to two different values of the count rate depending on whether
the source is in the rise or decay. This lack of hysteresis could be due to
the fact that, in most cases, the rise of the outburst is much badly
sampled than the decay (see Fig.~\ref{rate-time}). For \sw, \vtau, \hen,
and \gro, no or very few data points of the rise were observed. On the
other hand, in \ks, which is the source with the most complete coverage of
the rise, no indication of hysteresis is found.  To check this result, we
performed a colour analysis on the ASM light curves and we did not find
significant deviations between the rising and decaying tracks. However, it
should be noticed that the errors of the ASM colours are, in the best case,
roughly of the same size as the amplitude of the hysteresis effect. It is
also worth mentioning that while we did not observe significant hysteresis
in the HID of \exo, the characteristic frequency of the broad-band noise in
this source exhibits lower values during the rise than during the decay
(Fig.~\ref{freq-flux}).

One of the main findings of this work is that the transition from the HB to
the DB occurs when the source luminosity increases above a certain value,
which we find to be in the range $\sim0.06-0.3 L_{\rm Edd}$ ($\sim 1-4
\times 10^{37}$ erg s$^{-1}$). We observe that this value is different for
different sources. We suggest that the luminosity at which the transition
occurs is related to the critical luminosity and that each spectral branch
can be associated with one of the two modes of accretion proposed by
\citet{becker12}.  The DB, which corresponds to a high-luminosity state,
would be associated with the supercritical mode and the HB with the
subcritical one.

The case of \ks\ and \vtau\ would corroborate this association. These two
sources reached similar peak outburst luminosity. However, only \ks\ shows
the two branches. \ks\ reached a peak luminosity of $0.4L_{\rm Edd}$ and
made the transition from the HB to the DB at $\sim 0.06L_{\rm Edd}$ or
$\sim 1\times 10^{37}$ erg s$^{-1}$,  assuming a distance of $\sim10$ kpc
\citep{negueruela03}. In contrast, \vtau, with the same peak luminosity
stayed in the HB during the entire outburst. We conclude that the intensity
of the source is not the only parameter triggering the change of state and
some other parameter must play a role too. The magnetic field appears as
the best candidate to drive the transition. According to \citet{becker12},
the critical luminosity $L_{\rm crit}$ depends on the magnetic field, or
equivalently, on the energy of the cyclotron line. The higher the magnetic
field, the higher the critical luminosity. For typical neutron star
parameters \citep[see][for details]{becker12}, the critical luminosity can
be estimated as

\begin{equation}
L_{\rm crit} \approx 1.28 \times 10^{37} (E_{\rm cyc}/10{\rm keV})^{16/15} \, \,  {\rm erg
\, \,s^{-1}} 
\end{equation}

\noindent where $E_{\rm cyc}$ is the energy of the cyclotron line.  \vtau\
displays one of the highest energy cyclotron lines, at $\sim50$ keV 
\citep{caballero08}. Substituting in the above equation yields $ L_{\rm
crit}\approx 7\times 10^{37}$ erg s$^{-1}$, which is slightly larger than
the outburst peak luminosity. No cyclotron line has been reported for \ks,
which would indicate a lower magnetic field, in agreement with the lower
luminosity at which the spectral transition occurs. The luminosity at which
the transition is observed in \ks\ would imply a cyclotron line at around 8
keV. Unless the cyclotron line is very strong such a feature would be
difficult to detect because it would lie in a region where other spectral
features dominate. First, the fluorescence iron line at 6.4 keV and
possible its associated absorption edge. Second,  some X-ray pulsars show
significant wiggle residuals at 8-12 keV, whose origin is unclear
\citep{coburn02}.  Nevertheless, it is worth noticing that in the high-flux
average spectrum of \ks, an additional absorption component at about 9 keV
provides an improved fit with respect to the model used for the individual
spectra. The $\chi^2$ decreases from 104 for 72 degrees of freedom to 74
for 69 degrees of freedom. This component can be modelled with an
absorption Gaussian profile (GABS) and could be associated to a cyclotron
line that in the single spectra would be too weak to be unveiled. The
absorption could also be accounted for with an edge ($\chi^2$=71 for 70
degrees of freedom), although in this case the inspection of residuals
reveals that this component is unable to properly fit the entire spectral
region of interest. In short, it is difficult to tell whether the source
actually displays a cyclotron line at this energy, but if so, its low
characteristic energy would  agree with the critical luminosity reached by
the system.

The larger magnetic field in \vtau\ might also explain the reason that
previous studies of this source did not find a correlation between the
energy of the cyclotron line and the X-ray luminosity. \citet{terada06}
speculated that this energy might begin to change at a higher luminosity if
the object has a higher surface magnetic field.  Here we report
observations of \vtau\ at higher luminosities than previous studies
\citep{kendziorra94,wilson05,caballero07,terada06}. As it can be seen in
Fig.~\ref{cyc-flux}, the energy of the CRSF only begins to increase
significantly when the X-ray luminosity is above $\sim 3\times 10^{37}$ erg
s$^{-1}$.

\xte\ is another source whose peak luminosity is
close to its critical luminosity (see Table~\ref{info}). Although it does
not display a proper DB, an attempt to move to that branch can be observed
in Figs.~\ref{rate-SC} and \ref{gamma-flux}. In particular, note the upward
turn of the photon index at $L_X\sim0.1 L_{\rm Edd}$ in
Fig.~\ref{gamma-flux}.   

\citet{becker12} also showed that as the luminosity increases, the
height of  the emission zone inside the accretion column approaches the
neutron star  surface in the subcritical accretion state, while this
emission zones moves  upward in the supercritical accretion state.
Comptonisation occurs in the region between the radiative shock and
the neutron star surface (the sinking region). In the supercritical state
this region is relatively small, just a few tens of meters \citep[see
eq.~(40) in][]{becker12}, but its height increases as the luminosity
increases. In this region, the effective velocity of the comptonising
electrons is highly reduced because diffusion (outwards) and advection
(inwards) are almost balanced. Hence photons would not acquire enough
energy through bulk Comptonisation to populate the higher energy band and a
softening of the spectrum would be expected with increasing luminosity.
This agrees with the observations: in the DB, the spectrum becomes {\em
softer} (soft colour decreases, $\Gamma$ increases) as the luminosity
increases (Figs.~\ref{rate-SC} and \ref{gamma-flux}).  In the subcritical
mode, the typical emission height is a few kilometers  \citep[see eq.~(51)
in][]{becker12} but decreases as the X-ray luminosity increases. As the
size of the sinking region decreases with increasing luminosity, the
optical depth increases, which results in harder photons.  In the HB, the
spectrum becomes {\em harder} as the luminosity increases.

Finally, it is worth noticing that the four sources that display the two
states have short orbital ($P_{\rm orb} < 50$ days) {\em and} spin ($P_{\rm
spin} <<100$ s) periods, whereas the sources that do not transit to the DB
have, in general, $P_{\rm orb}> 100$ days {\em or} $P_{\rm spin} \simmore
100$ s (see Table~\ref{info}). The result that systems harbouring
fast-rotating neutron stars in narrow orbits display two branches in the HID
can be linked to the work by \citet{reig07}, who found that \bex\  with
short spin and orbital periods are more variable in the X-ray band, i.e.,
they exhibit higher amplitude of variability as measured by the
root-mean-square $rms$ over the long-term (years) X-ray light curves.
\citet{reig07} explained this result in the context of the viscous
decretion disc model \citep{okazaki01}. This model predicts the truncation
of the Be star's disc due to the tidal interaction exerted by the neutron
star. In the truncated disc model, the material lost from the Be star
accumulates and the disc becomes denser more rapidly than around an
isolated Be star \citep{okazaki02}. Truncation is favoured in systems with
short orbital periods and low eccentricities. In such systems mass transfer
would occur only when the disc is strongly disturbed. Accretion of large
amount of material from the distorted disc would give rise to very bright
X-ray outbursts.



\subsection{Correlation of spectral and timing parameters}
\label{spec-tim}

We searched for correlations between the different model parameters that
describe the energy and power spectra. Two approaches were taken to
investigate the relation between the X-ray emission and the rapid aperiodic
variability. In the first one, we tracked the spectral and timing parameters
following an X-ray outburst, in which case the general trend of the
relation with luminosity was inferred (Figs.~\ref{gamma-flux},
\ref{iron-flux},  \ref{cyc-flux}, \ref{freq-flux} and  \ref{rms1-flux}). In
the second approach, we directly compared parameters between themselves
(Figs.~\ref{ecut-gamma} and \ref{freq-gamma}). 
To quantify and assess the statistical significance of the various
relationships we obtained the Pearson's correlation coefficient. The
results are summarised in Table~\ref{pearson}.


Of particular interest are the correlations between spectral and timing
parameters. According to the model by \citet{becker07}, the
hard power-law component is due to bulk inverse Comptonisation in the
accretion column. Low-energy photons are upscattered in the shock and
eventually diffuse through the walls of the column.  Therefore, the photons
that populate the hard power-law tail come from a region close to the
surface of the neutron star (length scales $\sim10^6$ cm).

On the other hand, aperiodic variability must be generated at much larger
distance. The evidence in support of this statement comes from the
correlation between the characteristic frequency of the broad-band noise
and the X-ray flux and from the milliHertz range of frequencies of the QPO
seen in accreting X-ray pulsars.

 The radius of the magnetosphere,
$r_m$, depends on the mass accretion rate \citep{davies81}

\begin{equation}
r_m = 1.2 \times 10^9 \dot M_{15}^{-2/7} \, \mu_{30}^{4/7} \,
	\left(\frac{M}{\msun}\right)^{-1/7} \;\;\; {\rm cm} 
\end{equation}

\noindent where $\dot M_{15}$ is the mass accretion rate in units of
$10^{15}$ g s$^{-1}$ and $\mu_{30}$ the dipole magnetic moment in units of
$10^{30}$ G cm$^3$. As the X-ray flux increases, presumably as a
consequence of an increase in the mass accretion rate, the magnetospheric
radius decreases. If the processes that originate the broad-band noise are
linked to motion of matter outside the magnetosphere, then we would expect
that as the magnetosphere shrinks, the characteristic  frequency increases,
as observed (Fig.~\ref{freq-flux}). Thus this correlation  is consistent
with and extra-magnetospheric origin of the aperiodic variability.

Likewise, QPOs in HMXB lie typically in the milliHertz range
\citep{james10}. If one assumes that QPOs are produced as a result of
Keplerian motion of inhomogeneities in an accretion disc, this frequency
range (0.01-1 Hz) agrees with the QPO being originated outside (but near)
the magnetosphere, i.e., at length scales
$r_m\simmore 10^8$ cm.  Indeed, a Keplerian frequency of $\nu_{QPO}=100$
mHz would correspond to a radius $r=(GM/4\pi^2\nu_{QPO}^2)^{1/3}\approx 7.8
\times 10^8$ cm. Note also that in LMXBs, peaked (broad-band) noise have
been seen to developed into QPOs, indicating a physical relationship
between broad-band noise and QPOs \citep{klis06}. Thus it is reasonable to
assume then that the aperiodic variability originates from the same
physical region as the QPOs.  

Since the power-law photons and those responsible for the aperiodic
variability originate in very different physical regions, we would not
expect the spectral and timing parameters to correlate.  We have plotted
the relationship between the power-law photon index and the frequency of
the main noise component $L_1$ in Fig.~\ref{freq-gamma}. Despite the
different location of the production mechanisms,  a distinct correlation
between the power-law index and characteristic frequency is seen in almost
half of the sources.  Mimicking the transition from a negative to a
positive correlation seen in the $L_X-\Gamma$ diagram as the flux increases
(Fig.~\ref{gamma-flux}), the sign of the correlation  between $\nu_1$ and
$\Gamma$ also changes, from negative in the HB to positive in the DB
(Fig.~\ref{freq-gamma}). Whether this result is simply the consequence
of the strong correlation between the characteristic frequency and the X-ray
luminosity (Fig.~\ref{freq-flux}), {\it i.e.,} the frequency acts as a
proxy for luminosity, or else it has a more profound physical meaning remains
to be solved by further studies.

\section{Conclusions}
\label{con}

We have investigated the correlated X-ray timing and spectral variability
of accreting X-ray pulsars with Be-type companions during major X-ray
outbursts. The aim was to investigate whether this type of systems exhibit
source states and in case they do whether the X-ray timing and spectral
properties are strongly correlated  as it is seen in neutron-star low-mass
and black-hole X-ray binaries. Although there may be some discrepant
behaviour in particular sources, which substantiates the complexity of the
accretion process, we were able to extract general patterns of
variability. 

The evolution through the hardness-intensity diagram suggests that in the
early and late phases of the outburst,  \bex s undergo state transitions.
As the source evolves along the outbursts it transits from the horizontal
branch to the diagonal branch and back.  We show that the state
transitions occur when a critical luminosity is reached.  Only sources
whose peak luminosity is well above the critical limit do exhibit the two
branches. Lower than the critical luminosity sources display only the
horizontal branch. Because the value of the critical luminosity depends on
the pulsar magnetic field, the luminosity at which the transition takes
place varies across different sources.  For typical values of the magnetic
field in Be/X-ray pulsars, the critical luminosity is expected to be of the
order of a few times $10^{37}$ erg s$^{-1}$. 

\citet{becker12} showed that two different accretion regimes can
explain the bimodal behaviour of the cyclotron line with luminosity
displayed by different sources. We propose that the two branches correspond
to these two accretion regimes, likely to be related to the way in which
the accretion flow is decelerated in the accretion column:  radiation
pressure in the supercritical regime (diagonal branch) and coulomb
interactions in the subcritical regime (horizontal branch). In this work we
show that the continuum traces the two regimes as well. Despite the complex
spectral continuum, the power-law photon index correlates with X-ray
luminosity and  also exhibits two branches. When plotted as a function of
X-ray luminosity, the correlation changes sign in coincidence with the
change of state. In the horizontal branch, the power-law photon index
decreases with X-ray flux, while it increases  in the diagonal branch. We
speculate that this behaviour is due to the different dependence of the
accretion column height with flux in the two accretion regimes. 

In contrast, although a clear positive correlation between the frequency of
the broad-band noise components with flux is observed, the frequency does
not trace the two branches. This result would support the idea that the
aperiodic variability originates further away from the neutron star
surface, outside the accretion column.

We report for the first time a positive correlation between the energy of
the cyclotron resonant scattering feature in \vtau\ with luminosity on long
time scales. The energy of the cyclotron line is seen to start to increase
at a higher luminosity than other sources, presumably due to its larger
magnetic field. This is only the second X-ray pulsar (after Her X--1) to
display such positive correlation. 

Finally, some sources show a  correlation between the photon index and
the characteristic frequency of the aperiodic noise components, implying
that the accretion column, where energy spectra are generated somehow
communicates with the inner accretion disc, where the aperiodic variability
is supposed to originate. This result imposes a tight constrain to the
models that seek to explain the spectral and timing variability in
accretion-powered pulsars.



\begin{acknowledgements}

The authors thank P. Kretschmar, D. Klochkov and J. Wilms for their assistance
and useful comments. PR acknowledges support by the Programa Nacional de
Movilidad de Recursos Humanos de Investigaci\'on 2011 del Plan Nacional de
I-D+i 2008-2011 of the Spanish Ministry of Education, Culture and Sport. PR
also acknowledges partial support by the COST Action
ECOST-STSM-MP0905-020112-013371. EN acknowledges a ``VALi+d'' postdoctoral
grant from the ``Generalitat Valenciana" and was supported by the Spanish
Ministry of Economy and Competitiveness under contract AYA 2010-18352. 
This research has made use of NASA's Astrophysics Data System Bibliographic
Services and of the SIMBAD database, operated at the CDS, Strasbourg,
France. The ASM light curves were obtained from the quick-look results
provided by the ASM/RXTE team.

\end{acknowledgements}

\end{document}